\documentclass[5p,twocolumn]{elsarticle}

\usepackage{lineno,hyperref}
\usepackage{graphicx}	
\usepackage{subcaption} 
\usepackage{amsmath}	
\usepackage{amssymb}	
\usepackage[stable]{footmisc} 
\usepackage{colortbl}
\usepackage{caption}
\usepackage{rotating}

\journal{Physics of the Dark Universe}

\DeclareUnicodeCharacter{223C}{~}
\begin{document}

\begin{frontmatter}

\title{A single parameterization for dark energy and modified gravity models}

\author[1,2]{Mariana Jaber\fnref{orcidM}}
\ead{jaber@cft.edu.pl}

\author[3,4]{Gustavo Arciniega\fnref{orcidG}}
\ead{gustavo.arciniega@ciencias.unam.mx}

\author[4,5]{Luisa G. Jaime\fnref{orcidL}}
\ead{luisa@ciencias.unam.mx}

\author[6]{Omar Abel Rodr\'iguez-L\'opez\fnref{orcidO}}
\ead{oarodriguez.mx@gmail.com}

\fntext[orcidM]{ORCID: https://orcid.org/0000-0001-7507-9516}
\fntext[orcidG]{ORCID: https://orcid.org/0000-0002-9960-2882}
\fntext[orcidL]{ORCID: https://orcid.org/0000-0003-0159-8959}
\fntext[orcidO]{ORCID: https://orcid.org/0000-0002-3635-9248}

\address[1]{Institute for Astronomy, Faculty of Physics, Astronomy and Informatics, Nicolaus Copernicus University, Grudziadzka 5, 87-100 Toru\'n, Poland.}
\address[2]{Center for Theoretical Physics, Polish Academy of Sciences, Al. Lotników 32/46, 02-668 Warsaw, Poland.}
\address[3]{Centro Tecnol\'ogico Arag\'on, Universidad Nacional Aut\'onoma de M\'exico, 
Av. Rancho Seco S/N, Bosques de Arag\'on,\\ Nezahualc\'oyotl, 
Estado de M\'exico, 57130, M\'exico.}
\address[4]{Departamento de F\'{\i}sica, Facultad de Ciencias, Universidad Nacional Aut\'onoma de M\'exico, Apartado Postal 50-542, CDMX, 04510, M\'exico.}
\address[5]{Departamento de F\'{\i}sica, Instituto Nacional de Investigaciones Nucleares, Apartado Postal 18-1027, Col.  Escand\'on, CDMX, 11801, M\'exico.}
\address[6]{Instituto de F\'{\i}sica, Universidad Nacional Aut\'onoma de M\'exico, Apartado Postal 20-364, CDMX, 01000, M\'exico.}

\begin{abstract}
Perhaps the most explored hypothesis for the accelerated cosmic expansion rate arises in the context of extra fields or modifications to General Relativity. A prevalent approach is to parameterize the expansion history through the equation of state, $\omega(z)$. We present a parametric form for $\omega(z)$ that can reproduce the generic behavior of the most widely used physical models for accelerated expansion with infrared corrections. The present proposal has at most 3 free parameters which can be mapped back to specific archetypal models for dark energy. We analyze in detail how different combinations of data can constrain the specific cases embedded in our form for $\omega(z)$. We implement our parametric equation for $\omega(z)$ to observations from CMB, the luminous distance of SNeIa, cosmic chronometers,  and baryon acoustic oscillations identified in galaxies and in the Lymann-$\alpha$ forest. We find that the parameters can be well constrained by using different observational data sets. Our findings point to an oscillatory behavior consistent with an $f(R)$-like model or an unknown combination of scalar fields. When we let the three parameters vary freely, we find an EoS which oscillates around the phantom-dividing line, and, with over 99$\%$ of confidence, the cosmological constant solution is disfavored. 
\end{abstract}


\end{frontmatter}


\section{Introduction}

Ever since the discovery of the acceleration of the Universe \cite{Perlmutter:1998np, Riess:1998cb} (hinted previously in \cite{Roukema:1993yra}), cosmology has tried to answer the question of what makes the Universe accelerate.
Currently, the most accepted explanation by the scientific community is the cosmological constant, $\Lambda$, in the frame of General Relativity with an FLRW metric, which has become the concordance model known as $\Lambda$CDM \cite{Aghanim:2018eyx}. 
Several phenomena can be explained by using such a simple model; nevertheless, the physical nature of $\Lambda$ remains unaddressed. 

In the past few years, observations of different astrophysical sources have been used to measure the acceleration of the Universe. The results have brought with them even more uncertainty about the nature of dark energy. They show a discrepancy on the present value of the Hubble parameter derived when local measurements are used \cite{Riess:2019cxk, Wong:2019kwg, Shajib:2019toy} with the $H_0$ value when derived by fitting the cosmological parameters assuming the concordance model in the cosmic microwave background (CMB) \cite{Aghanim:2018eyx}. Different estimations of the discordance place the discrepancy as high as 4.4-$\sigma$ \cite{Riess:2019cxk} or even at the level of  5.3-$\sigma$, according to \cite{Wong:2019kwg} (see \cite{Verde:2019ivm} for a summary plot).

A systematic miscalculation may be behind this conundrum; nevertheless, the possibility of having some new physics is provocative. Many alternatives to the standard concordance model have been proposed (for a review of several of these alternatives see \cite{Zumalacarregui:2020cjh, Bamba:2012cp}, for a review about the current status of different alternative models see \cite{DiValentino:2021izs}). One alternative to explore the evolution of the Universe in a model-independent fashion way is by setting a parametric form for the equation of state (EoS), $\omega\equiv P/\rho$ of the dark energy component. 

In this framework, several parameterizations of the EoS have been proposed with the idea of simplifying the analysis of observations. However, we find either a lack of physical motivation or a severe dependence on a particular model.

Interestingly enough, in an observational effort carried out by  \cite{Zhao:2017cud}, a reconstruction of the EoS was presented. The evolution found by the authors shows oscillating behavior around the phantom line. 
It is well known that such evolution can not be provided by using a single scalar field (either phantom or quintessence like). 
Nevertheless, modified gravity or some unknown combination of multiple scalar fields could provide the reconstructed EoS. 
See \cite{Colgain:2021pmf} for an interesting discussion on dynamical dark energy.

Currently, different kinds of parameterizations provide dynamical dark energy. Some of them are motivated by scalar fields, and the work of \cite{Jaime:2018ftn} proposes an equation inspired by modified gravity.

We present a different parameterization that has the advantage of reproducing the generic behavior for both cases, depending on the parameters' choice.
Using this parameterization with current and future data, we could test the generic evolution of the equation of state of the accelerating mechanism.

As was pointed out in a recent article \cite{Linder:2022pel}, the cosmic expansion and growth history of large-scale structures can have independent behaviors for models beyond the standard one, then the analysis of each part should be performed separately. 
In the present work, we focus our analysis on the evolution of the background only. 
This way, we obtain an independent behavior that can later be connected with perturbations in several manners within different frameworks of the underlying physics. 
Unlike the background expansion, the perturbative regime is not describable in a unified manner, for the physical scenarios here considered.
\\
For example, the perturbations of models such as quintessence  do not cluster, while, in general, for models of a barotropic fluid, the perturbations do not diverge in the case of a constant equation of state, $\omega_X=$ constant (\cite{Amendola:2015ksp}). 
For models crossing the phantom dividing line, different ways to avoid divergences have been explored in the literature such as the parametrized post-Friedmann mechanism (see for instance \cite{Hu:2004kh}).
For the case of $f(R)$, the perturbed equations can be seen as in \cite{DeFelice:2011hq}, which are dependent on the Ricci scalar itself through the evolution of $H$, and there is no dependence on the equation of state of the dark geometric component. 
\\
It is evident therefore that the analysis of the perturbative regime should be performed in a case by case manner, depending on the physics of the accelerating mechanism.

In Section \ref{sec:parameterizations} we review the parametric approach to model cosmic expansion at late times. Our proposal is detailed in Section \ref{sec:bestiary}, where different sub-cases are discussed in detail. 
The implementation into the background equations is given in Section  \ref{sec:cosmology}, and Section  \ref{sec:methods} describes the methods used, including the data sets chosen and our statistical analysis. 
Our results are included in Section \ref{sec:results}, while Section  \ref{sec:aicbic} discusses the model comparison analyzes. 
Our conclusions can be found in Section \ref{sec:conclusions}. We left the propagation of uncertainties in the equation of state for \ref{app:deltaomega}, and the description of the numerical code in \ref{app:code}.


\section{Parameterising the cosmic expansion}
\label{sec:parameterizations}

The parameterizations of $\omega(z)$ in the literature are either mere mathematical descriptions, polynomials, or Taylor expansions around $a_0$ (or $z$), or an attempt to capture distinctive features for particular models. 
In \cite{Chevallier2001}, the authors perform a Hamiltonian analysis that provides physical tools to build their proposal for $\omega(z)$. 
They aim to explore slight deviations of the cosmological constant.
In light of the recent  Hubble tension, our motivation is reproducing predictions given by different alternative physical scenarios while avoiding inherent theoretical complications  for implementing certain models. 
This way, the parameterizations mimicking the well-supported evolution of the EoS can provide a more straightforward way to study complicated theories.

The standard approach is to take model by model and constrain the introduced free parameters against data to obtain conclusions for the chosen physical scenario.
Instead of choosing a parametric form for each model and performing the statistical analysis case by case, we propose a single framework. Our proposal allows for analyzing the generic behavior of the most widely used physical models for accelerated expansion with, at most, three parameters.  
Our proposal includes alternative models that make modifications for a late time while maintaining the EoS $\omega=-1$ at high redshift. 
With this in mind, we introduce our proposal for $\omega(z)$, explain its mathematical properties and describe its capabilities to mimic archetypal models for the accelerated expansion. A particularly interesting question for us is: do observations point to a $\omega(z)$ which crosses the phantom-line? 

Our proposal mimics and analyzes two paradigmatic scenarios: $f(R)$ modified gravity and quintessence/phantom models. However, we will see that the parameterization is not restricted to only these two cases.

We focus on $f(R)$ theories of gravity because they are very straightforward modifications of General Relativity and have been widely studied over the past twenty years (see for instance \cite{Jaime:2012gc} and references therein).
In this kind of modification, the Ricci scalar's dependence $R$ in the Hilbert-Einstein action is not linear; an arbitrary function of $R$ replaces it. 
Several $f(R)$ models have been proposed to provide an alternative explanation to the acceleration of the Universe. 
In \cite{Jaime:2013zwa} the definition of the EoS for the geometric dark energy is discussed, and it was shown that, in general, the generic evolution of $\omega(z)$ for the $f(R)$ models that are considered candidates for dark energy is oscillatory. 
In \cite{Jaime:2018ftn} it was presented a parameterization for the EoS in $f(R)$ that can reproduce in a very high precision the numerical results for some $f(R)$ models. Nevertheless, the proposal might suffer an oscillatory behavior at high-z values, which might introduce numerical errors in Boltzmann codes.
Using the present proposal, we avoid this numerical misbehaves while maintaining the generic behavior.
\\
Regarding scalar fields, the most popular proposal to provide an alternative explanation to the acceleration of the Universe is quintessence models. 
Such models can be separated into two kinds: ``thawing out'' and ``freezing in'', depending on if the slope when is going to $z=0$ is positive or negative \cite{Caldwell:2005tm}. 
In \cite{Roy:2018nce}, the authors presented a parameterization for several models directly in the scalar field. 
The present proposal recovers the generic evolution of the cases presented in \cite{Roy:2018nce}. Even more, as we elaborate in the following section, this can be accomplished by fixing one or two out of three free parameters.


\section{One parameterization to fit them all}
\label{sec:bestiary}
The parameterization we are proposing for the EoS, $\omega(z)$, is the following: 

\begin{equation}
    \label{eq:eos}
    \omega(z) = -1 - A\exp(-z)\left(z^{n}-z-C\right),
\end{equation}
where $A$, $n$ and $C$ are real numbers that can take positive or negative values. 

A quick inspection let us notice that the present-day value is given by $ \omega(z=0)\equiv \omega_0= -1 +AC$, while the high-redshift value rapidly converges to $\omega(z\gg0)=-1$, corresponding to a cosmological constant $\Lambda$ scenario avoiding high redshift divergences that can be present in other parameterizations \cite{Huterer:2000mj, Weller:2001gf, Jaber:2017bpx}.  
\\
Given the form of equation (\ref{eq:eos}), it is possible to mimic different dynamical dark-energy EoS, which can be characterized by, at most,  a single oscillation of the $\omega(z)$ at low redshifts.

We identify four well-posed cases: exponential, quintessence/phantom, $f(R)$, and general, besides the standard case of a cosmological constant. In the following subsections, we provide analytical analysis to explain the model's flexibility.

\subsection{Exponential: $n=C=1$  (one-free parameter).}
The simplest case of (\ref{eq:eos}) is when $n=1$, so that the parameterization reduce to the expression:
\begin{equation}
\label{CaseI}
\omega(z)=-1+ACe^{-z}.
\end{equation}
Without lost of generality, we can fix $C=1$, that is equivalent to rename $\tilde{A} \equiv AC$. Otherwise, the parametrization will present a degeneration effect given by the product $AC$.

Nevertheless, it is possible to avoid degeneration if the parameterization, for this particular case, takes just one parameter instead of two. 

In this case, the generic behavior is exponential. 

\begin{figure}
\begin{center}
\includegraphics[width=\linewidth]{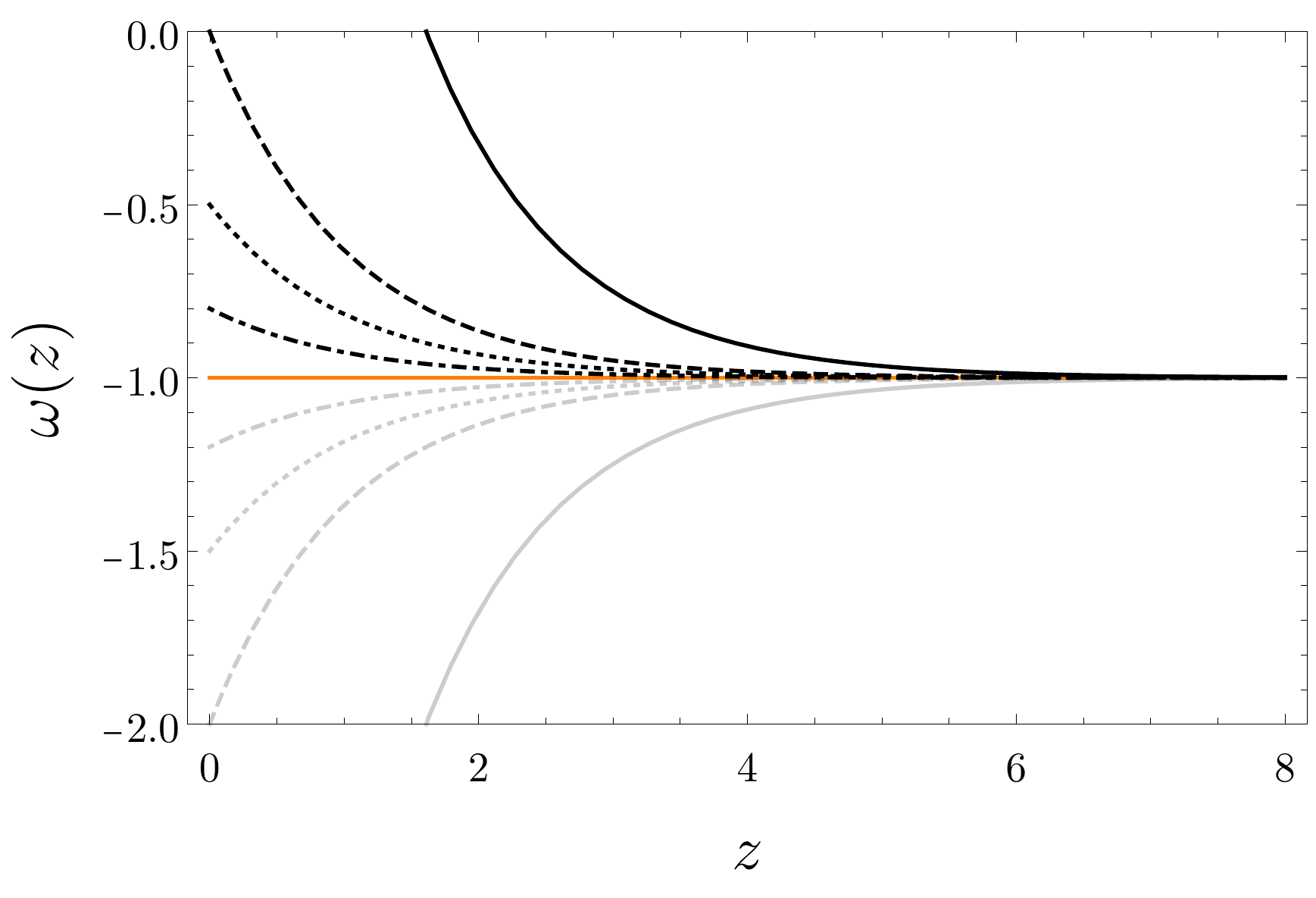}
\caption{ Exponential: $n=1$, equation (\ref{CaseI}) taking $\tilde{A}\equiv AC$.
Solid black line $\tilde{A}=5$, dashed black line $\tilde{A}=1$, dotted black line $\tilde{A}=0.5$, and dot-dashed black line $\tilde{A}=0.2$. 
Gray lines are the same as black lines but with $\tilde{A}\rightarrow -\tilde{A}$.}
\label{Fig:CaseI}
\end{center}
\end{figure}

Figure~\ref{Fig:CaseI} shows the evolution of the EoS for different values of the amplitude: $\tilde{A}= \pm 0.2, \, \pm 0.5, \, \pm 1, \, \pm 5$,  where the black lines represent the positive values of $\tilde{A}$, while the grey lines show the negative  $\tilde{A}$ values. 

From \ref{Fig:CaseI} we see that the value of $\omega$ at $z=0$ is different from $-1$ if $\tilde{A}\neq0$. 
Also, the evolution of $\omega(z)$ goes monotonically to the asymptotic value $\omega=-1$ for some $z>0$. 
How fast it converges to $-1$ depends on the value of $\tilde{A}$. 
This parameter controls the present value of the equation of state and the epoch $z$, for which $\omega(z)$ is practically $-1$.
In order to formally express the value of $z$ where we can consider that the EoS has reached the value $-1$, let us consider $|\epsilon | \ll 1$ so that $\omega(z)=-1+|\epsilon |=-1+\tilde{A}e^{-z}$.  
From here, we define $\tilde{z}\equiv ln(\tilde{A}/|\epsilon |)$ as the redshift  such that $\omega(\tilde{z})\simeq -1$, up to an $|\epsilon |$ for large $z$.

This characteristic will play a role in distinguishing this from other study cases. 
In particular, we anticipate that although the evolution within this model (Exponential) can be similar to the one obtained in the particular cases II and IV (compare figure~\ref{Fig:CaseI} to figures~ \ref{Fig:caseIIb} and \ref{fig:BeastIVa}), it is the value of $ \omega_{0}$ and the asymptotic relaxation to $\omega(z)=-1$ which can potentially distinguish between them observationally.


\subsection{Quintessence/Phantom-like: $n=0$ (two-free parameters).}

\begin{figure}
\begin{center}
\includegraphics[width=\linewidth]{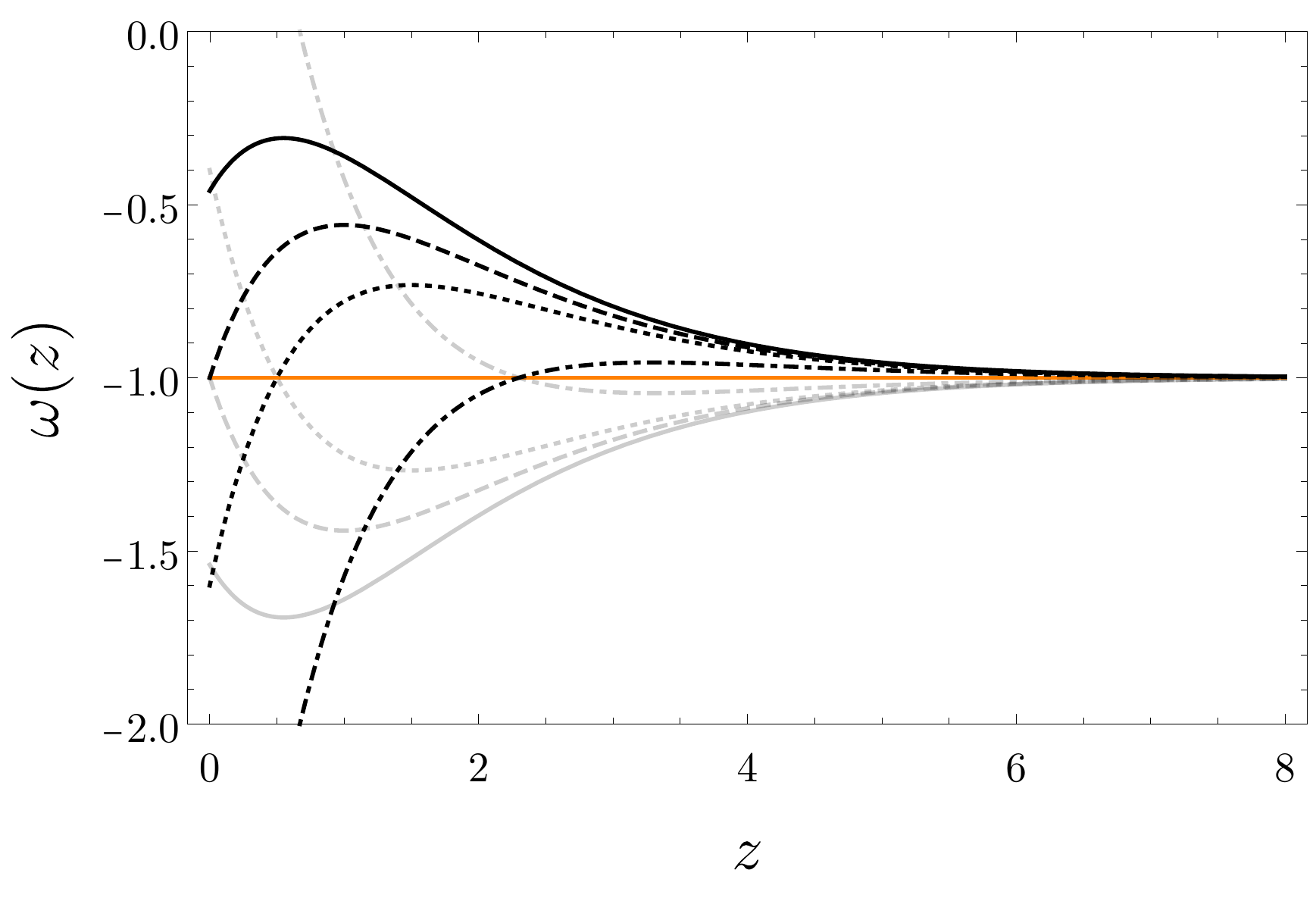}
\caption{Equation of state $\omega(z)$ for $n=0$, Quintessence/Phantom-like case: The amplitude for all the curves is  $A=1.2$. The solid, dashed, dotted, and dot-dashed black lines correspond to $C=1.45$, $C=1$, $C=0.5$ and $C=-1.3$, respectively. Gray lines are the equivalent to black lines for $A\rightarrow -A$.}
\label{Fig:CaseIIa}
\end{center}
\end{figure}

\begin{figure}
\begin{center}
\includegraphics[width=\linewidth]{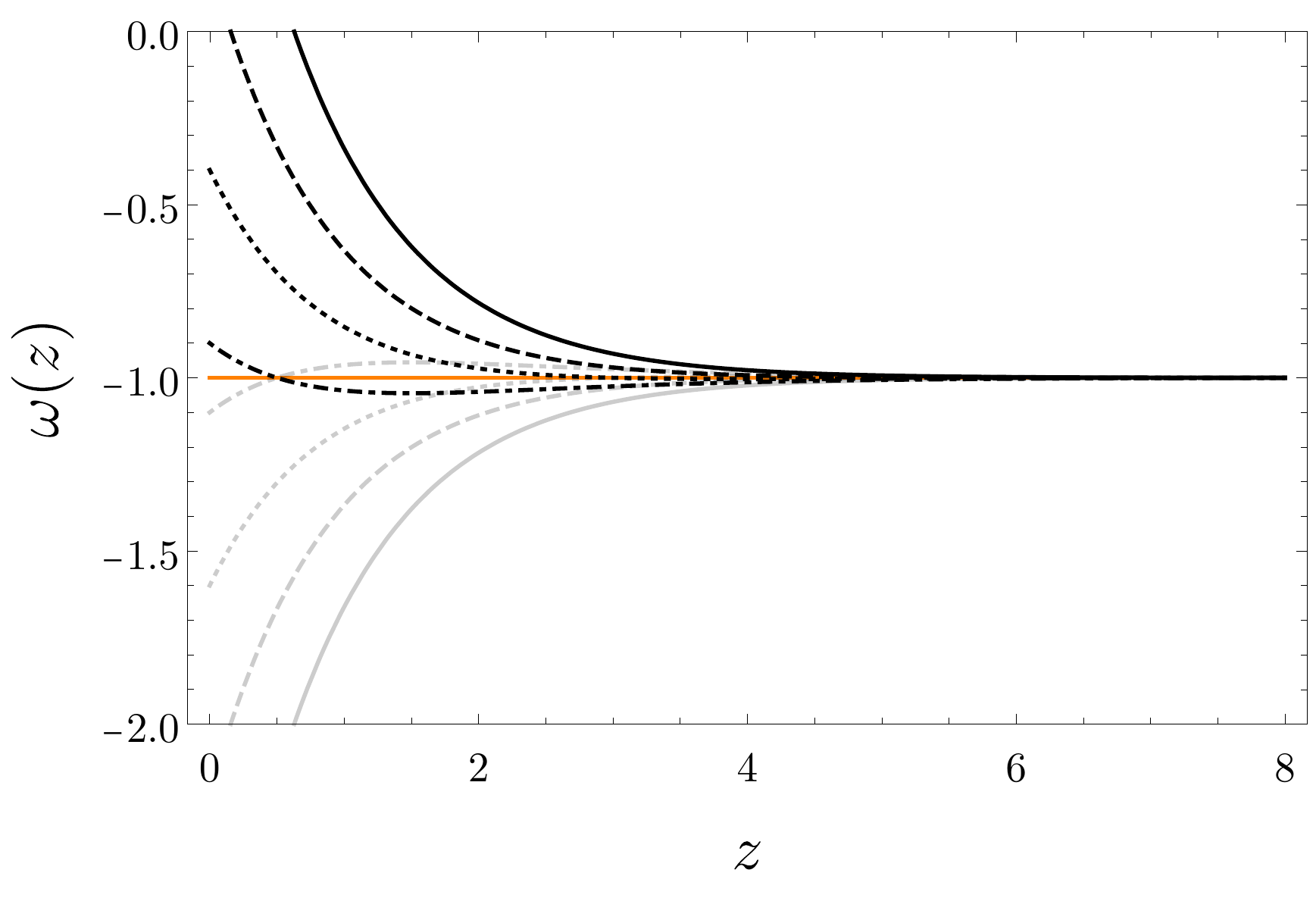}
\caption{Equation of state  $\omega(z)$ for the Quintessence/Phantom-like case, $n=0$: The amplitude $A$ is fixed for all the curves, $A=-0.2$. The solid black line is at $C=-9$, the dashed black line is at $C=-5$, the dotted black line is at $C=-2$, and the dot-dashed black line corresponds with $C=0.2$. Grey lines are the same as black lines but with $A\rightarrow -A$.}
\label{Fig:caseIIb}
\end{center}
\end{figure}

In order to mimic the shape of the EoS for Quintessence or Phantom models we fix the parameter $n=1$ in (\ref{eq:eos}), the EoS can be written then as,
\begin{equation}
\label{CaseII}
\omega(z)=-1-Ae^{-z}[(1-C)-z],
\end{equation} 
where $\omega(z)$ has a single minimum/maximum value located at $z=2-C$. The generic evolution of the EoS is depicted in figure~(\ref{Fig:CaseIIa}), where it can be seen how the parameterization is able to transit from a Quintessence-like profile \cite{Roy:2018nce} to a mixture of Quintessence and phantom-like fields, known as Quintom (for a review of this models, see \cite{Cai:2009zp}).
In order to visualize this case, in figure~\ref{Fig:CaseIIa}, we have fixed the amplitude for all the curves to $A=1.2$. 
The solid black line is when $C=1.45$, the dashed black line is when $C=1$, the dotted black line is when $C=0.5$, and the dot-dashed black line corresponds to $C=-1.3$. Grey lines are the same as black lines with $A\rightarrow -A$.
When $C\geq 2$, $\omega(z)$ looks monotonic in the range $z>0$, the minimum (maximum) will be located somewhere in the future $z<0$. 
This way, what we are observing, at $z>0$ is the evolution going up (down) from that minimum (maximum). 
In this case, the behavior will be similar to the one presented in the case $n=1$. 
It is important to remark that although the profiles depicted in figure~\ref{Fig:CaseI}, and \ref{Fig:caseIIb} look similar, the analytic form is entirely different, so it is guaranteed that there is no degeneration with the Exponential-like parameterization case. 
Even more, in this case $ \omega_{0}=-1$ if and only if $C=1$, so any deviation of $C=1$ will modify the value of $ \omega_{0}$ around $-1$. 
In the case that $C\leq 2$, the minimum (maximum) will be located at $z>0$ so the evolution of the EoS will cross the phantom line a single time, and the behavior will have the characteristic shape that is expected in the Quintom models (see the dot-dashed line in figure~\ref{Fig:caseIIb}).


\subsection{$f(R)$-like: $n\in(0,1)$ and $C=0$ (two-free parameters).}

In this case the parameter $n$ in (\ref{eq:eos}) takes any value between $0<n<1$, while the parameter $C$ is fixed to $C=0$. In this way we are imposing $ \omega_{0}=-1$. 
Equation (\ref{eq:eos}) can be written as 

\begin{equation}
\label{CasefR}
\omega(z)=-1-Ae^{-z}(z^n-z).
\end{equation}

If $n\neq \{0,1\}$, equation (\ref{CasefR}) will have at most two real roots. 
This allows an oscillatory behavior for $z>0$ with a fixed point $ (\omega_{0}=-1)$ that converges to $\omega(z)=-1$ for $z>>1$. 

\begin{figure}[h]
\begin{center}
\includegraphics[width=\linewidth]{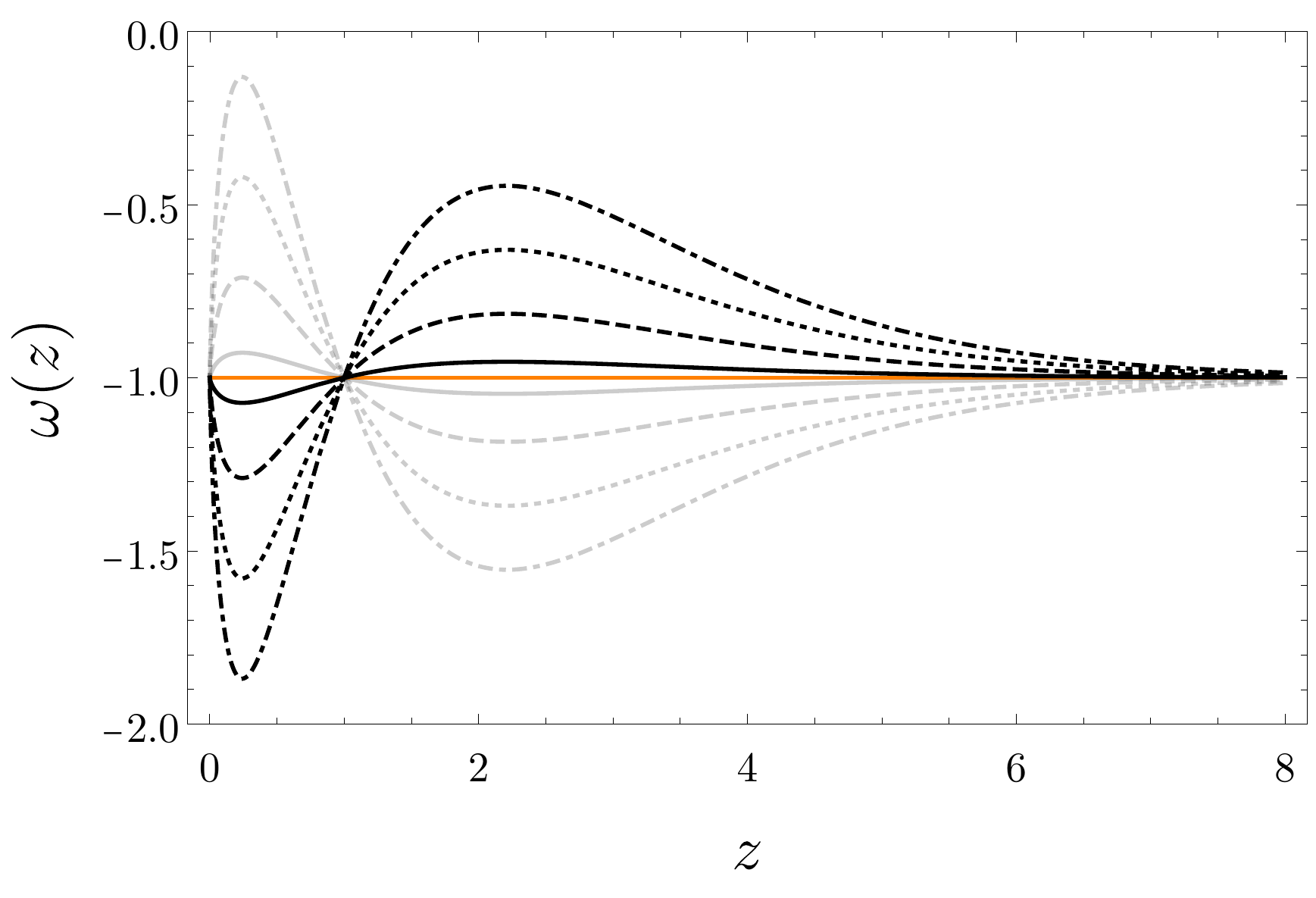}
\caption{Equation of state for the $f(R)$-like case: $C=0$. $n=0.9$ fixed for all the curves. Solid, dashed, dotted and dot-dashed black lines have $A=2.5$, $A=10$, $A=20$, and $A=30$ respectively. Gray lines are the same as black lines with $A\rightarrow -A$.}
\label{fig:BeastIIIa}
\end{center}
\end{figure}

\begin{figure}[h]
\begin{center}
\includegraphics[width=\linewidth]{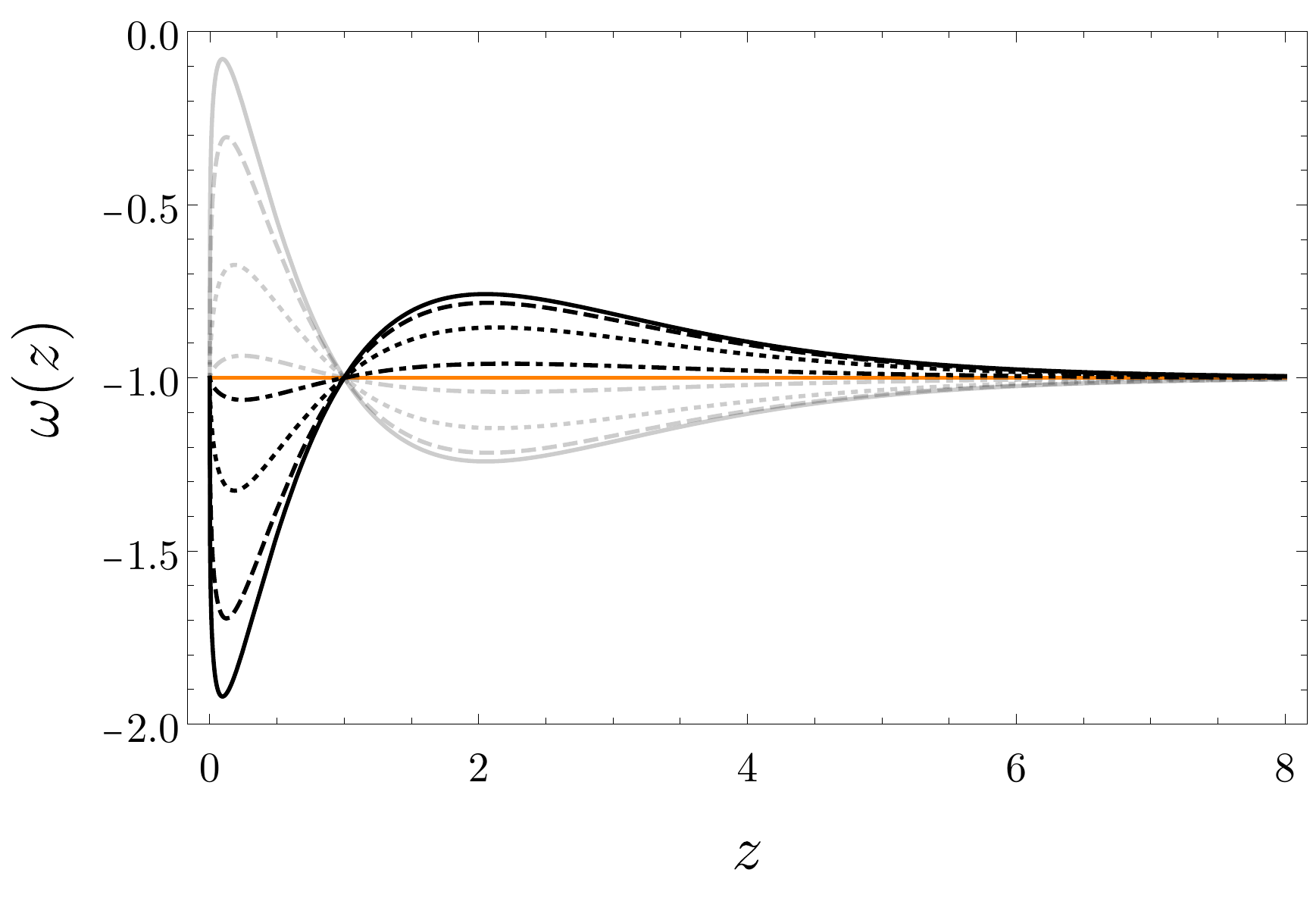}
\caption{Equation of state for the $f(R)-like$ case: $C=0$. The amplitude is $A=2.2$ for all the cases. Solid, dashed, dotted and dot-dashed black lines correspond to $n=0.25, \, 0.35, \, 0.6, \, 0.9$ respectively. Gray lines are the same as black lines with $A\rightarrow -A$.}
\label{fig:BeastIIIb}
\end{center}
\end{figure}

In order to depict a clear idea of this case, we make two plots of the generic behavior of the equation (\ref{CasefR}). 
In figure~\ref{fig:BeastIIIa} we fix $n=0.9$ while the amplitude has a value $A= \pm2.5, \, \pm10, \, \pm20, \, \pm30$. 
In figure~\ref{fig:BeastIIIb}, the amplitude is fixed to $A=\pm2.2$, and the parameter $n$ takes different values. 

The characteristic shape of $\omega_X$, in the $f(R)$ theories, oscillates around the phantom line, as is shown in figure 1 of reference \cite{Jaime:2013zwa}. This behaviour is generic for this kind of modified gravity in the frame of cosmology. In equation \ref{CasefR}, the evolution has the oscillatory feature that we are looking for in order to mimic $f(R)$ gravity  
with $\omega_0=-1$, depicted in figures \ref{fig:BeastIIIa} and \ref{fig:BeastIIIb}.

\subsection{General-model: $n\in(0,1)$ (three-free parameters) }\label{beastIV}

We will now consider the EoS (\ref{eq:eos}), for $n$ taking values in between the previous cases, \textit{i.e.}

\begin{equation}\label{eq:full-model}
\omega(z)=-1-Ae^{-z}\left( z^n-z-C\right),\quad 0<n<1.
\end{equation}

This equation (\ref{eq:full-model}) is the one we will be referring to as 'General-model' from now on. 

\begin{figure}[h]
\begin{center}
\includegraphics[width=\linewidth]{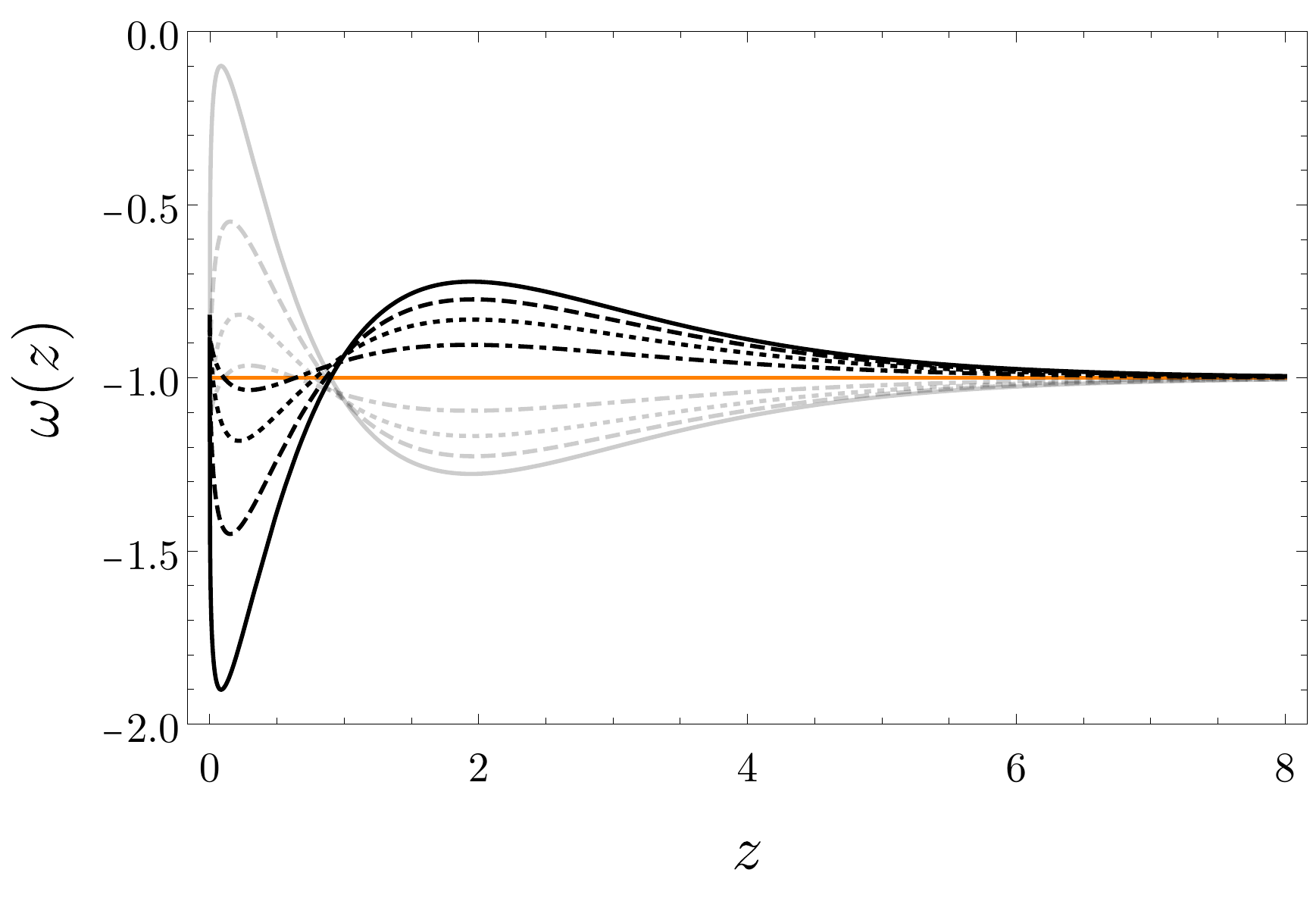}
\caption{General-model, equation (\ref{eq:full-model}). 
The amplitude and the constant value are fixed for all curves, $A=2.2$ and $c=0.08$. 
The solid black line corresponds with $n=0.2$, dashed black line is when $n=0.4$, dotted black line is when $n=0.6$, and dot-dashed black line is when $n=0.8$. 
Gray lines are the same as black lines but with $A\rightarrow -A$.}
\label{Fig:BeastIV}
\end{center}
\end{figure}

\begin{figure}[h]
\begin{center}
\includegraphics[width=\linewidth]{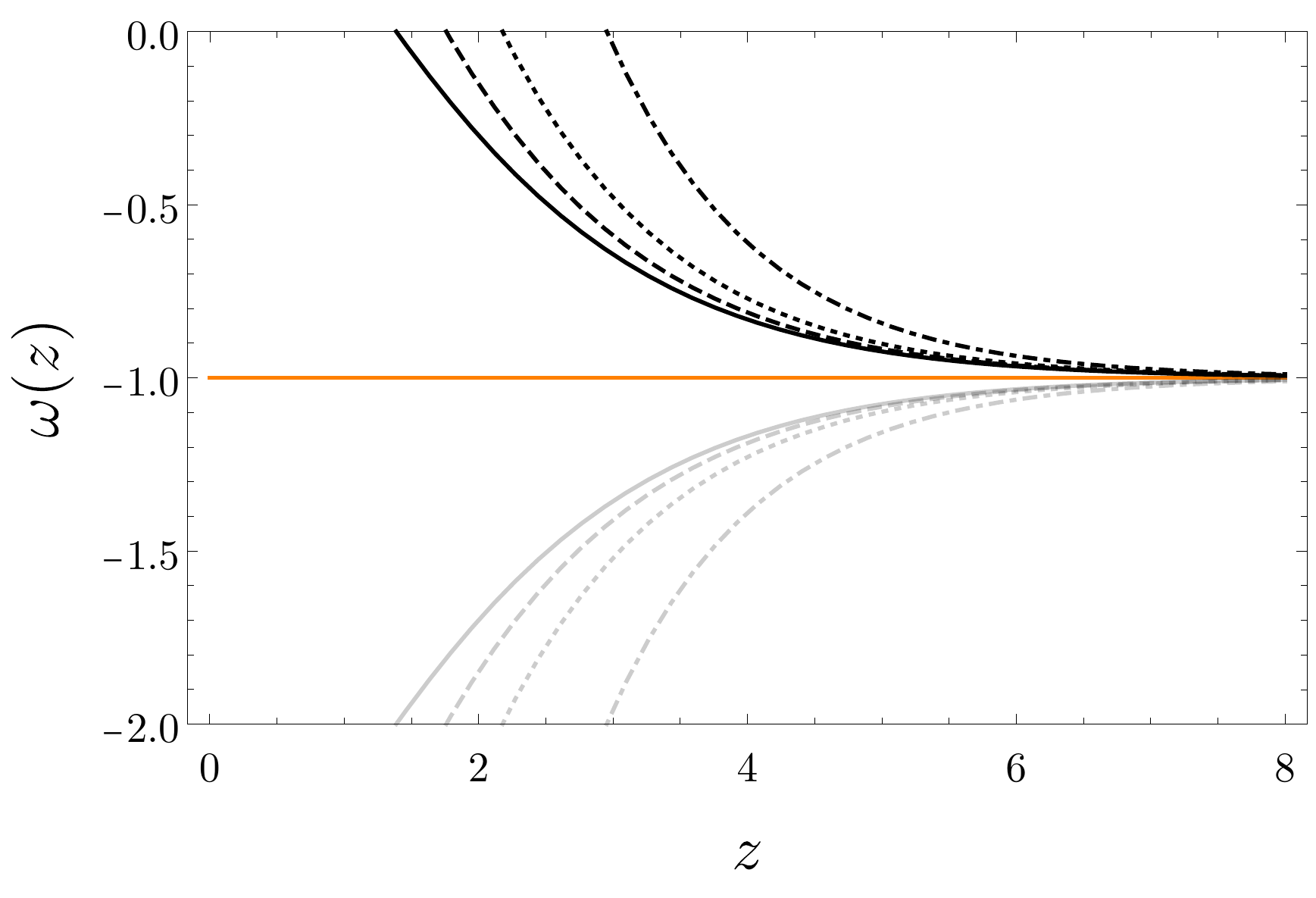}
\caption{General-model, equation (\ref{eq:full-model}). The monotonic behavior of $\omega(z)$ is achieved when the roots of $\omega'(z)$ are complex. In this case, $A=2.2$ and $n=0.2$ for all the curves. Solid black line has $C=1.5$, dashed black line has $C=2$, dotted black line has $C=3$, and dot-dashed black line has $C=7$. Gray are the same as black lines but with $A\rightarrow -A$.}
\label{fig:BeastIVa}
\end{center}
\end{figure}

\begin{figure}[h]
\begin{center}
\includegraphics[width=\linewidth]{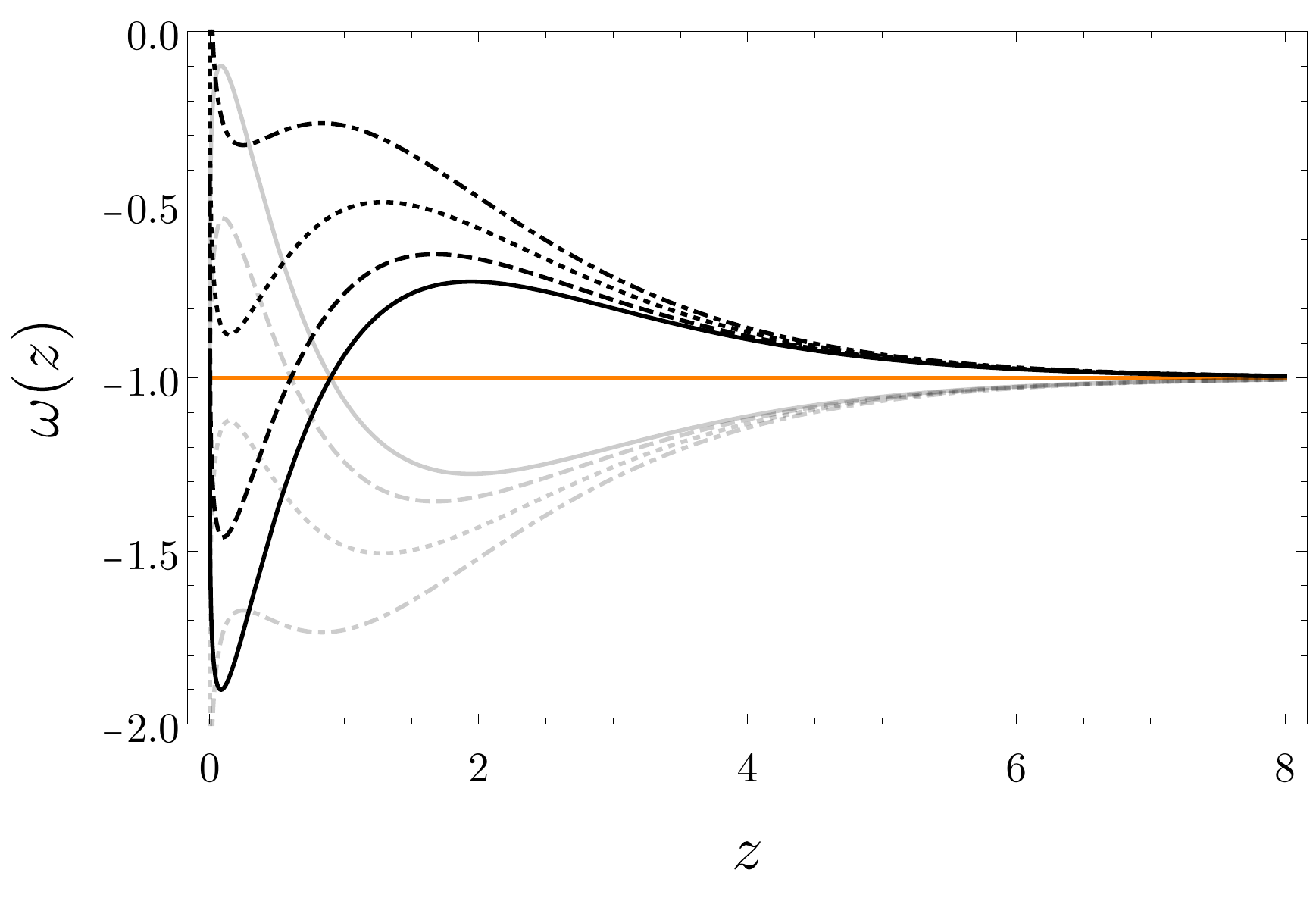}
\caption{General-model, equation (\ref{eq:full-model}). In this figure is depicted the general behavior of $\omega(z)$ when $C$ is varying. In this case, $A=2.2$ and $n=0.2$ for all curves. Solid black line is when $C=0.08$, dashed black line is when $C=0.3$, dotted black line is when $C=0.6$, and dot-dashed black line is when $C=0.9$. Gray lines are the same as black lines but with $A\rightarrow -A$.}
\label{fig:BeastIVb}
\end{center}
\end{figure}

In general, the EoS will present two characteristic behaviors: (1) two critical points with a local maximum and minimum (figure~\ref{Fig:BeastIV}), and (2) a monotone function (figure~\ref{fig:BeastIVb}). 
For all cases, the critical values are given by the two roots of the following expression: 

\begin{equation}
z^n-nz^{n-1}-z+(1-C)=0.
\label{eq:CaseII-min-max}
\end{equation}

It is not surprising that the critical $z$ values obtained from equation (\ref{eq:CaseII-min-max}) depend only on $n$ and $C$, because $A$ acts only as a homothety factor for the $\omega(z)+1$ function. 
However, it is worth mentioning that $\omega(z)$ has at most two real critical points, that we will name $z_1$ and $z_2$, no matter the value of $n\in(0,1)$.

It could happen that the roots, $z_1$ and $z_2$, are complex. This is the case when $\omega(z)$ is a monotonic function (figure~\ref{fig:BeastIVb}), that resembles the exponential case ($n=1$) (see figure~\ref{Fig:CaseI}), and the Quintessence/Phantom case ($n=0$) for $C\geq 2$ (see figure~\ref{Fig:CaseIIa}).

\begin{figure}[h]
\begin{center}
\includegraphics[width=\linewidth]{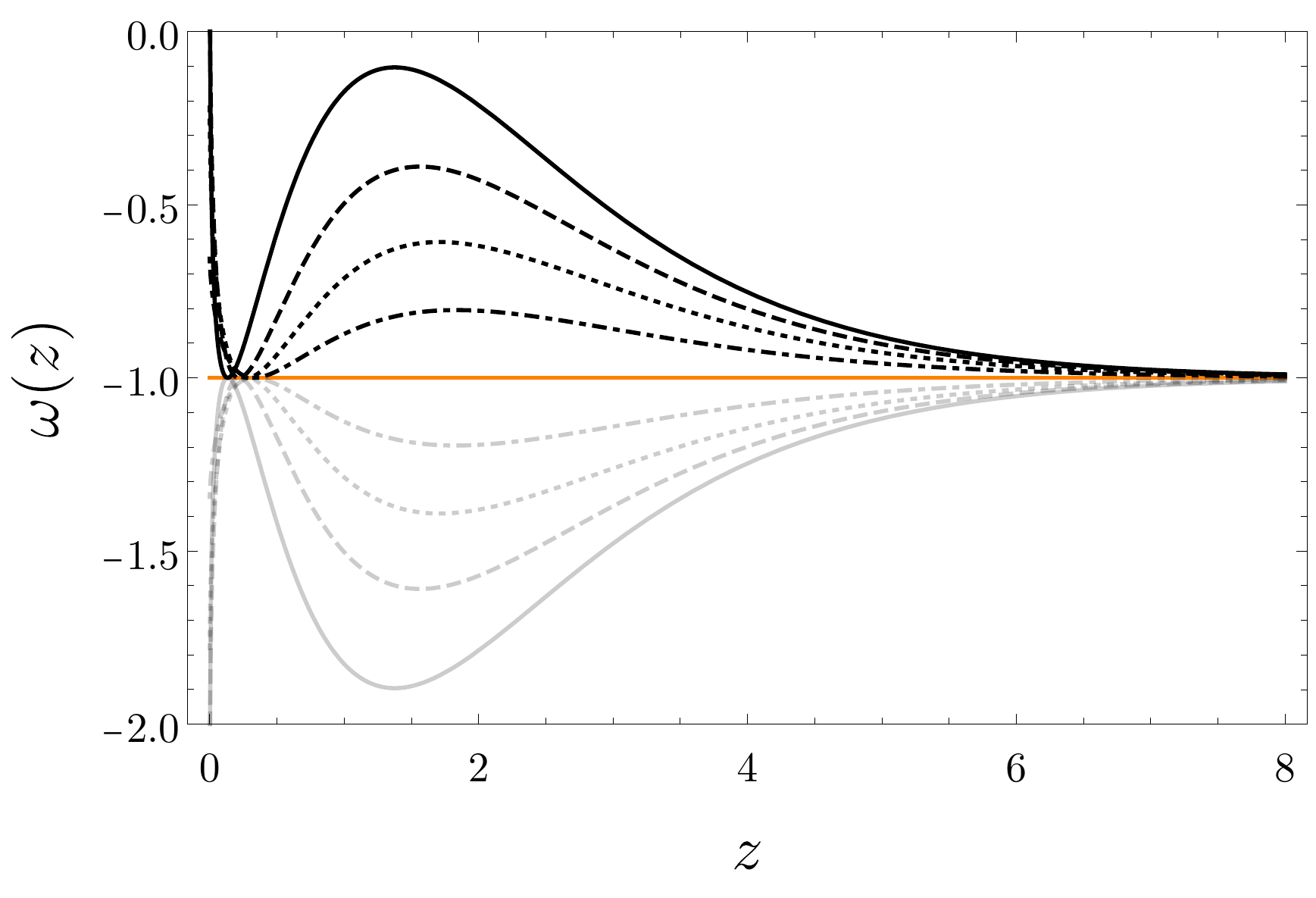}
\caption{General-model, equation (\ref{eq:full-model}). 
The amplitude and the constant value are fixed for all the curves, $A=4.2$ and $C=n^{n/(1-n)}-n^{1/(1-n)}$. Solid black line is when $n=0.2$, dashed black line is for $n=0.4$, dotted black line is when $n=0.6$, and dot-dashed black line is when $n=0.8$. Gray lines are the same as black lines but with $A\rightarrow -A$.}
\label{fig:BeastIVc}
\end{center}
\end{figure}

When the critical points, $z_1$ and $z_2$, are real numbers, we get the generic behavior of the EoS depicted in figure~\ref{Fig:BeastIV} with a local maximum and minimum. 
It will be helpful to understand how the parameter $C$ modifies the form of $\omega(z)$. 
As is shown in figure~\ref{fig:BeastIVb}, the parameter $C$ can either increase or lower the general $\omega(z)$ value alongside a subtle displacement into the $z$-axis direction. 
This feature allows us to put an anchor to the first critical point. 
Let us define $z_{c}=\rm{min}(z_{1},z_{2})$ and demand that $\omega(z)$ must be -1 at $z_{c}$. 
Under that condition, $z_c$ and $C$ are forced to be $z_c= n^{1/(1-n)}$ and $C= n^{n/(1-n)}-z_c$, where $z_c$ acts as the anchor point (figure~\ref{fig:BeastIVc}), and can be compared with case II (figure~\ref{Fig:CaseIIa}), when $n=0$ and $C\sim 1$, for example.

It is worth to mention that $\omega(z)=-1-Ae^{-z}(z^{n}-z-C)$, for $n>1$, has equivalent qualitative attributes to $n\in(0,1)$ case: (1) at most two critical points, and (2) monotone behavior when $z_{1}$, $z_{2} \in \mathbb{C}$. 
In particular, the $n>1$ case can mimic or behave too similar to $n\in(0,1)$ case. 
For this reason, we limit our analysis to $n\in(0,1)$.
By constraining these sub-cases separately, we can statistically study the constraints on a wide variety of models. 
By fitting the general form of our EoS against a collection of data ranging from $z_{\rm{cmb}}$ all the way down to  $z\approx 0.01$, we can answer the question of which one of the allowed scenarios is preferred by observations. 
Is the simplest case of a cosmological constant the favored model? Do different data sets point towards different dynamics of DE?


\section{Cosmological background}
\label{sec:cosmology}

We model the accelerated expansion of the Universe in terms of a barotropic fluid, $\rho_x$, described in terms of the equation of state $\omega_x \equiv p_x/\rho_x$.

Within the validity of General Relativity for a flat Universe and an FLRW metric, we can express the Friedmann equation as:
\begin{equation}
    \label{eq:H_of_z}
    H^2(z)/ H_0^2 = \Omega_m(1+z)^3+\Omega_r(1+z)^4+\Omega_{DE}F_{DE}(z), 
\end{equation}
where $\Omega_m+\Omega_r+\Omega_{DE}=1$, $H_0=100 h$ $km/s/Mpc$ is the Hubble constant, and $F_{DE}(z)$ is a function of redshift involving the specific form of $\omega_x$. 
For a Cosmological Constant, $\omega_x=-1$, and $F_{\Lambda}(z)=1$. In general, we have: 
\begin{equation}
\label{eq:F_de_of_z}
    F_{DE}(z)=\exp{\left(\int_0^z\frac{3(1+\omega_x(z'))}{1+z'}dz'\right)},
\end{equation}
where $\omega_x$ is the one given by (\ref{eq:eos}).

From equations (\ref{eq:H_of_z}) and (\ref{eq:F_de_of_z}) it is clear how data coming from cosmological distances can be used to constrain the free parameters in (\ref{eq:eos}).


\section{METHODS}
\label{sec:methods}

\subsection{Data}
\label{subsec:data}

In order to probe the parameters in equation (\ref{eq:eos}) we use different cosmological distance measurements, covering a wide range of redshifts: $0.02\lesssim z\lesssim1090$. 

\subsubsection{Baryon Acoustic Oscillations}

The Baryon Acoustic Oscillations feature is an imprint on the spatial distribution of galaxies and luminous tracers. It was detected for the first time by \cite{Colless:2003wz, Eisenstein:2005su} and has been explored with increasing detail becoming a po\-wer\-ful tool for cosmology.
It has consolidated as one of the most robust ways to probe late time dynamics of the Universe, as shown in several observational efforts like those carried by experiments like 6dF \cite{Beutler:2011hx}, WiggleZ  \cite{Kazin:2014qga}, Dark Energy Survey (DES) \cite{Abbott:2005bi} and the SDSS consortium \cite{Anderson:2013oza, Alam:2016hwk, Dawson:2015wdb}, finalizing with their latest and final report on \cite{Alam:2020sor}.
BAO is also one of the main features to be probed by experiments like the Dark Energy Spectroscopic Instrument (DESI) \cite{Levi:2013gra, Aghamousa:2016zmz, Aghamousa:2016sne}, and in the near future, Euclid \cite{laureijs2011euclid}.

In this work we use the spherically averaged BAO signature, in terms of the size $r_{BAO}(z)$:

\begin{equation}
    \label{eq:rbao}
	r_{BAO}(z) \equiv \frac{r_s(z_d)}{D_V(z)},
\end{equation}
where the comoving sound horizon at the baryon drag epoch is represented by $r_s(z_d)$,  and the dilation scale, $D_V(z)$, contains information about the cos\-mo\-lo\-gy used in $H(z)$:

\begin{eqnarray}
\label{eq:sd}
	r_s(z_d) & \equiv &  \int_{z_d}^{\infty} \frac{dz}{ H(z)  \sqrt{3(\tilde{R}(z)+1)}}, \\
\label{eq:DV}
	D_V(z) & \equiv & \left[\frac{z(1+z)^2}{H(z)}D_A(z)^2\right]^{1/3},
\end{eqnarray}
where $\tilde{R}(z)$ is the baryon to photon ratio, defined by $\tilde{R}(z)\equiv\frac{3\Omega_{\gamma}(z)}{4\Omega_b(z)}$, and the angular diameter distance, $D_A(z)$,  given by:
\begin{equation}
    \label{eq:daz}
	D_A(z)=\frac{1}{1+z} \int_0^z \frac{dz'}{H(z')},
\end{equation}
where we can clearly see how to use the BAO standard ruler to constrain the parameters in the equation  (\ref{eq:eos}). 
The sound horizon, $r_s(z_d)$, depends upon the physics prior to the recombination era,  given by  $z_d\approx1059$   \cite{Ade:2015xua}  and the baryon to photon ratio, $R(z)$.
However, the dilation scale, $D_V(z)$,  is sensitive to the physics of much lower redshifts, particularly to those probed by large scale structure experiments.

In this work, we make use of the observational points from the six-degree-field galaxy survey (6dFGS \cite{Beutler:2011hx}), the Main Galaxy Sample from Sloan Digital Sky Survey Data Release 7  BOSS-DR7 MGS \cite{Ross:2014qpa}) and the reconstructed value (SDSS(R) from \cite{Padmanabhan2pc}), as well as the uncorrelated values reported in the complete BOSS sample SDSS DR12 (BOSS-DR12 LRG) \cite{Alam:2016hwk}. 
We included the measurement done in the auto and cross-correlation of the Lyman-$\alpha$ Forest (Ly$\alpha$-F) measurements from the quasars sample of the 11th Data Release of the Baryon Oscillation Spectroscopic  (BOSS DR11) \cite{Delubac:2014aqe, Font-Ribera:2013wce}.  In total, we cover the redshift range $0.106<z<2.36$.
Since the volume surveyed by BOSS and WiggleZ \cite{Kazin:2014qga} partially overlap \cite{Beutler:2015tla}, we do not use data from the latter in this work. 
As in this case, all the measurements we are using are independent, we can write the $\chi^2_{BAO}$ in terms of the observed values $r_{BAO}^{obs}$ with their corresponding errors $\sigma_i$, and the predicted values $r_{BAO}^{th}$ as:

\begin{equation}
    \label{eq:chi2_bao}
	\chi^2_{BAO}=\sum_i \frac{\left(r_{BAO}^{obs}(z_i)-r_{BAO}^{th}(z_i)\right)^2}{\sigma_i^2}
	\end{equation}

\subsubsection{Cosmic Chronometers}

In \cite{Jimenez:2001gg}, the use of the relative ages of galaxies was proposed to track the expansion of the Universe. 
This method was coined "cosmic chronometers." 
In \cite{Moresco:2010wh} the authors presented a new methodology using the spectral properties of early-type galaxies. They showed that including the effect of metallicity impacts their results by less than $2-3\%$, even after considering different initial mass functions.

The Cosmic Chronometers (CC) data gives a measurement of the expansion rate, $H(z)$,  that does not depend on the cosmology model, unlike the case of BAO or Supernovae measurements. 
Another advantage lies in the fact that, unlike the distance measurements, we do not rely on the integral of $H(z)$ to constrain the parameters in the EoS (\ref{eq:eos}).  
It is convenient to write the expansion rate as
\begin{equation}
    \label{H(z)-CC}
    H(z)=\frac{\dot{a}}{a}=-\frac{1}{1+z}\frac{dz}{dt}.
\end{equation}

With the relation of the Hubble parameter written in this way, it is possible to use the redshift of the galaxies that are taken as chronometers because its redshift can be measured with high accuracy.
The differential expression for $dz$ and $dt$ helps to cancel out systematic errors and the possible effects that are given by the bias (see \cite{Moresco_2012} for a detailed revision of the method).

In this work, we use the sample compiled in \cite{Farooq:2013hq}, which covers the redshift range $0.07< z <2.3$, with 28 independent measurements of the Hubble parameter. 
The value of $\chi^2$ will be estimated as:
\begin{equation}
    \label{eq:chi2_cc}
    \chi_{CC}^2=\sum_i\frac{\left(H(z)_{CC}^{obs}-H(z)_{CC}^{th}\right)^2}{\sigma_i^2},
\end{equation}
where $(\,)^{obs}$, and $(\,)^{th}$, stands for observational values and predicted values of the theory, respectively. 

\subsubsection{Supernovae Ia}

Type-Ia supernovae (SNe Ia) were crucial for discovering the Universe's accelerated expansion and are angular cosmological probes. 
Ever since the discovery made by \cite{Perlmutter:1998np} and \cite{Riess:1998cb}, SNe Ia played a crucial role in discovering the cosmic acceleration and have consolidated as one of the most valuable and powerful tools to investigate the nature behind the cosmic acceleration.  

Several high quality samples  have been released over the past decade \cite{Kowalski:2008ez, Hicken:2009dk,Kessler:2009ys,Amanullah:2010vv,Conley:2011ku,Suzuki_2012,Betoule:2014frx,Scolnic:2017caz}. 

The $\chi^2$ function of Supernovae Ia can be expressed as
\begin{equation}
    \label{eq:chi2_sne}
	\chi^2_{SNe}=\Delta\mu^{T}\cdot C^{-1}\cdot\Delta\mu
\end{equation}
where $\Delta\mu\equiv\mu^{obs}-\mu^{th}$. 
We take $C=D_{stat}$ and $\mu^{obs}$ from the compilation Union 2.1 presented in \cite{Suzuki_2012}\footnote{Data can be found in \texttt{http://supernova.lbl.gov/Union/}.}, and estimate $\mu^{th}$, the distance modulus of the luminosity distance, as:

\begin{eqnarray}
\mu(z_i) = 5 \log_{10}\left[(1+z)H_0\int_{0}^{z}{dz'H^{-1}(z')}\right]+25,  
\end{eqnarray}
where $H(z)$ contains the free parameters of (\ref{eq:eos}) through equation (\ref{eq:H_of_z}). 
Even though SNe Ia provide a measurement of the luminosity distance as a function of redshift, their absolute luminosity is uncertain and is marginalized out, which also removes any constraints on $H_0$. 
For that reason, we omit $h$ as part of the parameter vector to be constrained during the analysis when we use only this sample, and we consider a given value $h=0.7$, as was done in \cite{Amanullah_2010, Suzuki_2012}.

This sample covers the range $0.028<z<1.03$ with a total of 557 data points.

\subsubsection{Cosmic Microwave Background}

In order to add information from the  CMB, we follow the strategy used by the Planck Collaboration in their Dark Energy and Modified Gravity paper \cite{planck15DE}, originally suggested in \cite{mukherjee}.
In \cite{mukherjee} it was shown how to compress the information of CMB power spectra within a few observable quantities such as the angular scale of the sound horizon at last scattering, $l_A\equiv\pi/\theta_*$,  the scaled distance to last scattering surface, $R \equiv \sqrt{\Omega_MH_0^2}d_{A}(z_*)$, the baryon density, $\Omega_bh^2$, and the scalar spectral index, $n_s$.

For correlated data, the $\chi^2$ estimator reads as

\begin{equation}
	\chi^2 = \Sigma_{ij}\left(D_i-y(x_i|\theta)\right)Q_{ij}\left(D_j-y(x_j|\theta)\right)
\end{equation}
where $Q_{ij}=C_{ij}^{-1}$, is the inverse of the covariance matrix of the data.

In the particular case of $\chi^2_{CMB}$, we have
\begin{equation}
	\chi^2_{CMB}=\vec{y}_{CMB}\cdot \mathbb{C}^{-1}_{CMB}\cdot\vec{y}_{CMB}
\end{equation}
where 
$\mathbb{C}^{-1}_{CMB}$ is the inverse of the covariance matrix and  $\vec{y}_{CMB} = D_i-y(x_i|\theta)$  given in terms of the data vector, $D_i = (R, l_A, \omega_b, n_s)$, and $y(x_i|\theta) = (R(z,\theta), l_A(z,\theta), \omega_b,n_s)$, the theoretical prediction that depends on the free parameters: $\vec{\theta} = \{A, n, C, h, \Omega_bh^2, \Omega_ch^2\}$.

In this case, the  inverse of the covariance matrix, $\mathbb{C}^{-1}$, is
\begin{equation}
\label{eq:covmatcmb}
\begin{footnotesize}
\mathbb{C}^{-1}=  
\bordermatrix
{~ & \mathbf{R} & \mathbf{l_A} & \mathbf{\omega_b} & \mathbf{n_s}\cr
	\mathbf{R} & 78470.9 & -41.8857 & 1.39247\times 10^7 & 77926.2\cr
	\mathbf{l_A} & -12169.8 & 76.7608 & 3.3485\times 10^6 & -1046.39 \cr
\mathbf{\omega_b} &15122.3 & 12.5159 & 2.82752\times 10^7 & -9366.69 \cr
\mathbf{n_s} &52165.5 & -2.41088 & -5.77371\times 10^6 & 94698.4 \cr
}
\end{footnotesize}
\end{equation}
where we have chosen the more conservative compressed likelihood values from \emph{Planck} TT +lowP marginalizing over the amplitude of the lensing power, $A_L$ as presented in \cite{planck15DE}.

The angle of the horizon at last scattering is defined to be
\begin{equation}
\label{eq:thetadec}
\theta_* \equiv \frac{r_s(z_*)}{d_A(z_*)},
\end{equation}
where $r_s(z_*)$ is the horizon size at the decoupling epoch ($z_*\approx1089.95$ according to Planck \cite{Ade:2015xua}), defined by the integral in equation (\ref{eq:sd}) evaluated from $z_*$ to $\infty$, and $d_A(z_*)$ is the comoving distance to last scattering surface:
\begin{equation}
\label{eq:dacmb}
d_A(z_*) = \int_{0}^{z_*}\frac{dz'}{H(z')}.
\end{equation}

Introduced in this way, we are using the position that corresponds to the sharply-defined acoustic angular scale on the sky and the relative heights of the successive peaks seen in the CMB power spectra.

\subsection{Statistical Analysis}
Our analysis combines the different measurements: BAO, CC, SNe, and CMB by adding their respective $\chi^2$ functions, as they are all independent of each other and are probing different cosmic epochs. 
In this manner, we write down the combination of all the data as:
\begin{equation}
    \label{eq:chitotal}
    \chi^2_{Total}=\chi^2_{BAO}+\chi^2_{CMB}+\chi^2_{CC}+\chi^2_{SNe},
\end{equation}
where each function is defined as explained in section \ref{subsec:data}. 

Furthermore, we are interested in the sample of standard rulers, fixed in the CMB and detected in the clustering of luminous tracers via the BAO. 
This will be defined as the combination: 
\begin{equation}
    \label{eq:chibc}
    \chi^2_{BAO-CMB}=\chi^2_{BAO}+\chi^2_{CMB},
\end{equation}
to explore the constraining power of acoustic oscillations. 

Additionally, we want to investigate the constrains coming from late time observations, and to that end we define the function:
\begin{equation}
    \label{eq:chibsh}
    \chi^2_{late}=\chi^2_{BAO}+\chi^2_{CC}+\chi^2_{SNe},
\end{equation}
where we ignore the CMB data. 

Even more, we investigate the constrains in our free parameters from the CC, $\chi^2_{CC}$ (\ref{eq:chi2_cc}), and the SNe samples (\ref{eq:chi2_sne}), $\chi^2_{SNe}$, independently. 

For $\Lambda CDM$, the energy density fraction for DE is constant and, we know that for a flat Universe, we can simply express it by the flatness condition, $\Omega_{\Lambda}=1-\Omega_m-\Omega_r$. 
However, with different dynamics for dark energy, this cannot be assumed to be equal to the fiducial value provided by the Planck collaboration \cite{Aghanim:2018eyx}, for instance, for $\Lambda CDM$. 
This means that, in addition to $A$, $n$, and $C$, the free pa\-ra\-me\-ters in (\ref{eq:eos}), we let the physical densities, $\Omega_ch^2$, $\Omega_bh^2$, and the reduced Hubble constant, $h$, free. 

The free parameters were varied within uniform priors: $A\in[-50, 50]$, $n\in[0,1]$, $C\in[-20, 20]$, $\Omega_ch^2\in[0.001, 0.2]$,  $\Omega_bh^2\in[0.005, 0.045]$, and the Hubble parameter $h\in[0.5,0.8]$.

However, not all data samples have the same constraining power over different cosmological parameters. 
In particular, if CMB data is not included in the fitting process, we fix $\Omega_bh^2 = 0.0222$ to the value set by the \emph{Planck} TT + lowP likelihood \cite{Ade:2015xua}. 

We individually optimize the parameters in each case by minimizing the $\chi^2$ statistic. Details about the numerical implementation are given in \ref{app:code}.


\section{Results}
\label{sec:results}


\begin{figure}[h]
    \centering
    \begin{subfigure}[b]{0.48\textwidth}
        \includegraphics[width=\textwidth]{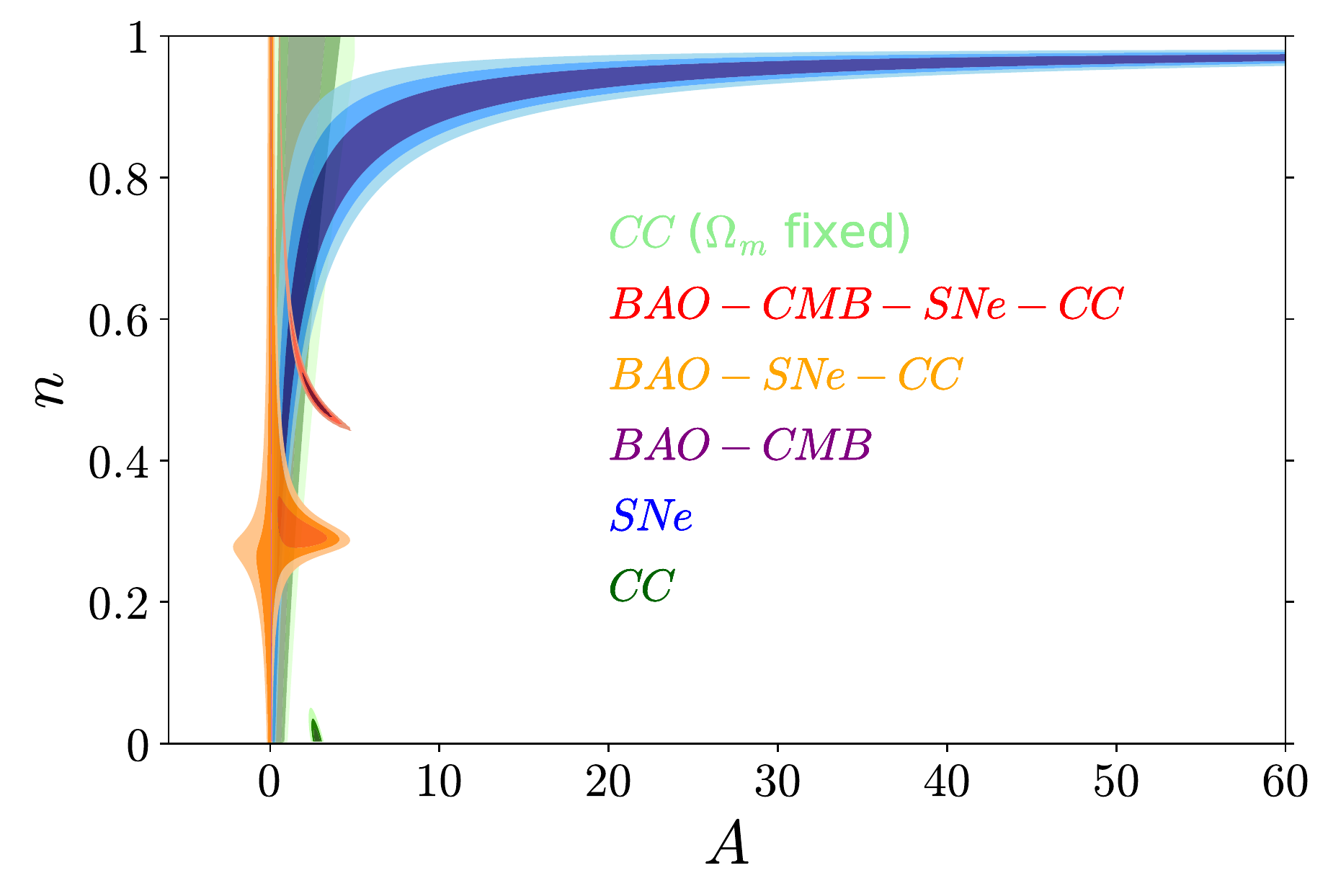}
        \caption{Full explored parameter space $A-n$ within the most general case of the parameterization, (\ref{beastIV}).}
        \label{fig:A-n_a}
    \end{subfigure}
   
    \begin{subfigure}[b]{0.48\textwidth}
        \includegraphics[width=\textwidth]{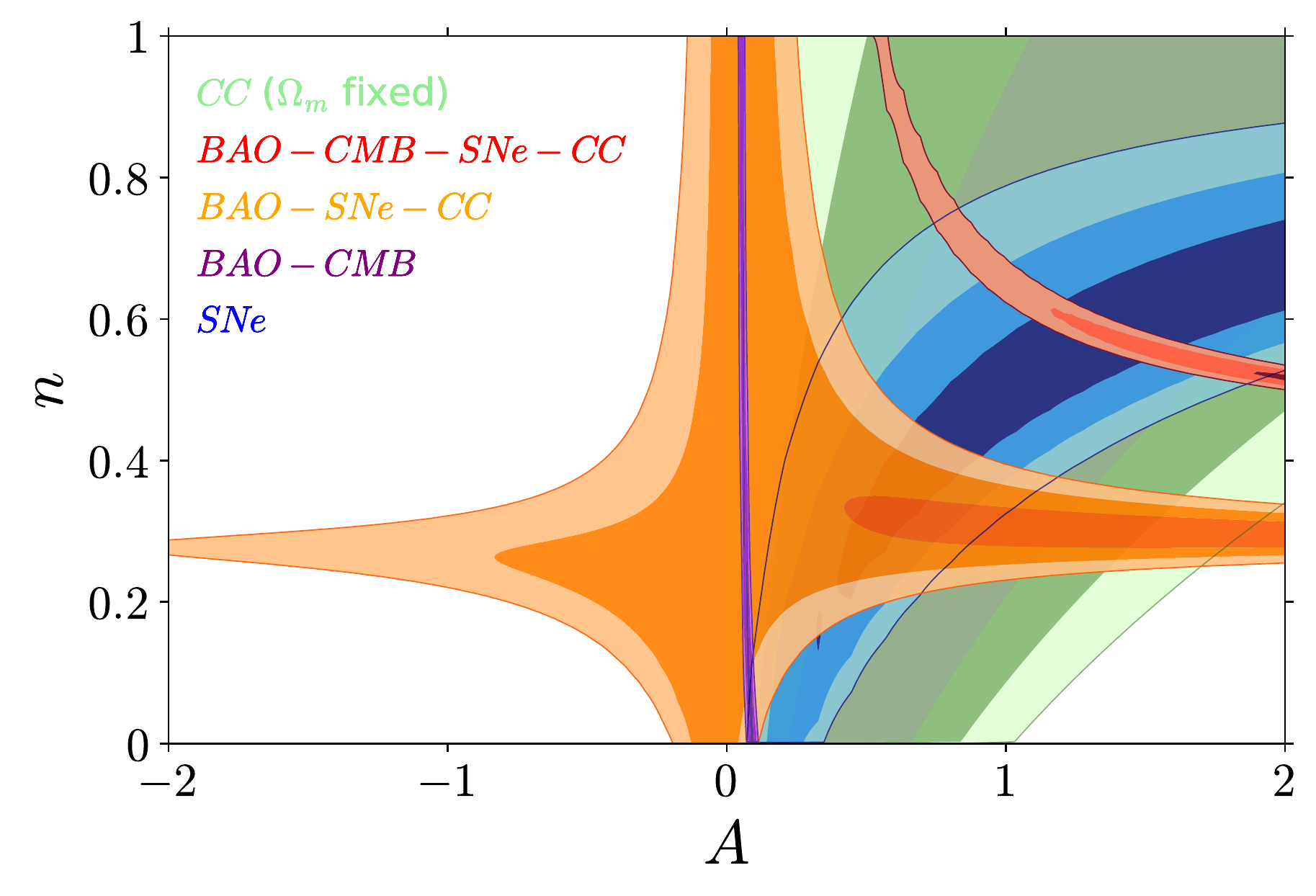}
        \caption{Close up in the region $A\in[-2, 2]$}
        \label{fig:A-n_zoom}
    \end{subfigure}
    \caption{1,2,3$\sigma$ confidence levels in the parameter space $A-n$ for the general model (\ref{beastIV}) using the likelihoods described in  (\ref{eq:chi2_cc}), (\ref{eq:chi2_sne}), (\ref{eq:chitotal}), (\ref{eq:chibc}), and (\ref{eq:chibsh}).}\label{fig:A-n}
\end{figure}

\begin{figure}[h]
    \centering
    \begin{subfigure}[b]{0.48\textwidth}
        \includegraphics[width=\textwidth]{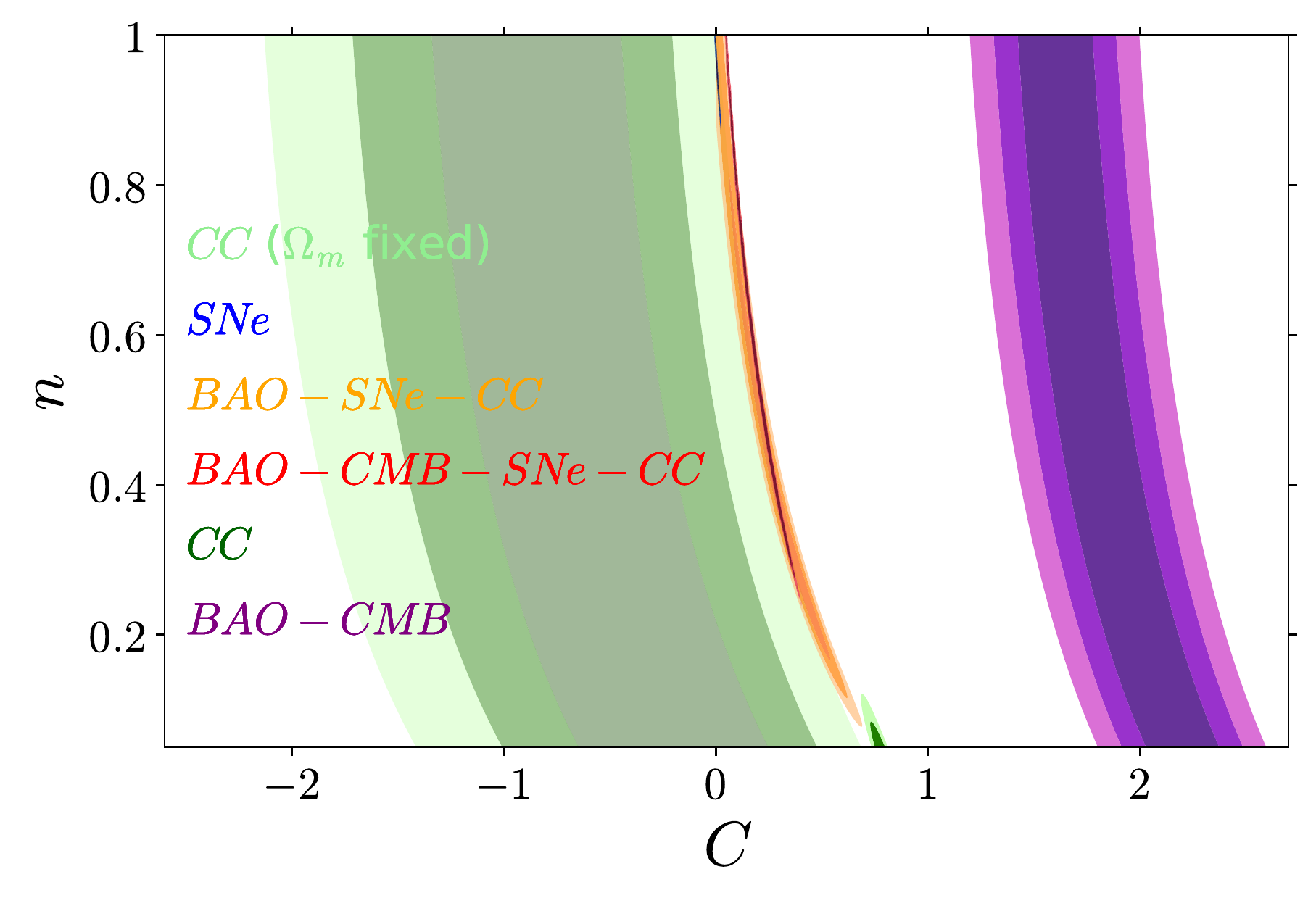}
        \caption{Full explored parameter space $n-C$ within the most general case of the parameterization, (\ref{beastIV}).}
        \label{fig:n-C_a}
    \end{subfigure}
   
    \begin{subfigure}[b]{0.48\textwidth}
        \includegraphics[width=\textwidth]{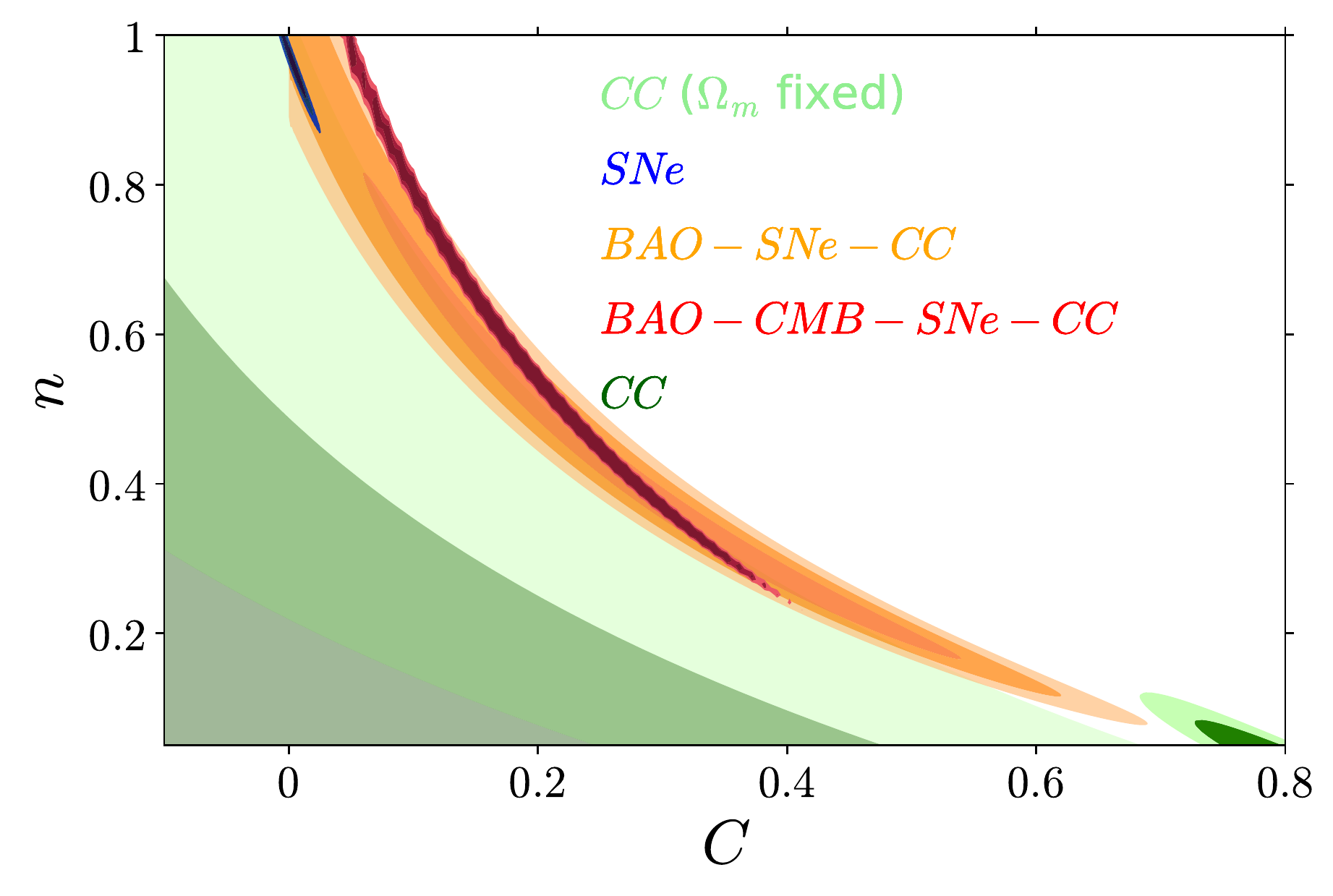}
        \caption{Close up in the region $C\in[-0.1,0.8]$}
        \label{fig:n-C_zoom}
    \end{subfigure}
    \caption{1,2,3$\sigma$ confidence levels in the parameter space $n-C$ for the general model (\ref{beastIV}) using the likelihoods described in  (\ref{eq:chi2_cc}), (\ref{eq:chi2_sne}), (\ref{eq:chitotal}), (\ref{eq:chibc}), and (\ref{eq:chibsh}).}\label{fig:n-C}
\end{figure}

Given that we obtained a good fit for all our models and likelihoods (see the last column of Table \ref{Table:best-fit-values}) of order unity (close to 1), we proceed by discussing our results as follows.

We report the 1-3 $\sigma$ confidence intervals for different combinations in parameter space: the parameters of equation (\ref{eq:eos}) $A-n$, $n-C$ (figures~\ref{fig:A-n}-\ref{fig:n-C}), and the  parameters, $\Omega_bh^2-h$, and $\Omega_ch^2-h$ (figures~\ref{fig:omc-vs-h} and \ref{fig:omegab-h-contours}, respectively). 
Also, we report the individual uncertainties after marginalisation over the other dimensions, and these can be found in Table \ref{Table:best-fit-values}. 
Once we know the constraints on individual parameters, we report the resulting dynamical behavior of $\omega(z)$  (\ref{eq:full-model}). 
These are shown in Table \ref{Table:Plots-EOS-dataset}.

We can see how sensitive the parameterization (\ref{eq:full-model}) is to a different set of observational data when we fit all three parameters, $A$, $n$ and $C$ simultaneously. 
To this end we discuss  the  joint constraints on the $A-n$ and $n-C$ parameter spaces for the unrestricted model, (\ref{eq:full-model}), using the different data sets as described in section \ref{subsec:data}. 

\subsubsection*{$A$-$n$ contour plots}

Figure~\ref{fig:A-n} shows the 1-3$\sigma$ joint confidence levels (CL) for the parameters $n$ and $A$ in the general model, (\ref{eq:full-model}), fitted by each set of observations.  
It is noticeable how different observations constrain the behavior of $n$ differently. 
In particular, from the Cosmic Chronometers (CC) sample, its value is tightly constrained around $n\approx0$, whereas for the joint likelihood BAO-CMB-SNe-CC, it is consistent with $n\geq 0.4$.
In the same figure, for the case of equation (\ref{eq:full-model}) constrained with the CC sample, we see that the resulting dynamics agrees with that of a cosmic fluid with a dust-like equation of state $\omega \approx 0$ (see the figure depicted in the last column, the second row of Table \ref{Table:Plots-EOS-dataset}),  which in turn is consistent with a low value for $\Omega_m$. 
To further test this hypothesis, we reanalyzed the sample fixing the value of $\Omega_m$ to the one reported by \cite{Aghanim:2018eyx} ($\Omega_ch^2=0.1197$), confirming its impact on the resulting dynamics for the EoS. This result is shown in the lighter green contours of figures~\ref{fig:A-n}-\ref{fig:n-C}.

Figure~\ref{fig:A-n_zoom} shows a close up to the region $A\in[-1,2]$ in the parameter space $A-n$.
Here we can appreciate better the fact that the BAO-CMB joint likelihood constraints tightly the value of $A$ around the value $A = 0.053\pm{0.01}$. It is important to recall that, at 1-$\sigma$ level, the value $A=0$ is excluded by all data sets and data combinations, which corresponds to and EoS $\omega=-1$, which recovers a cosmological constant model.

\subsubsection*{$C$-$n$  contour plots}

From the  $n-C$ CL, figure~\ref{fig:n-C}, we notice that the value $n=0$ is excluded at the 3-$\sigma$ level by the  SNe sample and the joint likelihoods BAO-SNe-CC and BAO-CMB-SNe-CC. This is, all the data sets analyzed that included the SNe sample. 
On the other hand, the CC sample imposes very tight constraints on the value of $n\approx 0$. 
Let us point out again that for this particular result, we recover a dust-like EoS with almost no matter ($\Omega_M \approx0.04$), and that, in order to understand the effect of the parameters of (\ref{eq:full-model}) on the value of $\Omega_M$, we rerun the analysis taking $\Omega_ch^2=0.1197$ (\emph{Planck} TT+lowP).
In this case, we find that the value of $n$ is not constrained, allowing a uniform variation along the $n$ axis. Similarly, for the joint acoustic oscillations sample, BAO-CMB, we find that these are insensitive to the value of $n$. 

Looking at the $C$ axis of figure~\ref{fig:n-C} we see that the joint likelihood BAO-CMB constrains $C$ around $C\approx 2$, while the sample of CC with the prior on $\Omega_{m}$ from Planck, imposes the weakest constraints on this parameter around $C\approx0$.
However, from the same sample, without fixing the value for $\Omega_m$, we find very tight constraints for $C\approx 0.8$. 
The supernovae sample, on its own, constrains $C$ around $C=0$, as we can see from the figure~\ref{fig:n-C_zoom}. 

When used in combination with other data sets, as in BAO-SNe-CC, we find that the value $C=0$ is not excluded at the 3-$\sigma$ level.
For the joint analysis of all the data sets, we find that, even when $C=0$ is excluded with $99.7\%$ of confidence, we obtain a value for $C<0.5$, which in turn implies a present value for $\omega$ close to $\omega=-1$. 
The case $C=0$, as discussed in section \ref{CasefR}, gives a dynamics that is consistent with an $f(R)$-like expansion for $ \omega_{0}=-1$. 

The parameter $n$ controls whether the parameterization depicts one, two, or no oscillations at all.  
It is worth noticing that the value $n=0$ was excluded with 99.7$\%$ of confidence by the full joint likelihood (BAO-CMB-SNe-CC), the SNe sample, and the late time observations (BAO-SNe-CC likelihood). 
The acoustic oscillations joint likelihood, BAO-CMB,  and the Cosmic Clocks sample with $\Omega_m$ fixed does not constrain $n$ within the explored range, $n\in[0,1]$.
On the other hand, the Cosmic Clocks sample, by itself, fixes $n=0.003^{+0.025}_{-0.003}$.

\subsubsection*{$\Omega_bh^2$-$h$ and $\Omega_ch^2$-$h$ contours.}
To explore more carefully these possibilities, we perform the same analysis for the other three cases of our proposal: the particular case $n=1$ for a non-oscillatory EoS (I: Exponential case, figure~\ref{Fig:CaseI}), the case $n=0$ which allows $\omega$ to cross only once the phantom dividing line, $\omega=-1$, (II: Quintessence/Phantom or Quintom, figures~\ref{Fig:CaseIIa} and \ref{Fig:caseIIb}), and the case $C=0$, which presents an oscillatory behavior around $\omega=-1$ (III: $f(R)$, figures~\ref{fig:BeastIIIa} and \ref{fig:BeastIIIb}). 

\begin{figure}[h]
\begin{center}
      \includegraphics[width=\linewidth]{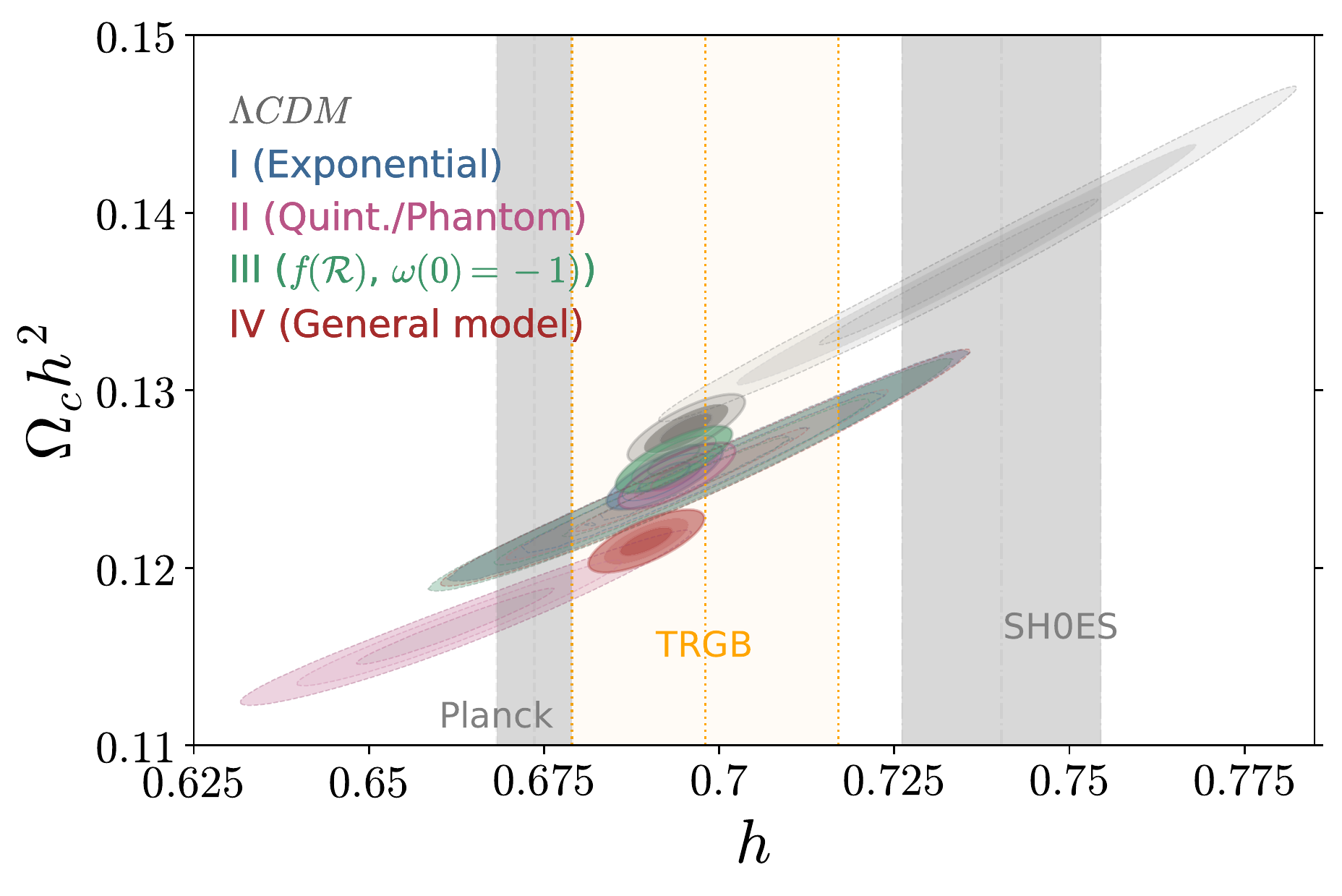}
        \caption{Confidence intervals for the space $\Omega_ch^2-h$ for the models $\Lambda$CDM ($A=0$), Exponential ($n=C=1$), Quintessence/Phantom ($n=0$), $f(R)$ ($C=0$), and the general form of our EoS ($n\in[0,1]$).  
        Joint likelihoods: $BAO-CMB$ are showed in dotted contour lines and $Total$ with solid contour lines. 
        Vertical shaded zones show the different $H_0$ values from the CMB reported by Planck \cite{Aghanim:2018eyx}, the TRGB determination \cite{Freedman:2019jwv}, and the SH0ES experiment \cite{Riess:2019cxk}.}
        \label{fig:omc-vs-h}
\end{center}
\end{figure}

\begin{figure}[h]
    \centering
    \begin{subfigure}[t]{0.48\textwidth}
        \includegraphics[width=\linewidth]{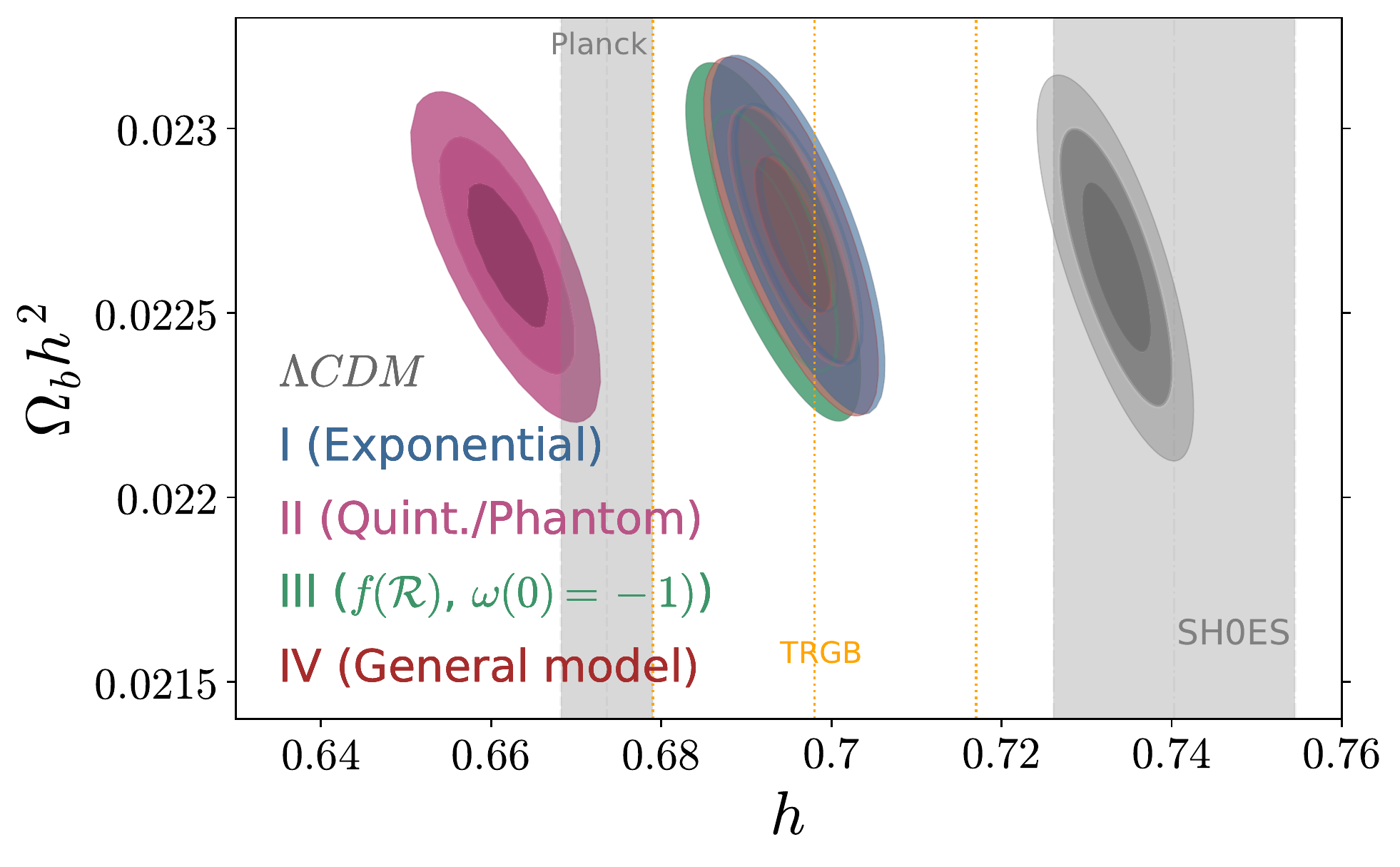}
        \caption{CL in $\Omega_bh^2-h$ for the models I-IV and $\Lambda$CDM from BAO-CMB joint likelihood.}
        \label{fig:omb-vs-h_baocmb}
    \end{subfigure}
   
     \begin{subfigure}[t]{0.48\textwidth}
        \includegraphics[width=\linewidth]{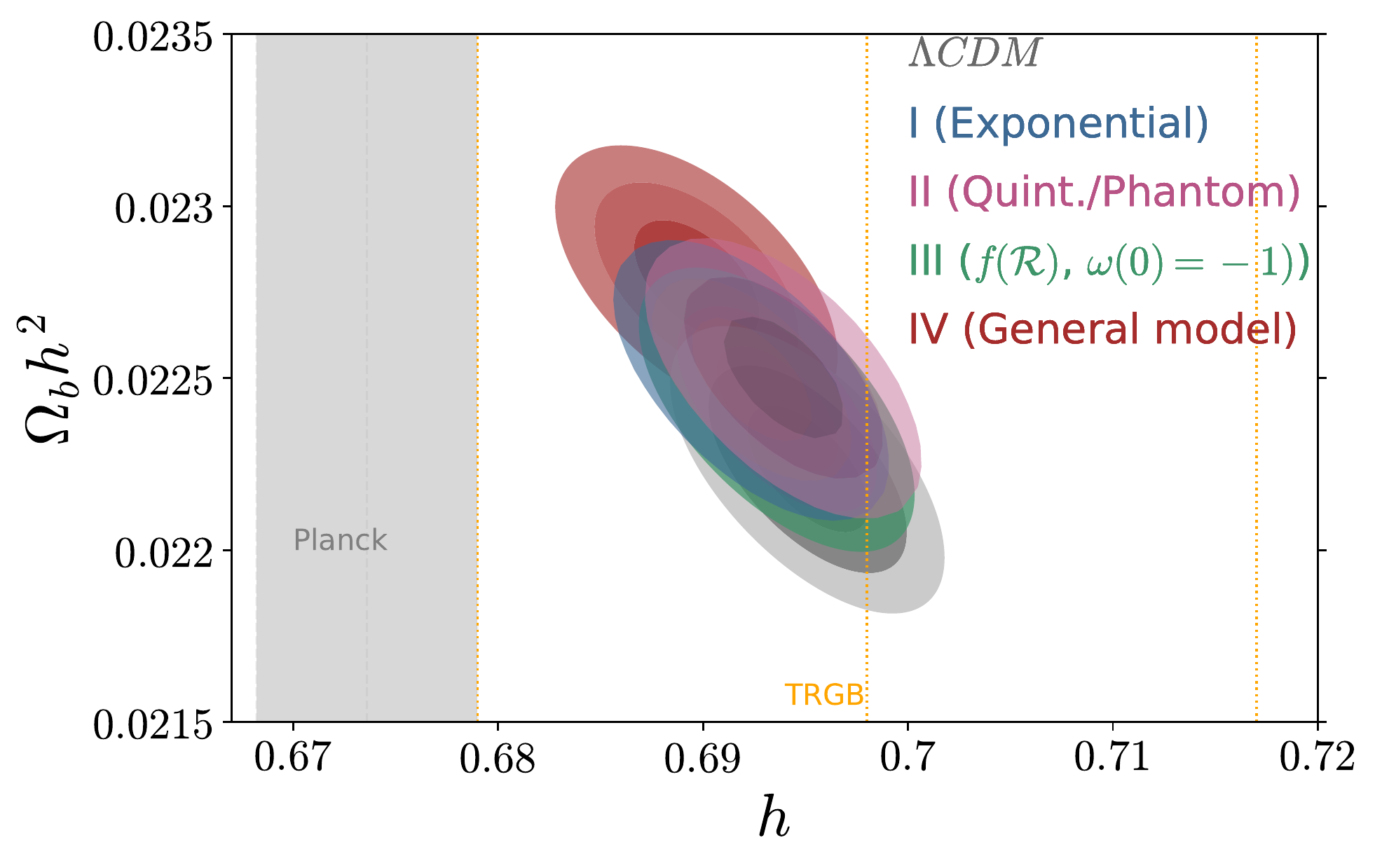}
        \caption{CL in $\Omega_bh^2-h$ for the models I-IV and $\Lambda$CDM from BAO-CMB-SNe-CC joint likelihood.}
        \label{fig:omb-vs-h_total}
    \end{subfigure}

    \caption{Confidence intervals for the space $\Omega_bh^2-h$ for the models $\Lambda$CDM ($A=0$), Exponential ($n=C=1$), Quintessence/Phantom ($n=0$), $f(R)$ ($C=0$), and the general form of our EoS ($n\in[0,1]$).  
     Vertical shaded zones show the different $H_0$ values from the CMB reported by Planck \cite{Aghanim:2018eyx}, the TRGB determination \cite{Freedman:2019jwv}, and the SH0ES experiment \cite{Riess:2019cxk}.}
    \label{fig:omegab-h-contours}
\end{figure}

By performing this analysis, we can investigate if some of the features marked by the value of the parameters have a statistical preference. We compare cases I, II, III against the general model (IV) and with the concordance $\Lambda CDM$ scenario. 
Figures~\ref{fig:omc-vs-h} and \ref{fig:omegab-h-contours} summarize our results for this part of the analysis, along with the figures in table~\ref{Table:Plots-EOS-dataset}. 
In both figures we present the constrains on the different models using the total likelihood, \emph{i.e.}, $\chi^2_{Total} $ (\ref{eq:chitotal}), and the acoustic oscillations observations, \emph{i.e.}, BAO-CMB (\ref{eq:chibc}). 
As it was detailed in section \ref{sec:bestiary},  each particular case of (\ref{eq:eos}) is referred to as a different model since each choice is motivated by a specific dynamical behavior. 
  
Figure~\ref{fig:omc-vs-h} presents the parameter space for the physical density of cold dark matter, $\Omega_ch^2$, and the Hubble parameter, $h$.  In this case, the first thing we notice is that the constraints are more extended for the BAO-CMB likelihood (represented by dotted contour lines) than for the combination of all data sets (shown in solid contour lines). 
Moving away from that observation to more specific, we see that different models agree with different values of $\Omega_ch^2$ and $h$. 

Focusing first on the constraints from BAO-CMB, we see that $\Lambda$CDM gives a higher $h$ and large amount of matter while the model  Quintessence/Phantom, on the contrary, is consistent with a lower $h$ and smaller amount of matter. 
Using BAO-CMB data sets, on the space of cosmological parameters $\Omega_ch^2$-$h$, it is impossible to distinguish the Exponential model from the general form of the parameterization, or from the $f(R)$-like background expansion.
In other words, the exponential ($n=C=1$), $f(R)$-like ($C=0$), and the general form of the EoS are fully compatible with each other in the parameter space $\Omega_ch^2-h$.
However, we must remember that each one is quite distinctive from the other in the space of their respective parameters, $n$, and $C$. 
low $H_0$ value reported by the Planck collaboration. In contrast, the rest of the models (I Exponential, III $f(R)$, and IV, the general model) have an $H_0$ value consistent with the determination for the Hubble parameter using the Tip of the Red Giants Branch (TRGB) done by \cite{Freedman:2019jwv}, which sits midway in the range defined by the current Hubble tension (and indicated by the orange shaded area around $h= 0.698$). 

Now, from the joint constraints of all data sets, BAO-CMB-SNe-CC, we see that the confidence regions are smaller in $\Omega_c h^2 - h$ space than those obtained only from the acoustic oscillations. In particular, we find that the general form of the EoS (model IV) is consistent with a lower $\Omega_c h^2$, while $\Lambda$CDM prefers a slightly higher value.
Exponential (I) and Quintessence/Phantom (II) models agree with each other at the 1-$\sigma$ level, as well as the Exponential (I) and $f(R)$-like models. 
The Quintessence/Phantom (II) and $f(R)$-like (III) models are consistent with each other only at the 2-$\sigma$ level.  However, all the resulting CL lie within the uncertainties of the TRGB determination of $H_0$ \cite{Freedman:2019jwv}.


For the sake of clarity, we present the CL in $\Omega_bh^2-h$ space in two separate figures. 
Figure~\ref{fig:omb-vs-h_baocmb} shows the resulting contours from the acoustic oscillations, BAO-CMB, while the resulting constraints from the combination of all data sets, BAO-CMB-SNe-CC, can be seen in figure~\ref{fig:omb-vs-h_total}. 
Same for the $\Omega_ch^2-h$ contours.
From figure~\ref{fig:omb-vs-h_baocmb} (top panel of \ref{fig:omegab-h-contours}) we find that the Quintessence/Phantom model (case II, $n=0$) is consistent with a low value of $h$ which lies within Planck's determination of $H_0$. 
In contrast, $\Lambda$CDM is consistent with a higher value of $h$, which coincides with the local determination of $H_0$ made by SH0ES \cite{Riess:2019cxk}.
Models I (Exponential, $n=C=1$), III ($f(R)$ with $C=0$), and IV (general form of (\ref{eq:full-model}) with $n\in[0,1]$), are consistent within each other with 99.7$\%$ of confidence. They share a value of $h$ in agreement with  the TRGB central value. Even when these three models cannot be discerned in the $\Omega_bh^2-h$ space using BAO-CMB data, they are quite distinctive in the values of their respective parameters ($n$ and $C$). 
All the CL shown in \ref{fig:omb-vs-h_baocmb} lie around a central value for $\Omega_bh^2\approx0.0226$, close to Planck's value for the baryonic content. 

The lower panel of figure~\ref{fig:omegab-h-contours} shows the constraints obtained by using the full combination of data, BAO-CMB-SNe-CC, $\chi^2_{Total}$. 
There is more dispersion around the $\Omega_bh^2$ value compared to the BAO-CMB constraints. 
However, we see that the five models agree within 1-$\sigma$ level with each other, both in the $h$ and in the $\Omega_bh^2$ dimensions. 
At 3-$\sigma$ level, the constraints for $h$ lie within the uncertainties from the TRGB determination of $H_0$. 
More in detail, we observe that, at the 1-$\sigma$ level, $\Lambda$CDM (gray contour) and the general model (red contour) do not overlap in figure~\ref{fig:omb-vs-h_total}.


\subsubsection*{Goodness of the fit}


\begin{table*}
    \centering
    \caption{Best fit values by data set. Column 1 refers the model: I-Exponential, II-Quintessence/Phantom, III-$f(R)$ with $\omega(0)=-1$, IV-General case and we compare with $\Lambda$CDM in the last row for each observational set. Columns 2, 3 and 4 show the best fit values for the $A$, $n$ and $C$ parameters. Columns 5, 6 and 7 correspond to the values $ \omega_0$, $\Omega_0^m$ and $H_0$ respectively. Column 8 shows the value of the reduced $\chi^2$}
\label{Table:best-fit-values}
    {\setlength{\extrarowheight}{3pt}
\begin{tabular}{|p{0.17\linewidth}|p{0.10\linewidth}|p{0.10\linewidth}|p{0.10\linewidth}|p{0.10\linewidth}|p{0.10\linewidth}|p{0.09\linewidth}|p{0.05\linewidth}|}
\hline 
   & A $(99.7\%)$ & n $(99.7\%)$& C $(99.7\%)$ & $\omega_0$ $(68\%)$ & $\Omega_M^{(0)}$ $(68\%)$ & $H_0$ $(68\%)$  & $\chi_{red}^2$ \\
\hline
\rowcolor[gray]{0.95}   SNe  &  &  &  &  &  &  &    \\
\hline 
  I Exponential  & $-0.018^{+0.104}_{-0.107}$ & 1.00 & 1.00 & $ -1.018 {\scriptstyle \pm 0.035} $ & $0.277^{+0.014}_{-0.013}$ & 70.00 & 0.978 \\
        
  II Quint./Phant. & $-3.196^{+1.423}_{-1.641}$ & 0.00 & $1.012^{+0.058}_{-0.056}$ & -1.039$^{+0.067}_{-0.068}$ & $0.390 {\scriptstyle \pm 0.011}$  & 70.00   & 0.978 \\
        
  III $f(R)$, $\omega_0=-1$ & $6.036^{+5.741}_{-5.443}$ & $0.906^{+0.084}_{-0.075}$ & 0.00 & -1.00 & $0.316 {\scriptstyle \pm 0.013}$ & 70.00 &  0.979 \\
        
  IV General & $39.400^{+10.596}_{-25.005}$ & $0.962 {\scriptstyle \pm  0.015}$ & $0.003 {\scriptstyle \pm0.003}$ & $-0.870^{+0.073}_{-0.072}$ & $0.336 {\scriptstyle \pm 0.012}$ & 70.00 & 0.980 \\
  
  $\Lambda$CDM  & 0.00 & $-$ & $-$ & -1.00 & $0.270^{+0.014}_{-0.013}$ & 70.00 & 0.976 \\
\hline 
\rowcolor[gray]{0.95}    CC    &  &  &  &  &  &  &  \\
\hline
  I Exponential & $-0.202^{+0.351}_{-0.482}$ & 1.00 & 1.00 & $-1.20^{+0.351}_{-0.142}$ & $0.274^{+0.004}_{-0.003}$& $70.83^{+2.47}_{-2.49}$ & 0.646 \\
        
  II Quint./Phant. & $2.799^{+0.211}_{-0.234}$ & 0.00 & $0.841^{+0.047}_{-0.052}$ & $-1.44^{+0.057}_{-0.058}$ & $0.039{\scriptstyle \pm 0.001}$ & $77.25{\scriptstyle \pm 1.52}$ &  0.594 \\
        
  III $f(R)$, $\omega_0=-1$ & $3.120^{+9.173}_{-6.537}$ & $0.820^{+0.182}_{-0.323}$ & 0.00 & -1.00 & $0.275{\scriptstyle \pm 0.003}$ & $70.30^{+2.41}_{-2.44}$ &  0.669 \\
        
  IV General & $2.811^{+0.306}_{-0.345}$ & $0.003^{+0.025}_{-0.003}$ & $0.837{}^{+0.035}_{-0.038}$ & $1.35^{+0.122}_{-0.127}$ &$0.039^{+0.014}_{-0.001}$ & $77.11{\scriptstyle \pm 1.52}$ &  0.620 \\
  
 IV$^{*}$ General & $0.493^{+0.451}_{-0.497}$ & $0.082^{+0.929}_{-0.071}$ & $-0.202^{+0.756}_{-0.781}$ & -1.10$^{+0.173}_{-0.186}$ & $0.258 {\scriptstyle \pm 0.019}$ & $74.11^{+2.73}_{-2.76}$ & 0.677 \\
  
  $\Lambda$CDM & 0.00 & $-$ & $-$ & -1.00 & $0.288{\scriptstyle \pm 0.002}$ & $68.04^{+2.24}_{-2.26}$ &  0.630 \\
\hline 
\rowcolor[gray]{0.95}    BAO-CMB  &  &  &  &  &  &  &   \\
\hline
  I Exponential & $0.084^{+0.018}_{-0.019}$ & 1.00 & 1.00 & -0.92${\scriptstyle \pm 0.006}$ & $0.305{\scriptstyle \pm 0.001}$  & $69.63{\scriptstyle \pm 0.21}$ &  1.763 \\
        
  II Quint./Phant. & $-0.363^{+0.061}_{-0.062}$ & 0.00 & $0.313^{+0.171}_{-0.167} $ & $-0.75^{+0.034}_{-0.035}$ & $0.318{\scriptstyle \pm 0.001}$ & $66.21{\scriptstyle \pm 0.23}$ &  2.076 \\
        
  III $f(R)$, $\omega_0=-1$ & $-1.239{\scriptstyle \pm 0.272}$ & $0.750^{+0.047}_{-0.042}$ & 0.00 & -1.00 & $0.306{\scriptstyle \pm 0.001}$ & $69.34{\scriptstyle \pm 0.21}$ &  2.094 \\
        
  IV General & $0.053^{+0.011}_{-0.012}$ & $0.979^{+0.017}_{-0.728}$ & $1.604^{+0.346}_{-0.351}$ & $-0.91{\scriptstyle \pm 0.012}$ & $0.305{\scriptstyle  \pm 0.001}$ & $69.55^{+0.20}_{-0.21}$ &  2.645  \\
  
  $\Lambda$CDM & 0.00 & $-$ & $-$ & -1.00 & $0.296{\scriptstyle \pm 0.0004}$ & $73.36{\scriptstyle \pm 0.19}$ &  1.563  \\
\hline 
\rowcolor[gray]{0.95}    SNe-CC-BAO  &  &  &  &  &  &  &   \\
\hline
  I Exponential & $-0.003^{+0.077}_{-0.080}$ & 1.00 & 1.00 & -1.00${\scriptstyle \pm 0.026}$ & $0.276^{+0.007}_{-0.006}$ & $69.92^{+0.26}_{-0.25}$ & 0.973 \\
        
  II Quint./Phant. & $0.503^{+0.437}_{-0.485}$ & 0.00 & $0.806^{+0.318}_{-0.342}$ & $-1.10{\scriptstyle  \pm 0.085}$ & $0.269^{+0.007}_{-0.006}$ & $70.27{\scriptstyle \pm 0.26}$ &  0.973 \\
        
  III $f(R)$, $\omega_0=-1$ & $0.198^{+2.766}_{-2.652}$ & $0.885^{+0.115}_{-0.647}$ & 0.00 & -1.00 & $0.276{\scriptstyle \pm 0.007}$ & $69.95{\scriptstyle \pm 0.26}$ &  0.974 \\
        
  IV General & $1.992^{+2.221}_{-3.217}$ & $0.295^{+0.037}_{-0.035}$ & $0.375^{+0.038}_{-0.039}$ & $-0.25^{+0.340}_{-0.383}$ & $0.273^{+0.007}_{-0.006}$ & $69.63^{+0.26}_{-0.25}$ &  0.972 \\
  
  $\Lambda$CDM  & 0.00 & $-$ & $-$ & -1.00 & $0.277{\scriptstyle \pm 0.007}$ & $69.87{\scriptstyle \pm 0.26}$ &  0.971 \\
\hline 
\rowcolor[gray]{0.95}    SNe-CC-BAO-CMB  &  &  &  &   &  &  &  \\
\hline
  I Exponential & $0.056{\scriptstyle \pm 0.019}$ & 1.00 & 1.00 & -1.05${\scriptstyle \pm 0.006}$ & $0.308{\scriptstyle \pm 0.0003}$ & $69.25{\scriptstyle \pm 0.16}$ &  0.988 \\
        
  II Quint./Phant. & $0.159^{+0.058}_{-0.059}$ & $0.00$ & $1.061 {\scriptstyle \pm 0.139}$ & $-0.99^{+0.008}_{-0.009}$ & $0.307{\scriptstyle \pm 0.0003}$ & $69.40{\scriptstyle \pm 0.16}$ &  0.989 \\
        
  III $f(R)$, $\omega_0=-1$ & $-2.914^{+1.540}_{-1.521}$ & $0.945^{+0.028}_{-0.026}$ & 0.00 & -1.00 & $0.308{\scriptstyle \pm 0.0005}$ & $69.39{\scriptstyle \pm 0.16}$ &  0.991 \\
        
  IV General & $2.689^{+0.321}_{-0.337}$ & $0.486{\scriptstyle \pm 0.011}$ & $0.228{\scriptstyle \pm 0.006}$ & $-0.39^{+0.030}_{-0.031}$ & $0.303{\scriptstyle \pm 0.0007}$ & $69.01{\scriptstyle \pm 0.17}$ & 0.984 \\
  
  $\Lambda$CDM & 0.00 & $-$ & $-$ & -1.00 & $0.310{\scriptstyle \pm 0.0002}$ & $69.50{\scriptstyle \pm 0.16}$ &  0.989 \\
\hline
\end{tabular}
}

\end{table*}


Table \ref{Table:best-fit-values} shows the BFV for all models separated by observational data set. Columns 2, 3, 4 give the BFV for the parameters $A$, $n$ and $C$ respectively with uncertainties at 3-$\sigma$, while columns 5, 6, 7, show the values $\omega_0$, $\Omega_m^0$, and $H_0$ within 1-$\sigma$. 
The last column of this table shows the $\chi^2_{red}$.

From the $\chi^2_{red}$ values,  we point out that all fits were very close or approximately of order unity. 
Also, we find minor discrepancies between each other, and when analyzing data sets individually, we find better fits for our model than for $\Lambda CDM$.  

On the other hand, we notice that from all cases, the best fit was obtained for model IV ($f(R)$-like) fitting the full likelihood, $\chi^2_{Total}$, equation  (\ref{eq:chitotal}). 
From Table \ref{Table:best-fit-values}  we see that this has a value of $\chi^2_{red} = 0.991$. 

From all the cases analyzed, the poorest 
fits were obtained for BAO-CMB likelihood, equation (\ref{eq:chibc}). 
In this case, the less favorable result corresponds to model IV (General) with $\chi^2_{BAO-CMB}$ for which we find $\chi^2_{red}=2.645$. 
In this particular situation, we need to keep in mind that the number of data points in our acoustic oscillations data set is of the same order of magnitude as the number of free parameters. 

Similarly, in the case of the CC sample, we notice an over-fitting of the data points, resulting in $\chi^2_{red} = 0.594$, for model II (Quintessence/Phantom).
%
Even in the simplest model,  \emph{i.e.} $\Lambda CDM$, the $\chi^2_{red}$ is of the same order. 
This is to be expected, given the size or the error bars for this sample (see section \ref{sec:comparisonObs}).

Taking a closer look at results from Table \ref{Table:best-fit-values} each data set at a time, we find that: 
\begin{itemize}
    \item The model IV (General) was the best fit for data sets SNe and the CC sample.
    \item The III ($f(R)$-like) model was the best fit for local data, BAO-SNe-CC, and also for the combination of all data sets.
    \item $\Lambda CDM$ was the best fit only in the case of the acoustic oscillations sample, BAO-CMB.  
\end{itemize}

To conservatively report our uncertainties for $A$, $n$, and $C$, we quote them within a 3-$\sigma$ level, whereas the cosmological parameters $\Omega_M^{(0)}$ and $h$ are reported at 1$\sigma$ to facilitate the comparison with other works. The full 1,2,3 $\sigma$ contours have already been discussed. 

The exceptionally compact constraints obtained from the CC sample are a consequence of the profile for the EoS obtained in this case.  Such profile behaves as dust ($\omega\sim 0$) during its evolution, making this fluid not negligible and hence, being able to constrain the values of the parameters in Equation (\ref{eq:full-model}) very strictly.  
We can notice this in the figure portrayed in the last column and second row of Table \ref{Table:Plots-EOS-dataset}, where we notice $\omega(z)\approx0$ during $z\in[0.5, 1.5]$. As the density for a dust-like component is non-negligible during this epoch, their dynamics can be better constrained. 
As a counter-example, we point out the dynamics we obtained for the supernovae sample under model II (Quintessence/Phantom). This can be seen in detail in the first row and second column of Table \ref{Table:Plots-EOS-dataset}. In this case, the resulting dynamics for the EoS is that of a phantom component: for $z\geq0$, we have $\omega(z)<-1$. 
Since this results in a highly sub-dominant component, $\rho\approx (\rho^{(0)})(a/a_0)^{-3(1+\omega)}$, the involved parameters are much less tightly constrained.

To directly show the resulting dynamics of our EoS for a given model and how it is constrained by different data sets, in Table \ref{Table:Plots-EOS-dataset} we depict the evolution of the EoS for each model according to its best fit values. 

Table \ref{Table:Plots-EOS-dataset} shows all the different profiles for $\omega(z)$ that we obtain within its 3-$\sigma$  uncertainties (see appendix \ref{app:deltaomega}). It is organized as follows: Columns 1, 2, 3, and 4, show the evolution for cases Exponential, Quintessence/Phantom,  $f(R)$ with $\omega_0 =-1$, and the General model, respectively. 
Each row shows the resulting constraints from the different data sets and their combinations.

We particularly stress the general case and how the observational data sets shape the evolution of the EoS. (Column 4)
In the first row, we notice that, for the $z-$range of the SNe sample, the EoS prefers values $\omega(z)<-1$. In the second row, interestingly,  the CC allows the EoS to take values very close to $\omega \approx 0$, this way, the parameterization could mimic the EoS for dust, \textit{i.e.} matter, and we obtain $\Omega_m^0=0.039$. For this reason, in the contour plots of $A$ vs $n$ (Fig \ref{fig:A-n}) and $n$ vs $C$ (Fig. \ref{fig:n-C}) we include one case where we fix the value of $\Omega_m^0=0.258$. When the parameterization is fitted by using BAO and the reduced CMB, it is evident that the best fit of the EoS goes very close to $\omega(z)=-1$ but still $\omega_0\neq-1$.  For the late-time collection (BAO-CC-SNe), the evolution of $\omega(z)$ crosses twice the phantom-line and shows an oscillatory behavior. When the CMB is included (last row), the behavior is very similar, and the uncertainty is dramatically reduced.


\begin{table*}
    \centering
\caption{Equation of state of each model by data set. Column 1 corresponds to model I-Exponential, column 2 shows the EoS for model II-Quintessence/Phantom, column 3 is model III-$f(R)$ with $\omega_0=-1$, and column 4 shows the General case, model IV. 
Each row stands for each data set combination (SNe, CC, BAO-CMB, SNe-CC-BAO, and SNe-CC-BAO-CMB). The solid lines depict $\omega(z)$ and the contours around them are the $3\sigma$ error propagation (see appendix \ref{app:deltaomega}). 
The color vertical shadow show the range of $z$ for each data set. In the plots vertical axis is $\omega(z) \in [-2.5,0]$ and the horizontal axis corresponds to the redshift $z \in [0,3]$.}
\label{Table:Plots-EOS-dataset}
\begin{tabular}{|p{0.22\linewidth}|p{0.22\linewidth}|p{0.22\linewidth}|p{0.22\linewidth}|}
\hline
        I & II & III & IV \\
       \text{(Exponential)} & \text{(Quint./Phantom)} & \text{($f(R)$, $\omega_0=-1$)} & \text{(General)}  \\
\hline
\rowcolor[gray]{.9} SNe &  &  & \\ 
\hline
       \includegraphics[width=4cm]{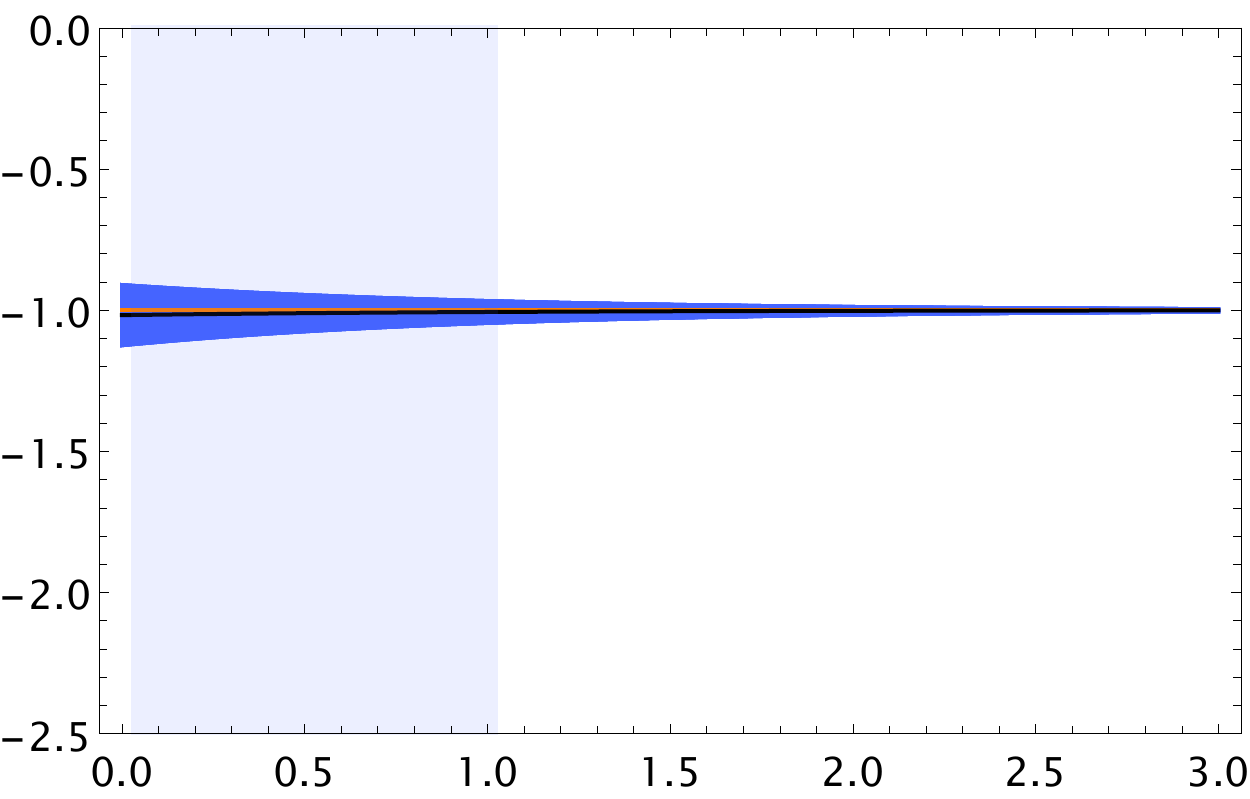} &
       \includegraphics[width=4cm]{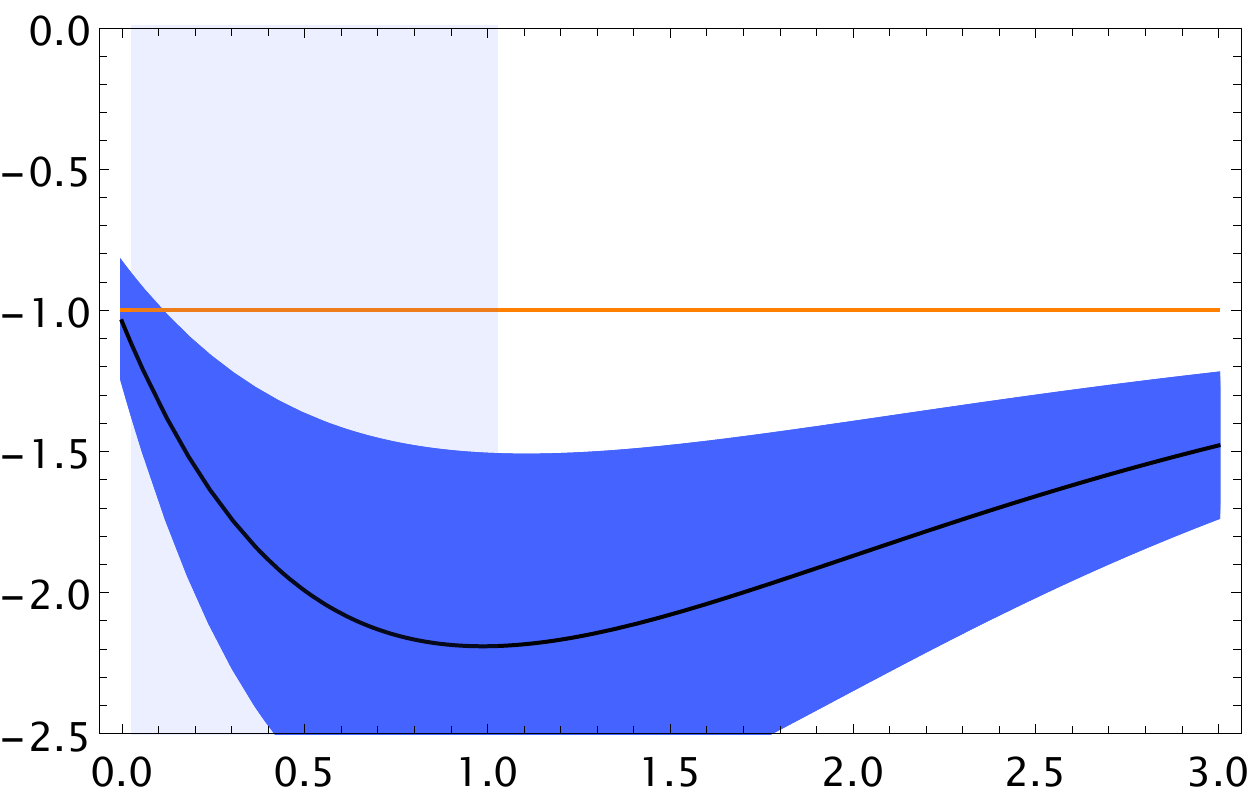} &
       \includegraphics[width=4cm]{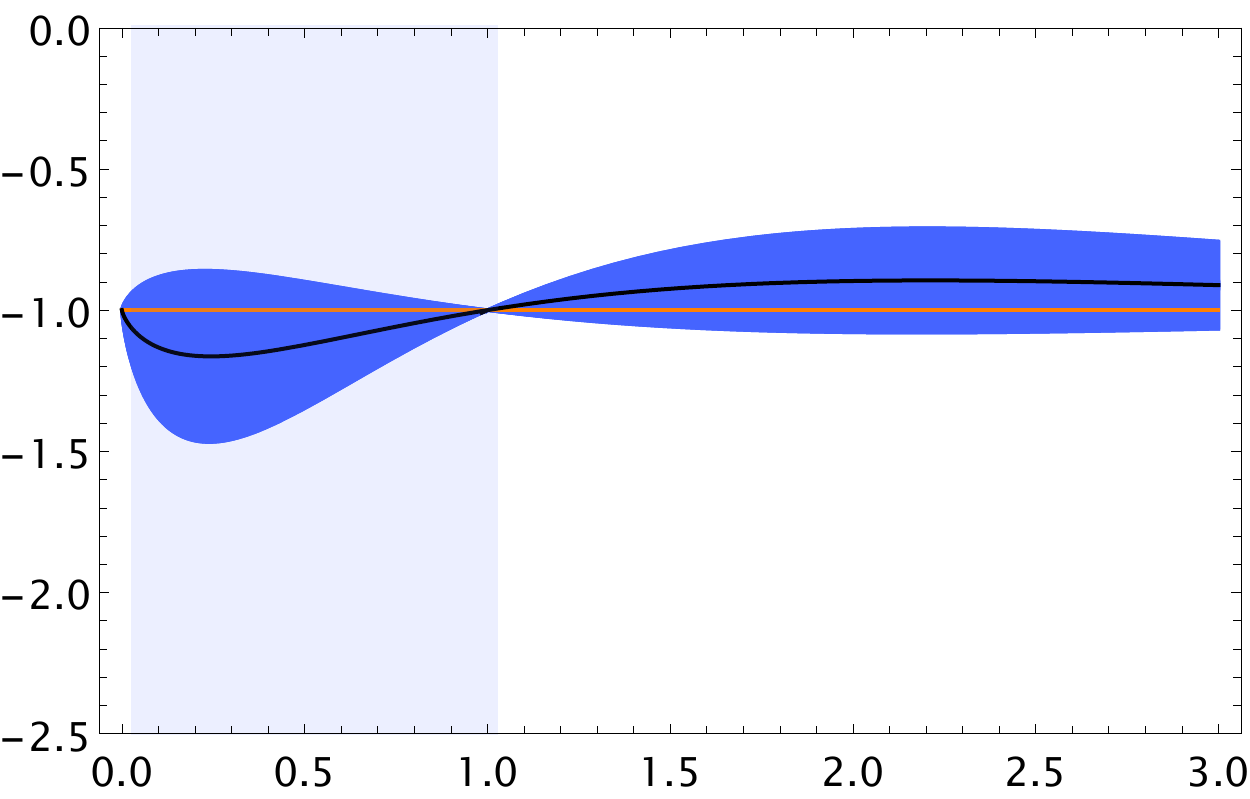} &
       \includegraphics[width=4cm]{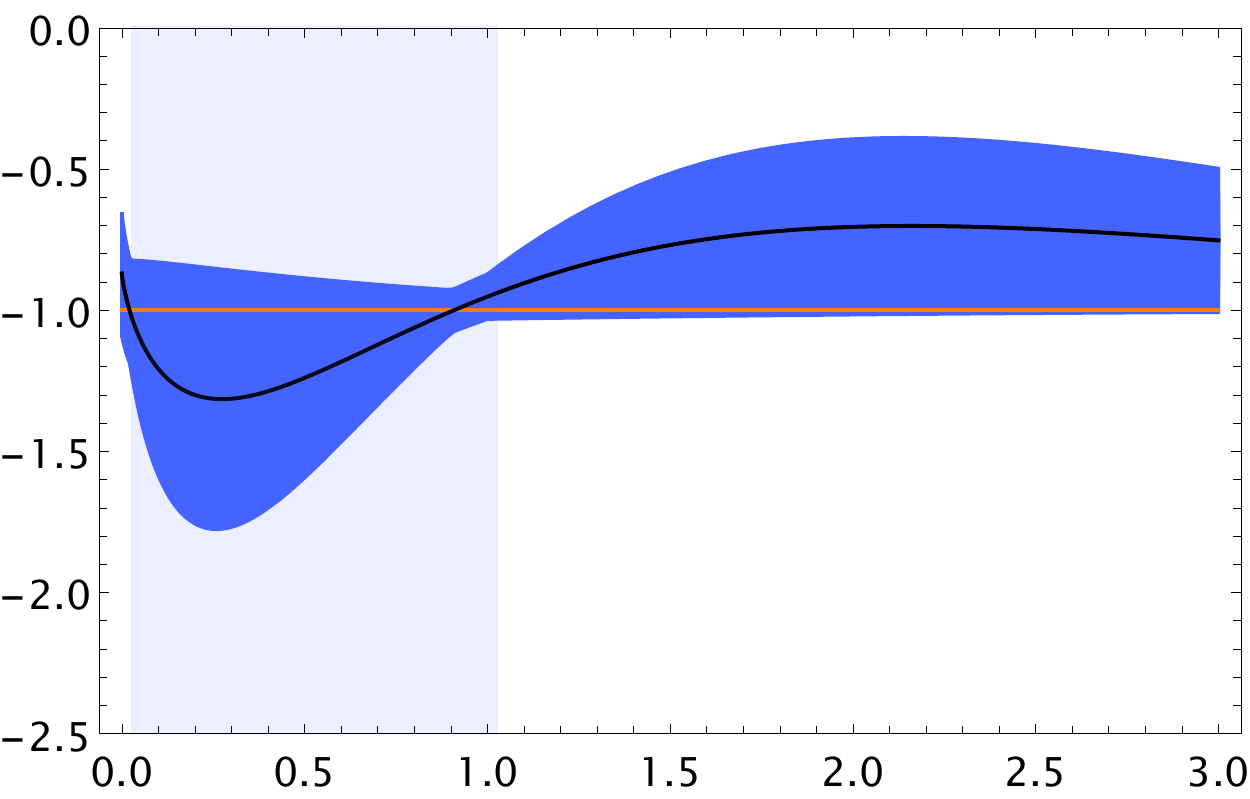}  \\ 
\hline
\rowcolor[gray]{.9} CC & & & \\
 \hline 
       \includegraphics[width=4cm]{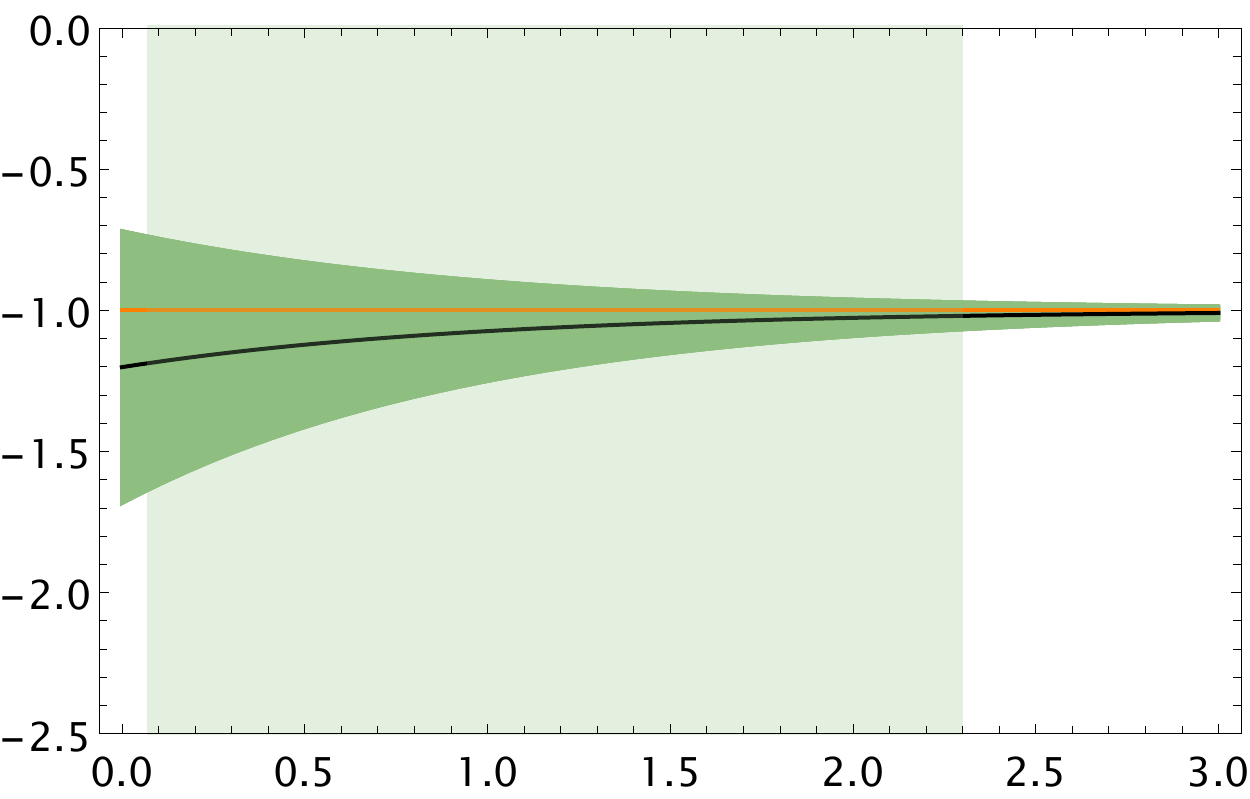} &
       \includegraphics[width=4cm]{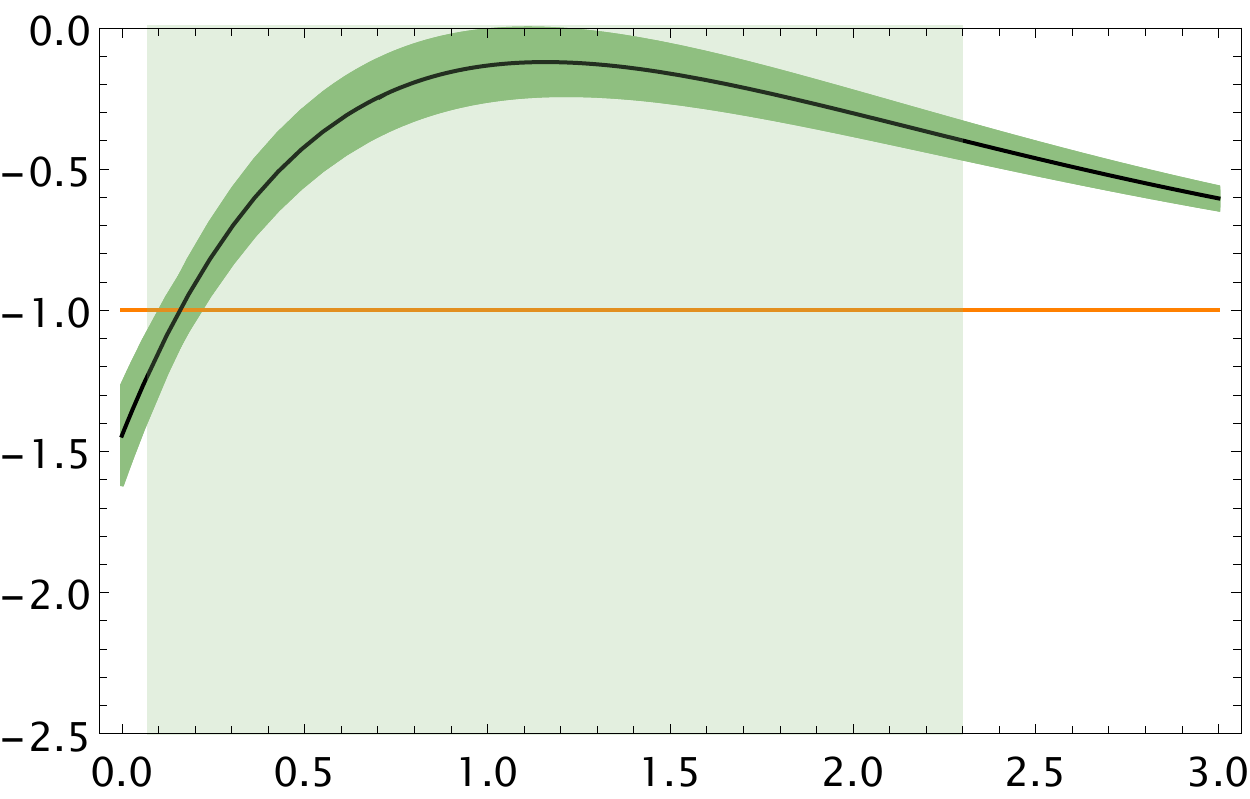} &
       \includegraphics[width=4cm]{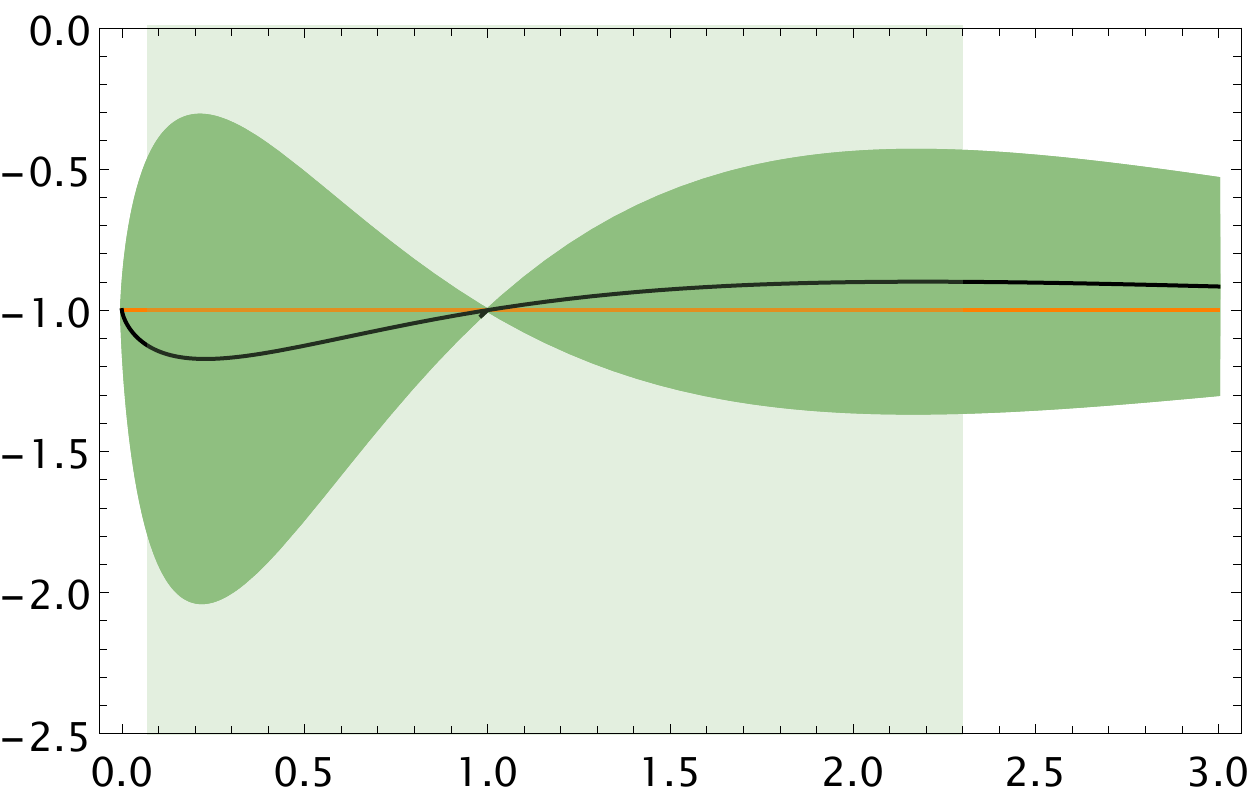} &
       \includegraphics[width=4cm]{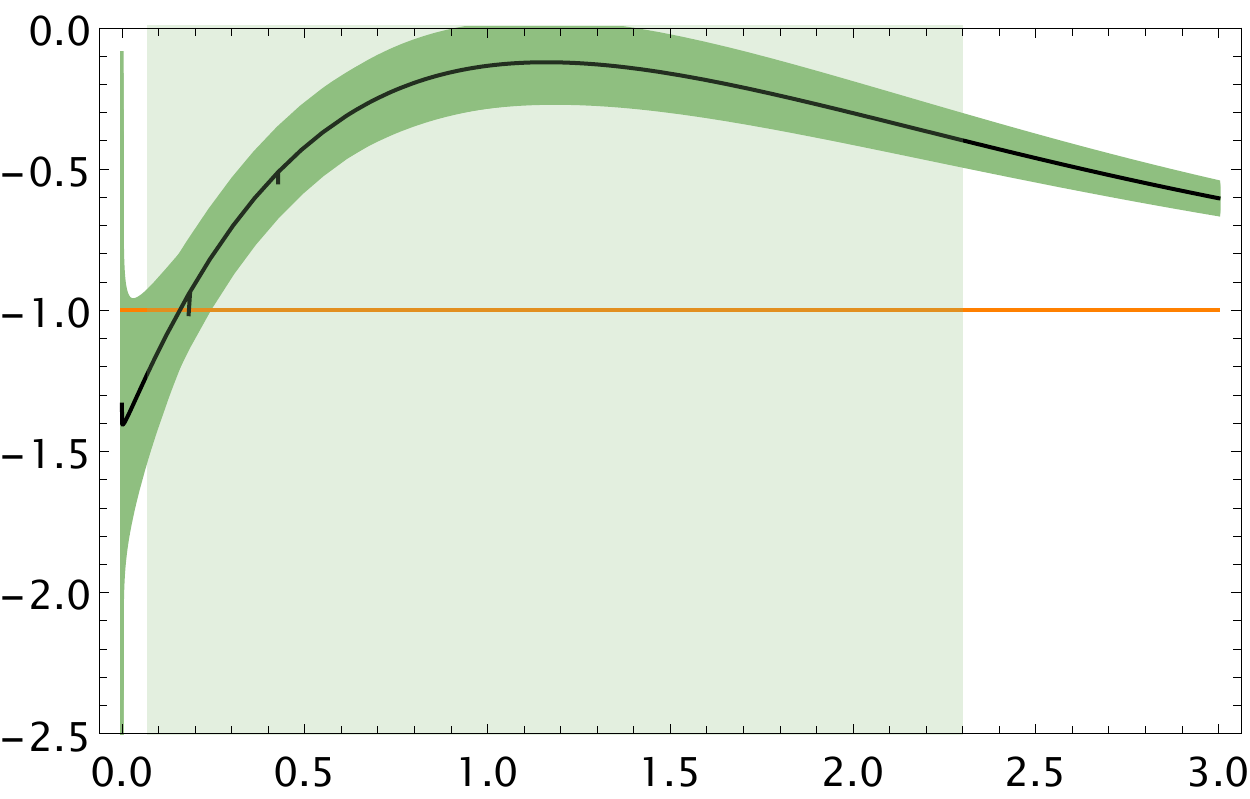}  \\ 
\hline
\rowcolor[gray]{.9} \text{BAO-CMB} & & & \\
\hline
       \includegraphics[width=4cm]{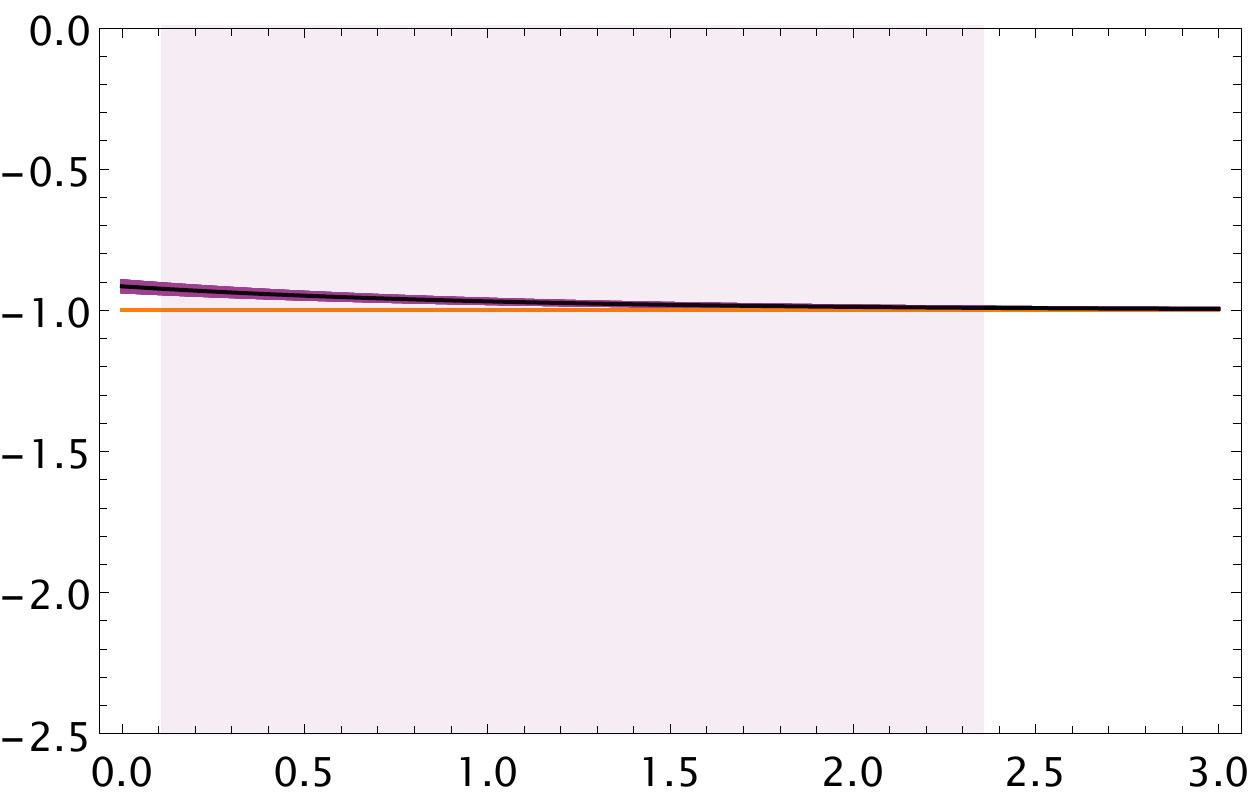} &
       \includegraphics[width=4cm]{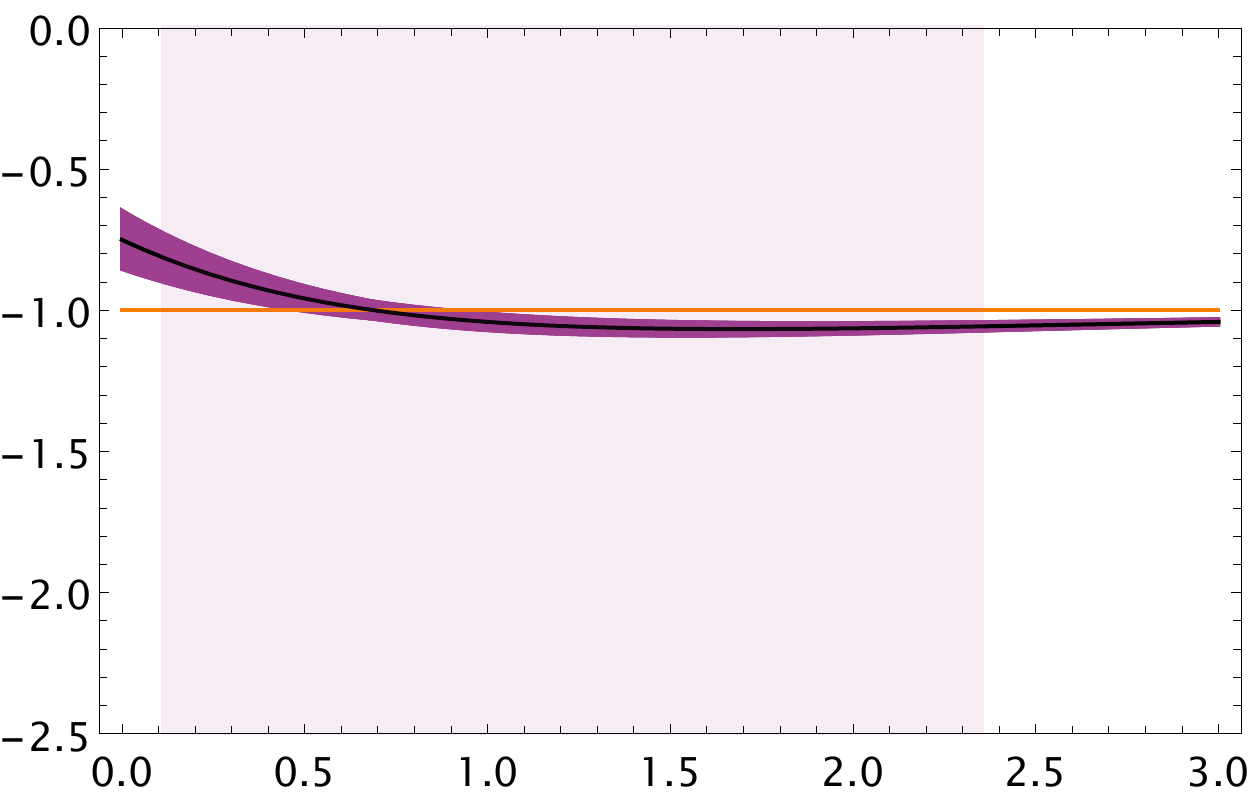} &
       \includegraphics[width=4cm]{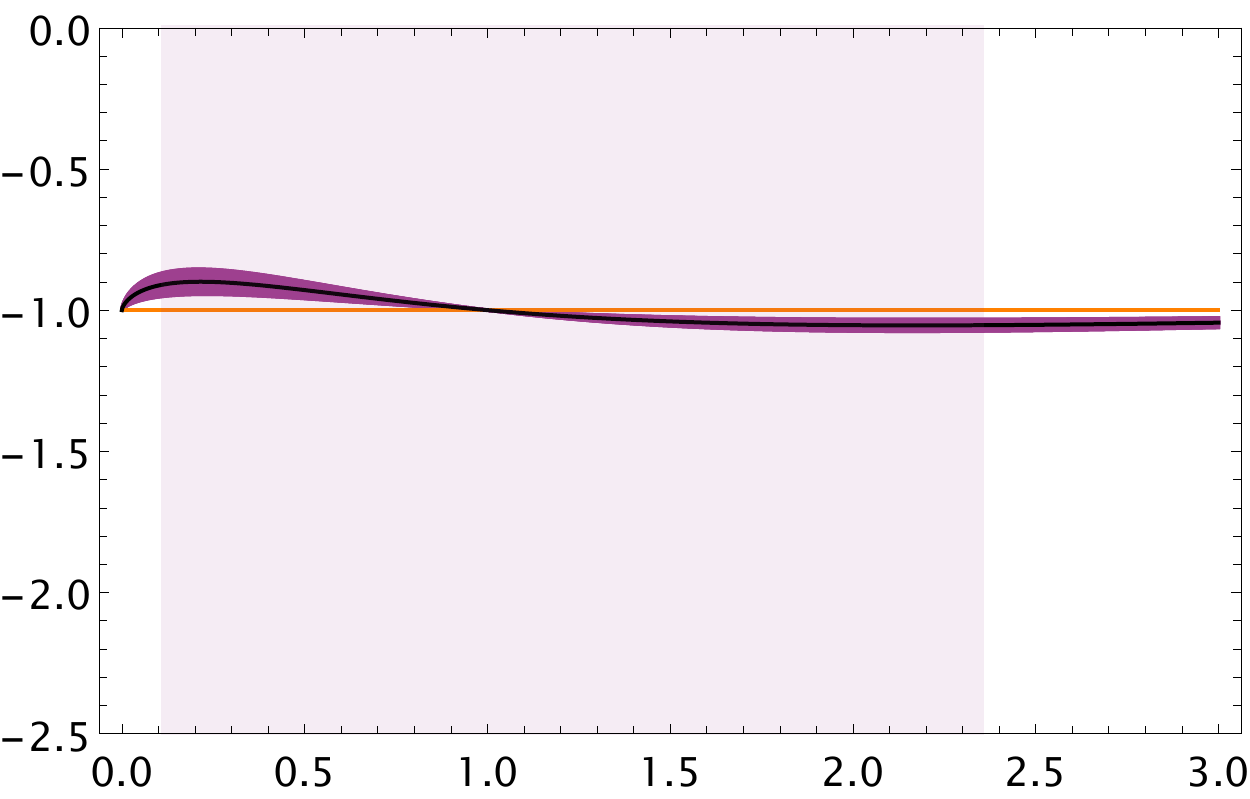} &
       \includegraphics[width=4cm]{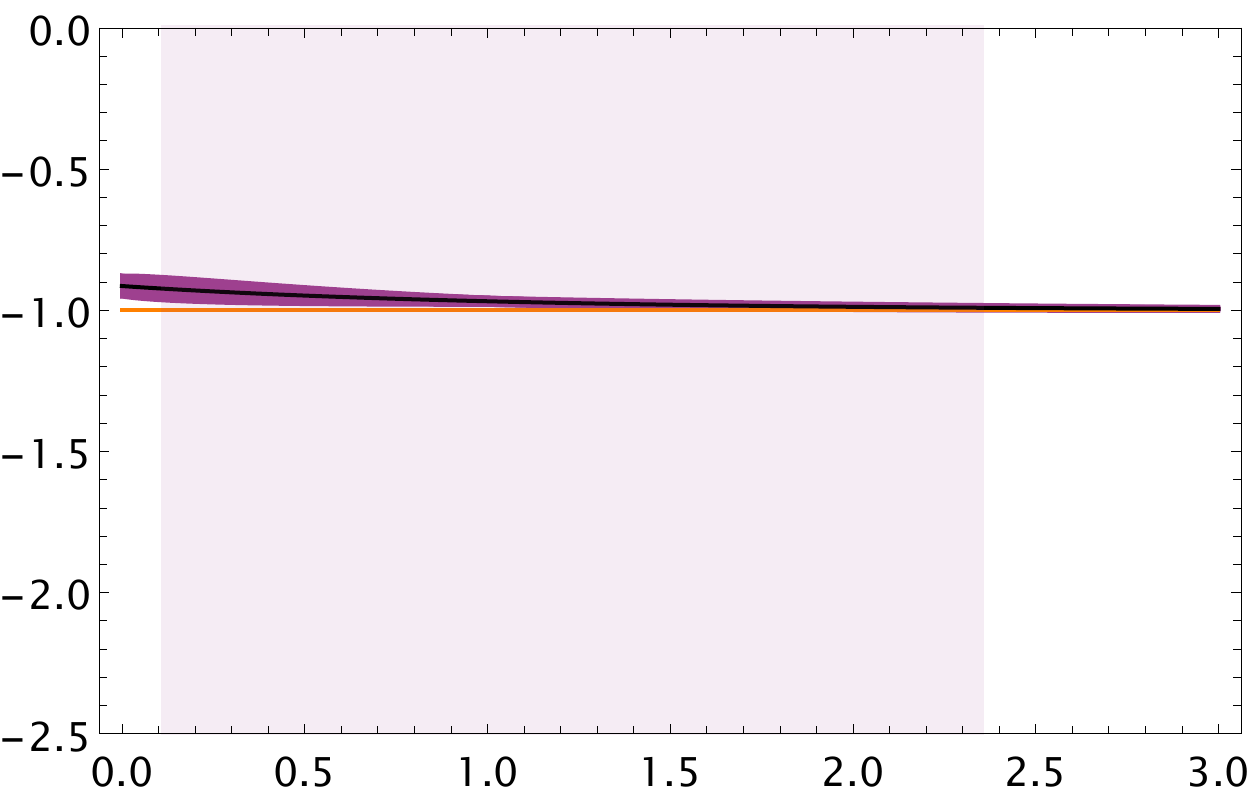}  \\ 
\hline
\rowcolor[gray]{.9} \text{BAO-CC-SNe}   & & & \\
 \hline
       \includegraphics[width=4cm]{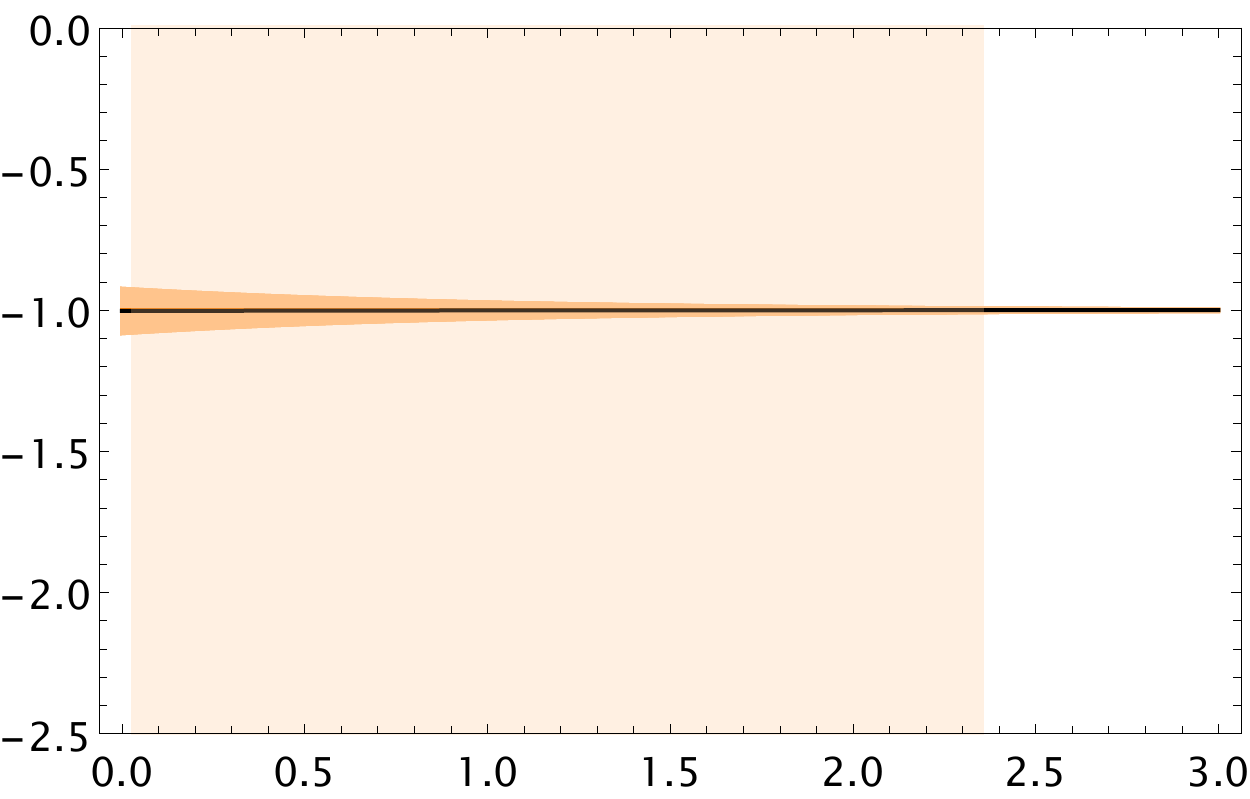} &
       \includegraphics[width=4cm]{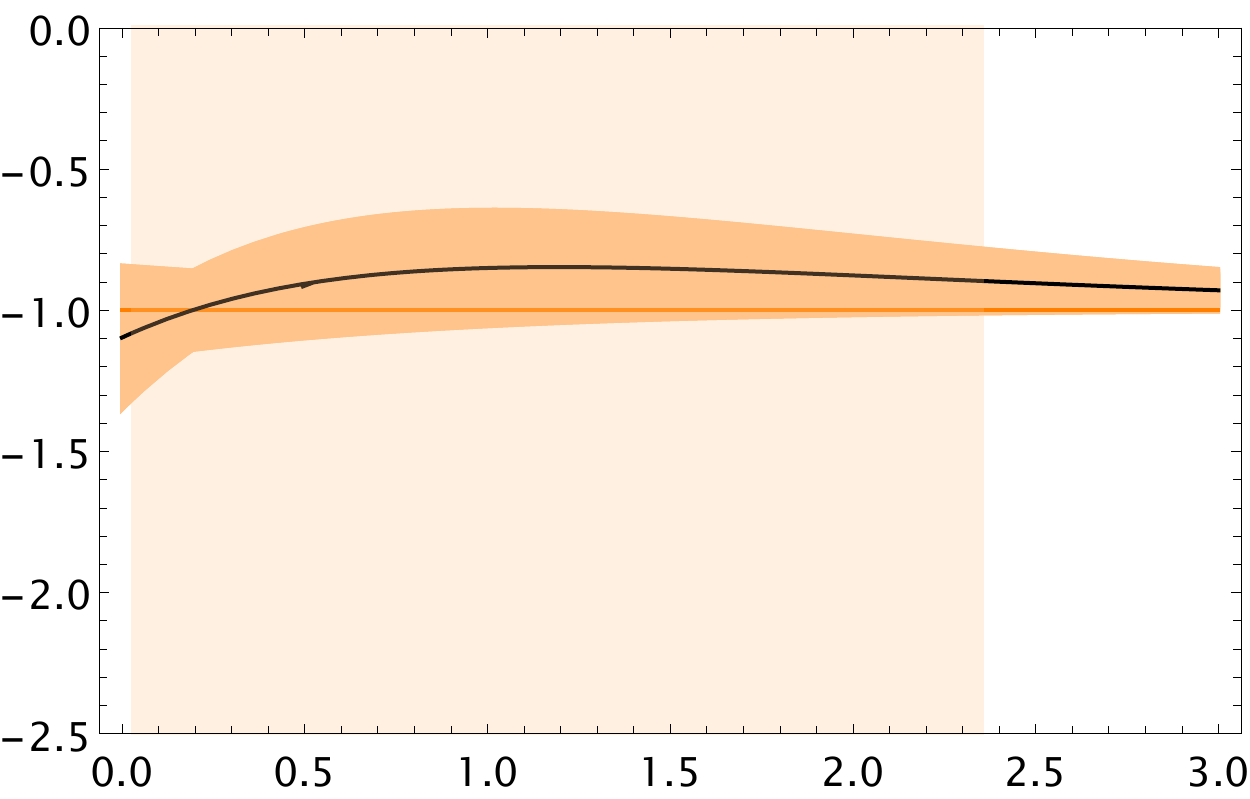} &
       \includegraphics[width=4cm]{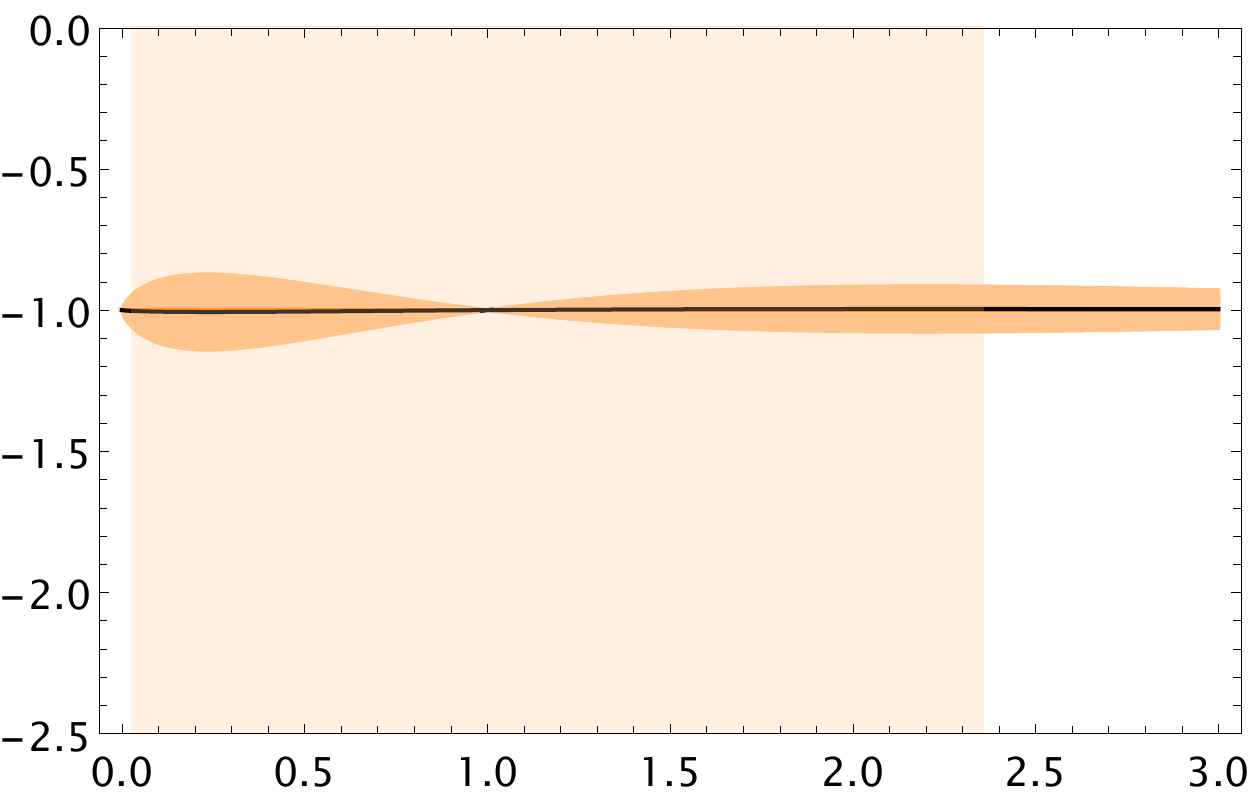} &
       \includegraphics[width=4cm]{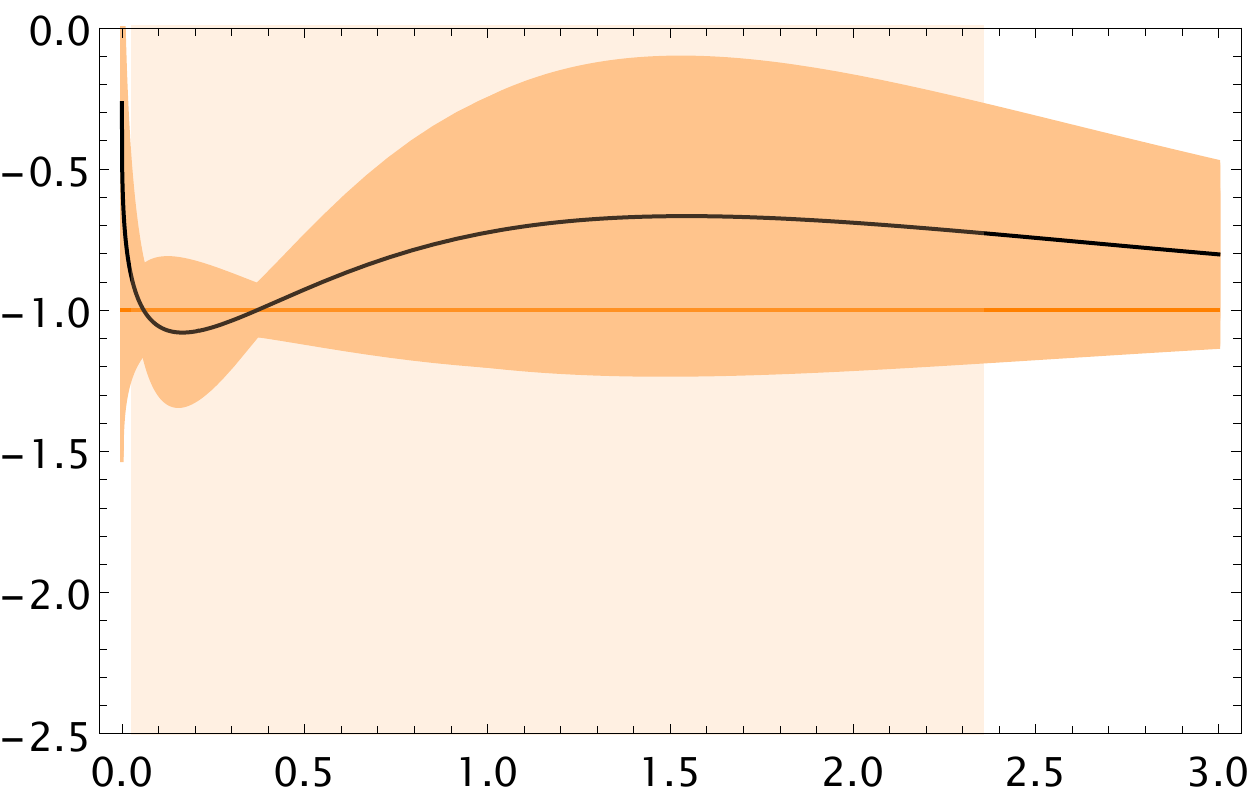}  \\ 
\hline
\rowcolor[gray]{.9} Total    &  & & \\
 \hline
       \includegraphics[width=4cm]{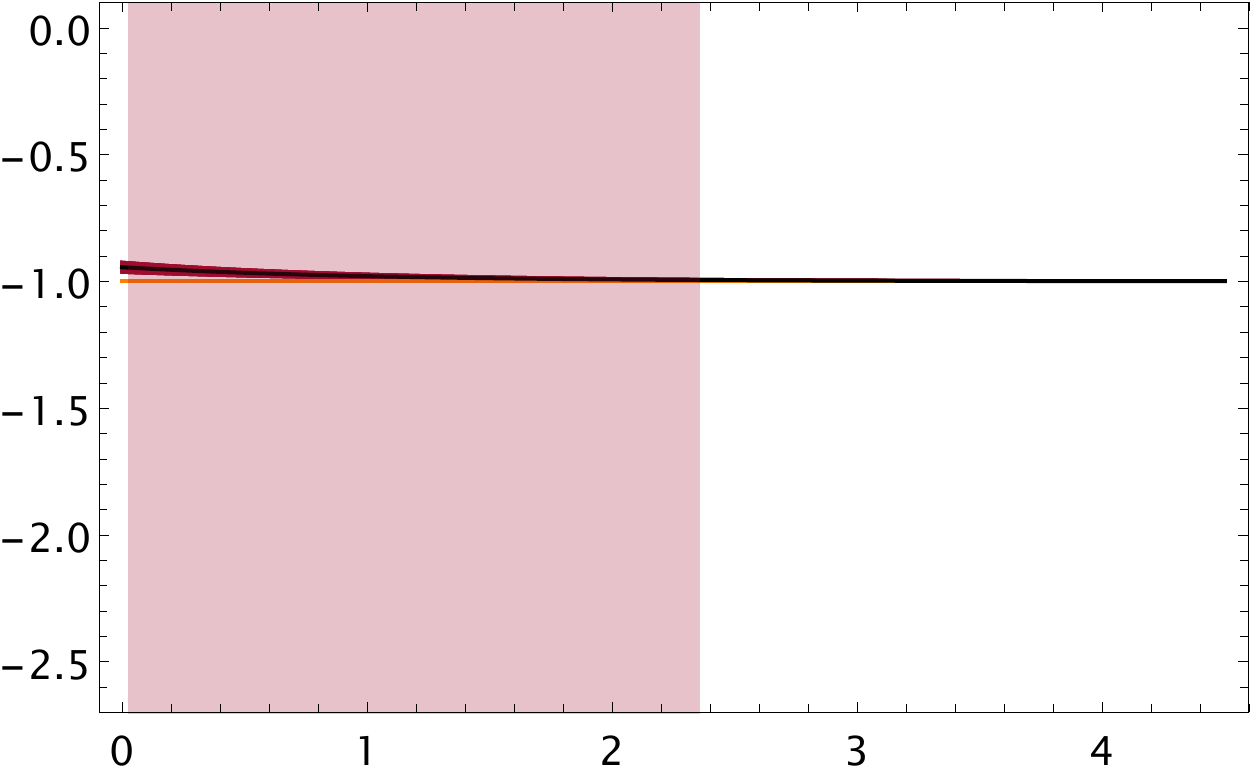} &
       \includegraphics[width=4cm]{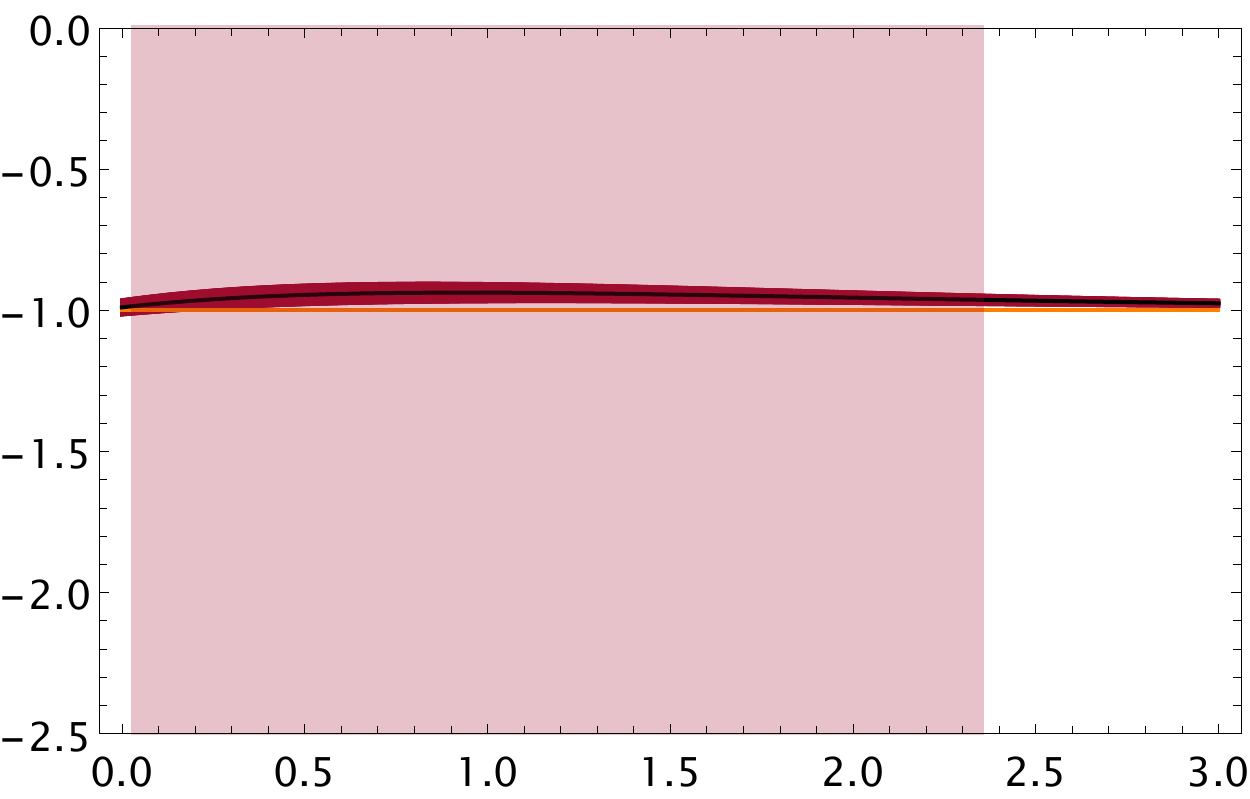} &
       \includegraphics[width=4cm]{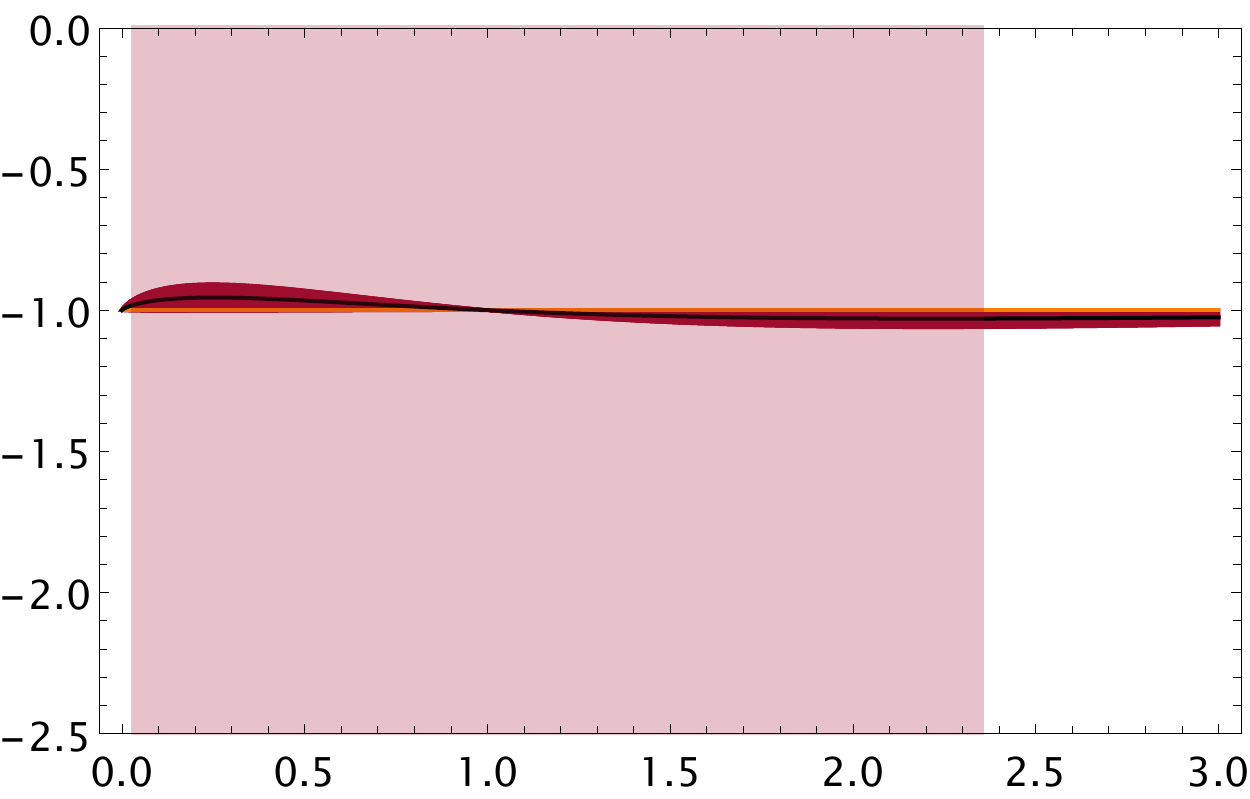} &
       \includegraphics[width=4cm]{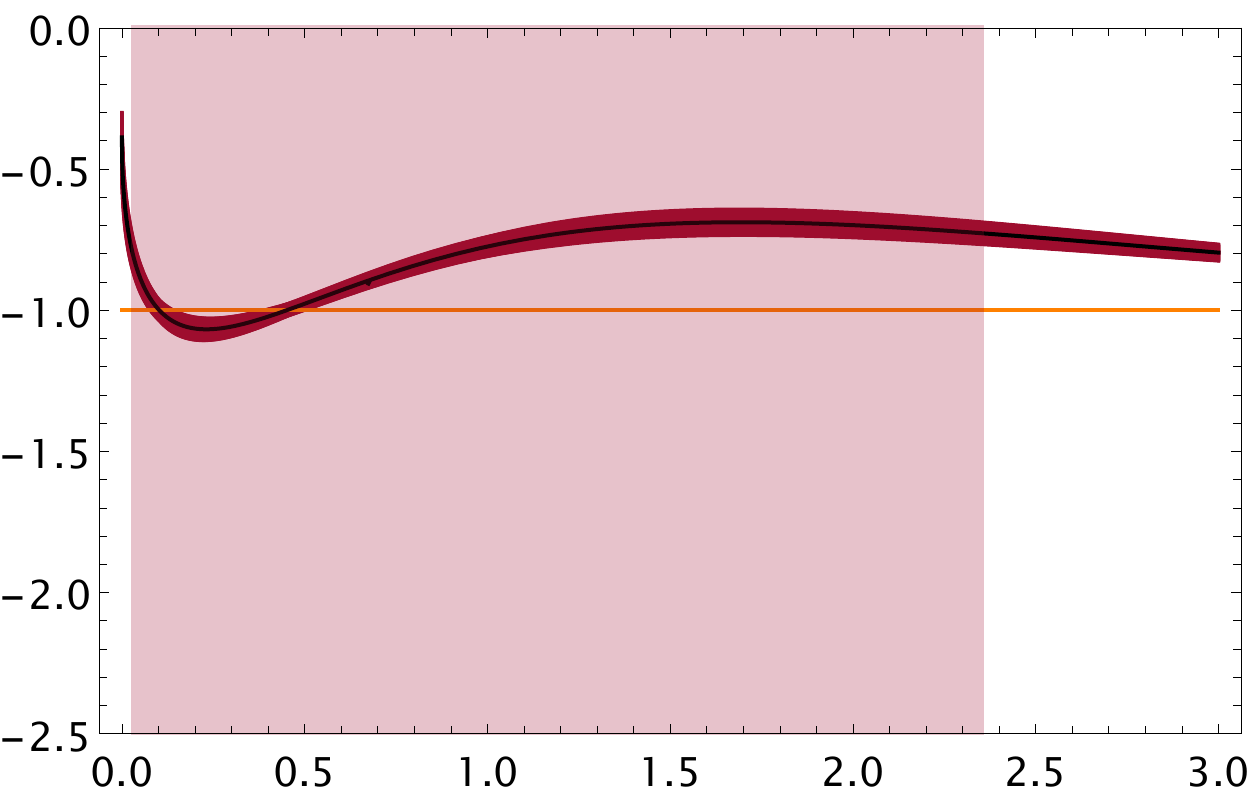}  \\
\hline

\end{tabular}

\end{table*}



\subsection{Direct comparison to observations}\label{sec:comparisonObs}

\begin{figure}[h]
\begin{center}
\includegraphics[width=\linewidth]{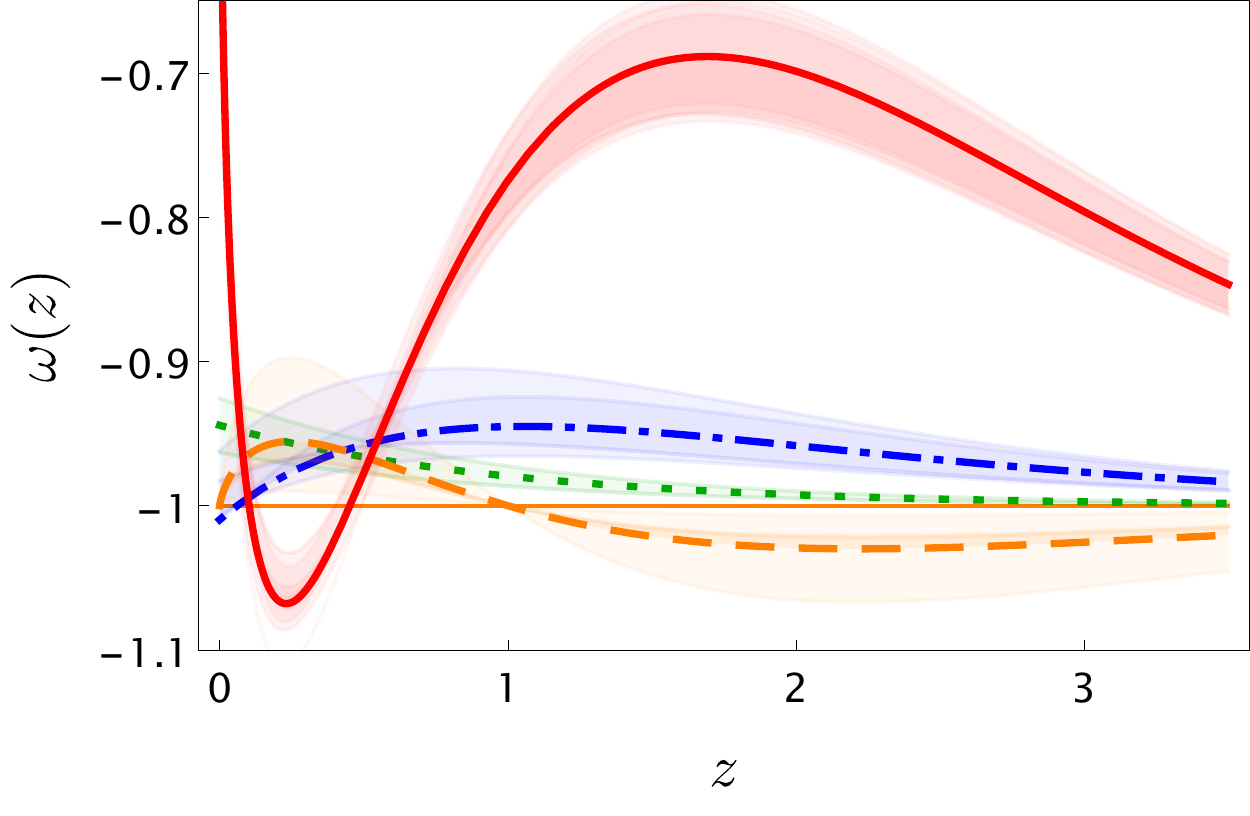}
\caption{(Color on line) Equation of state for the four models constrained by the full observational data set at 99.7$\%$ of CL. Dotted (green) line shows the evolution of model I (Exponential). Dotted-dashed (blue) line depicts the behavior of model II (Quint./Phantom). Dashed (orange) line shows the evolution for model III, $f(R)$, $\omega_0=-1$. Finally, the full model is depicted in the solid (red) line.}
\label{Fig:EOS-all}
\end{center}
\end{figure}

Figure~\ref{Fig:EOS-all} shows the EoS for all the models constrained by the different observations used at $99.7\%$ CL of the BFV. The evolution in the exponential case (green dotted line) goes very close to $\omega(z)=-1$, nevertheless as it approaches $z=0$, the EoS goes toward higher values going to $\omega_0=-1.05\pm {0.006}$. 
In model II, Quintessence/Phantom (blue dotted-dashed line) the EoS reach a value of $\omega_0=-0.99^{+0.008}_{-0.009}$. In this case, the EoS does not cross the phantom line, which is consistent with a single scalar field as in standard quintessence models.
In the case of Model III, $f(R)$, $\omega_0=-1$ (orange dashed line), the value of the EoS was fixed to $\omega_0=-1$, the evolution has the characteristic behavior of $f(R)$ and the $\chi^2_{red}$ value is closer to $1$ than any other model of the parameterization, including the case $A=0$ which corresponds to the standard $\Lambda$CDM model. Finally, the general model (solid red line), with three free parameters, shows an oscillatory behavior that goes to $\omega_0=-0.39^{+0.030}_{-0.031}$ at $z=0$, crossing the phantom line twice.

\begin{figure}
    \centering
   
    \begin{subfigure}[t]{0.49\textwidth}
        \includegraphics[width=\linewidth]{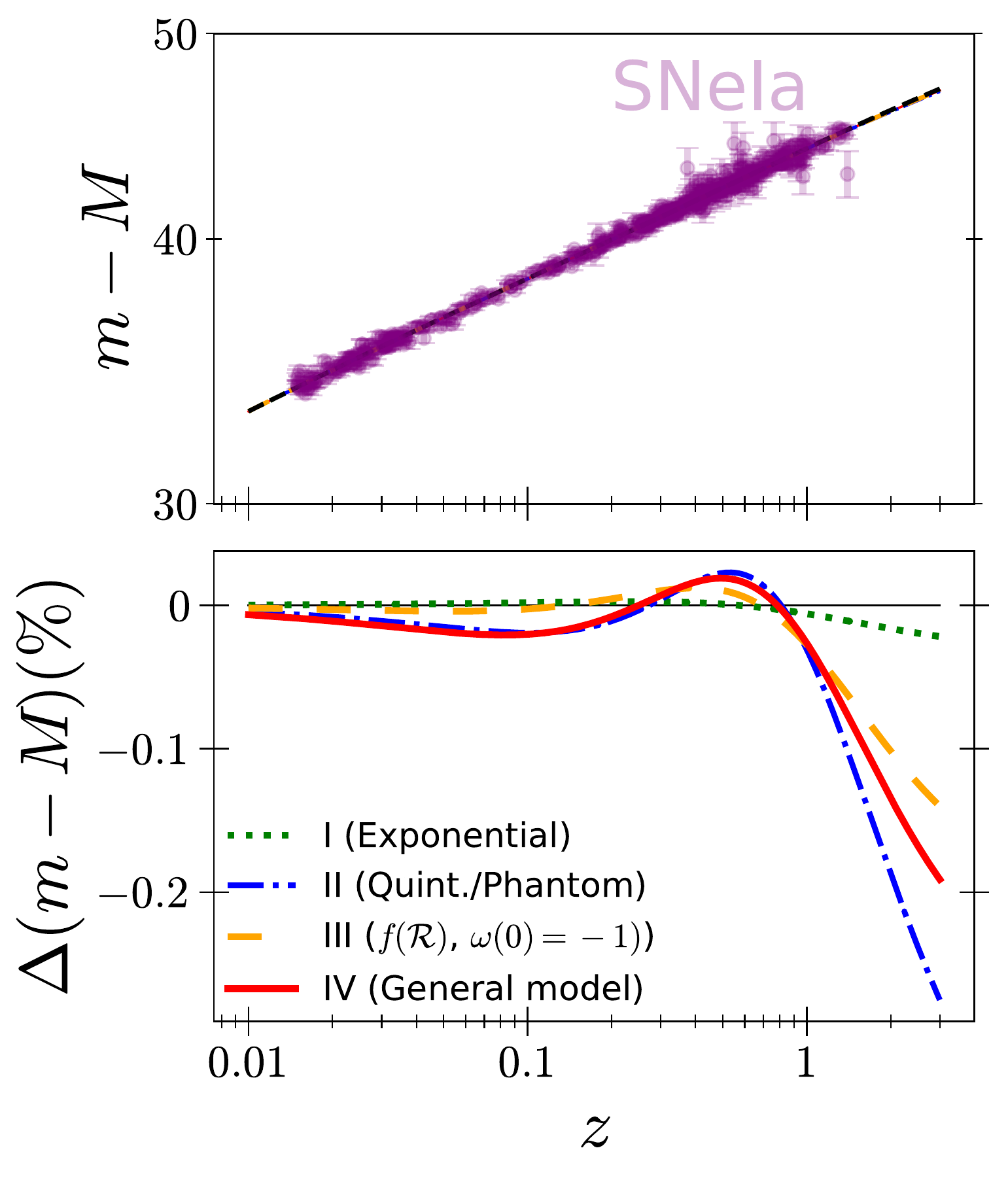}
        \caption{$\mu(z)$}
        \label{fig:muratio}
    \end{subfigure}  
    
     \begin{subfigure}[t]{0.49\textwidth}
        \includegraphics[width=\linewidth]{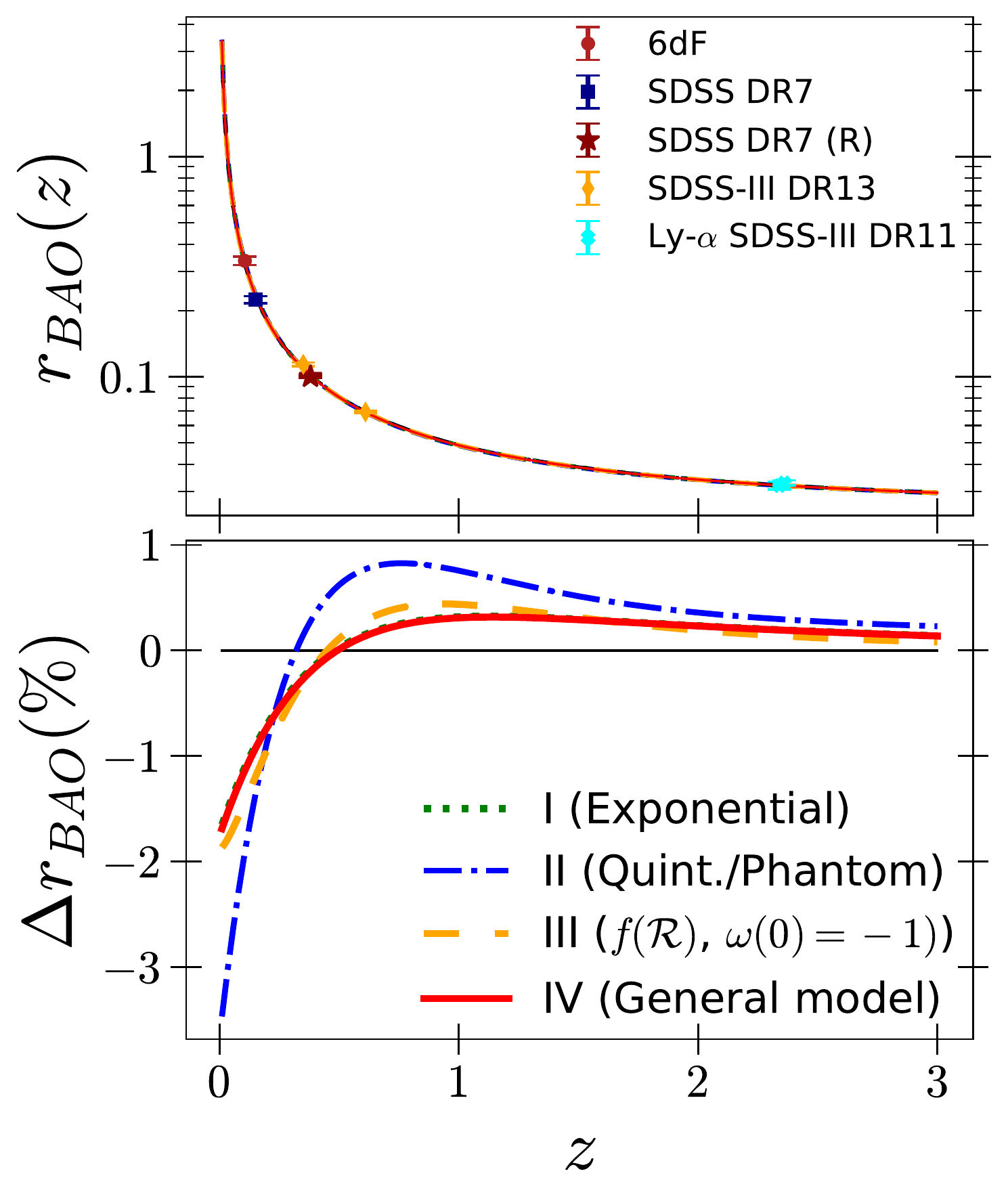}
        \caption{$r_{BAO}(z)$}
        \label{fig:rbaoratio}
    \end{subfigure}
    \caption{Cosmic distances: $\mu(z)=m-M$ , and $r_{BAO}(z)$, along with the data points used to constraint the model. On top we show the best fitting models. The bottom panel shows the relative ratio from each model with respect to $\Lambda CDM$. For the SNe sample, relative differences are below 1$\%$, while for BAO, below 5$\%$. }\label{fig:ratios}
\end{figure}

However, $\omega(z)$ is not directly observable. Hence, to explore how distinguishable the models are from each other,  we compare their fits to each set of data individually. 
Figure~\ref{fig:ratios} shows the prediction for the cosmic distances $\mu(z)=m-M$, for supernovae, and $r_{BAO}(z)$ according with the best fit values obtained constraining the respective data sample assuming each one of the models from table \ref{Table:best-fit-values}. 
In each case we show the direct prediction for the observable quantity (either $\mu(z)$, $r_{BAO}(z)$, or $H(z)/(1+z)$), and the ratio between the best fit for models I-IV to $\Lambda CDM$, $\Delta D = (D-D_{\Lambda CDM})/D_{\Lambda CDM}$. 
In figure~\ref{fig:muratio} we show $\mu(z)=m-M$  for the best fit obtained of all cases. We focus only on the fits done to Union 2.1 supernovae sample. 
The upper panel of the figure shows the predictions for $\mu = m-M$ vs $z$ according to the best fit values obtained for the models, along with the observational points with error bars. 
The bottom panel shows the ratio between  each model's prediction for $\mu(z|A,n, C)$, and $\mu(z)_{\Lambda CDM}$, this is, $\Delta \mu(z) = (\mu(z) - \mu(z)_{\Lambda CDM})/\mu(z)_{\Lambda CDM}$. 
In this case we find differences of the order $\Delta \mu(z) = 0 - 0.3\%$. 
Figure~\ref{fig:rbaoratio} shows the evolution of $r_{BAO}(z)$ vs $z$ for our models, along with the observations we used in this work. The bottom panel contains the ratio $\Delta r_{BAO}(z)$, from where we see that the  Quintessence/Phantom model differs the most from the $\Lambda CDM$ prescription. Nevertheless, the difference is below $\Delta r_{BAO}\approx1\%$ in all cases, reaching the maximum discrepancy around $z\approx0.1$.  
This range of redshift ($0.05<z<0.4$) will be accurately measured using BAO in the Bright Galaxy Survey (BGS) \cite{Ruiz_Macias_2020} by DESI, making it possible to accurately differentiate between these models at low redshifts.

\begin{figure}[h] 
        \includegraphics[width=\linewidth]{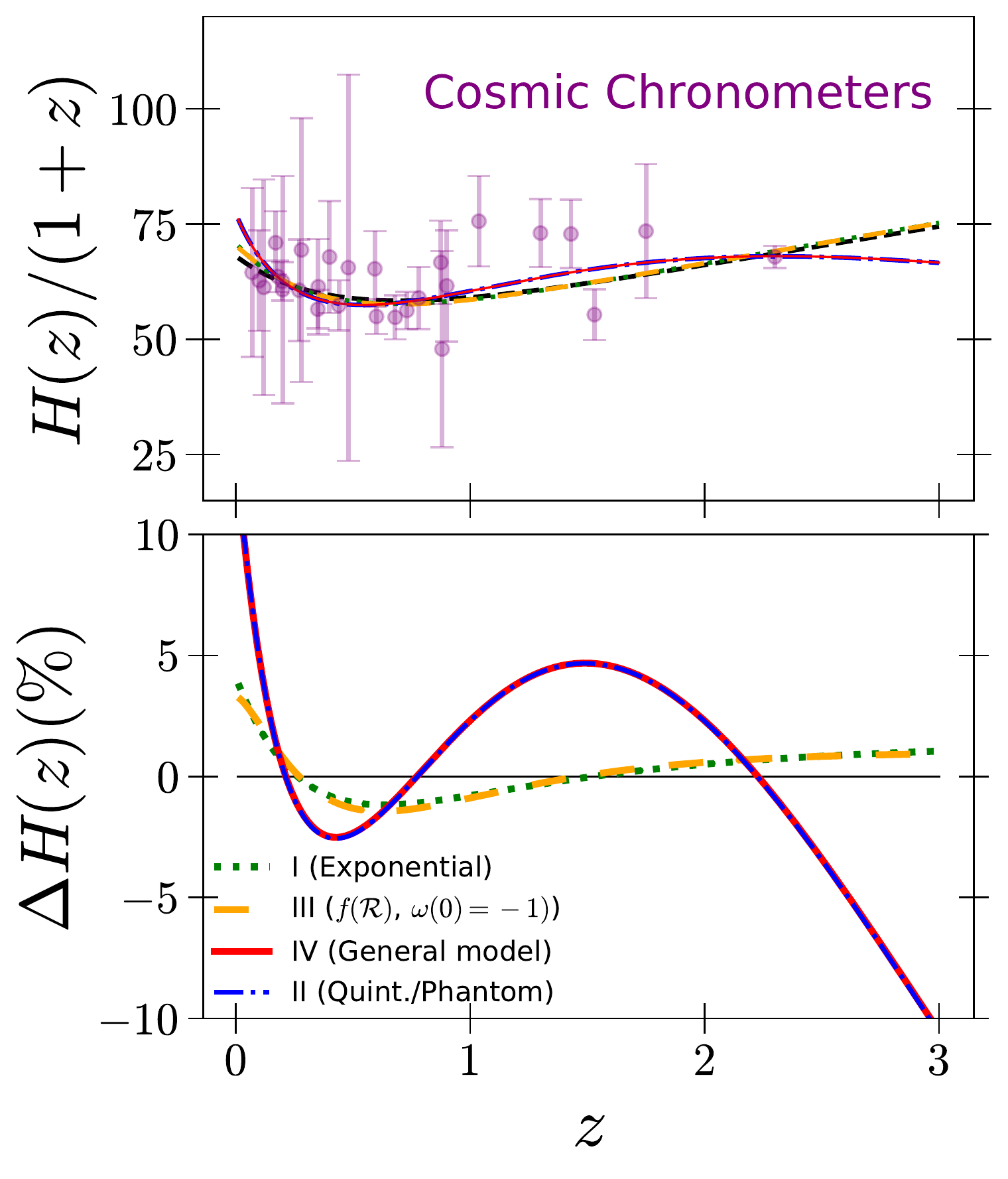}
        \caption{Hubble function, $H(z)/(1+z)$, along with the cosmic chronometers data set used. The bottom panel shows the relative ratio from each model with respect to $\Lambda CDM$. Relative differences reach $10\%$. }
        \label{fig:hzratio}
\end{figure}

Figure~\ref{fig:hzratio} shows the evolution of $H(z)/(1+z)$ vs $z$ for our models and their best fit to the CC sample.
We superimpose the observations with error bars. 
In this case, we notice bigger discrepancies between our models' predictions for the quantity $H(z)/(1+z)$, and $\Lambda CDM$. $\Delta H(z)$ reaches a value of $\approx 10\%$. 
The exponential (case I) and  $f(R)$-like models (case III) differ from $\Lambda$CDM in the same proportion, with up to a $\Delta H(z)\approx3\%$ at $z=0.1$, $\Delta H(z)\approx-1.5\%$ around $z\approx0.5$, and $\Delta H(z)\approx+1.5\%$ at $z\approx3$. 
On the other hand, Quintessence/Phantom  (case II) and the General model show the same pattern in their discrepancies  from $\Lambda$CDM, but with  $\Delta H(z) \approx H(z)5\%$ at $z=0.1$, $\Delta H(z)\approx-3\%$ around $z\approx0.3$, $\Delta H(z)\approx+5\%$ at $z\approx1.5$, and peaking at $z\approx3$, with $\Delta H(z)\approx-10\%$. 
Something worth to be mentioned is the change in sign of $\Delta H(z)$, so even when the errors for this particular observable are systematically dominated and hence, bigger than for the other type of observations, the oscillatory behavior in $\Delta H(z)$ can potentially help distinguishing them. 

\section{Model comparison}
\label{sec:aicbic}
We introduced a single parametric equation which is able to describe the expansion history of the Universe according to a number of models with different physical origins. This is expressed in the possibility for different dynamical features in $\omega(z)$.  Such dynamics can be realized with different number of parameters.

In view of the flexibility of our proposal and the number of models encoded, a natural question would be how to select among them.

Rather than determining the values of parameters sets according with some data, the aim of model selection is to make an objective comparison between competing models (which may feature different numbers of parameters) against the same data set.

Following other works in the literature (see for example \cite{Liddle:2004nh, Liddle:2007fy}), we compute the Akaike Information Criterion (AIC), and the Bayesian Information Criterion (BIC). Both criteria confront the peak of the likelihood distribution, $\hat{L}$, with an additional term penalizing the complexity of each model. Given a collection of models for the data, they estimate the quality of each model, relative to each of the other models.

AIC is founded on information theory and it is defined as:

\begin{equation}
    \mathrm {AIC} = 2k-2\ln({\hat {L}}),
\end{equation}
where $k$ is the number of estimated parameters in the model. 

BIC on the other hand, is  rooted in Bayesian inference, which rely on the computation of Bayes factors as a Bayesian alternative to classical hypothesis testing (see for instance \cite{Trotta:2005ar}). However, the information criterion approach provides a simple objective way of deciding for the inclusion of new parameters. The BIC is formally defined as 
\begin{equation}
     \mathrm {BIC} =k\ln(n)-2\ln({\hat {L}}).
\end{equation}
 penalizing the complexity of the model is $k\cdot\ln(n)$, with $n$ the dimension of the data set and $k$, the number of free parameters.

In both cases, $\hat{L} =p(x\mid {\widehat {\theta }},M)$ where $ {\displaystyle {\widehat {\theta }}}$ are the parameter values that maximize the likelihood function.
The BIC generally penalizes free parameters more strongly than the AIC, although it depends on the size of $n$ and the relative magnitude of $n$ and $k$.
The preferred model is the one with the lowest  AIC/BIC value, against which the rest are to be compared.  

If we are agnostic about  the nature of the mechanism behind the accelerated cosmic expansion, we can compare the models contained in our parameterization along with other prescriptions found in the literature. We include for this analysis three more cases to compare our proposal with other parameterizations found in the literature: JJE (\cite{Jaime:2018ftn}), CPL ($\omega(a) = w_0 + w_a(1-a)$) (\cite{Chevallier2001,Linder2003}), and wCDM ($\omega = $ constant).

\begin{figure}[h]
    \centering
    \begin{subfigure}[t]{0.48\textwidth}
        \includegraphics[width=\linewidth]{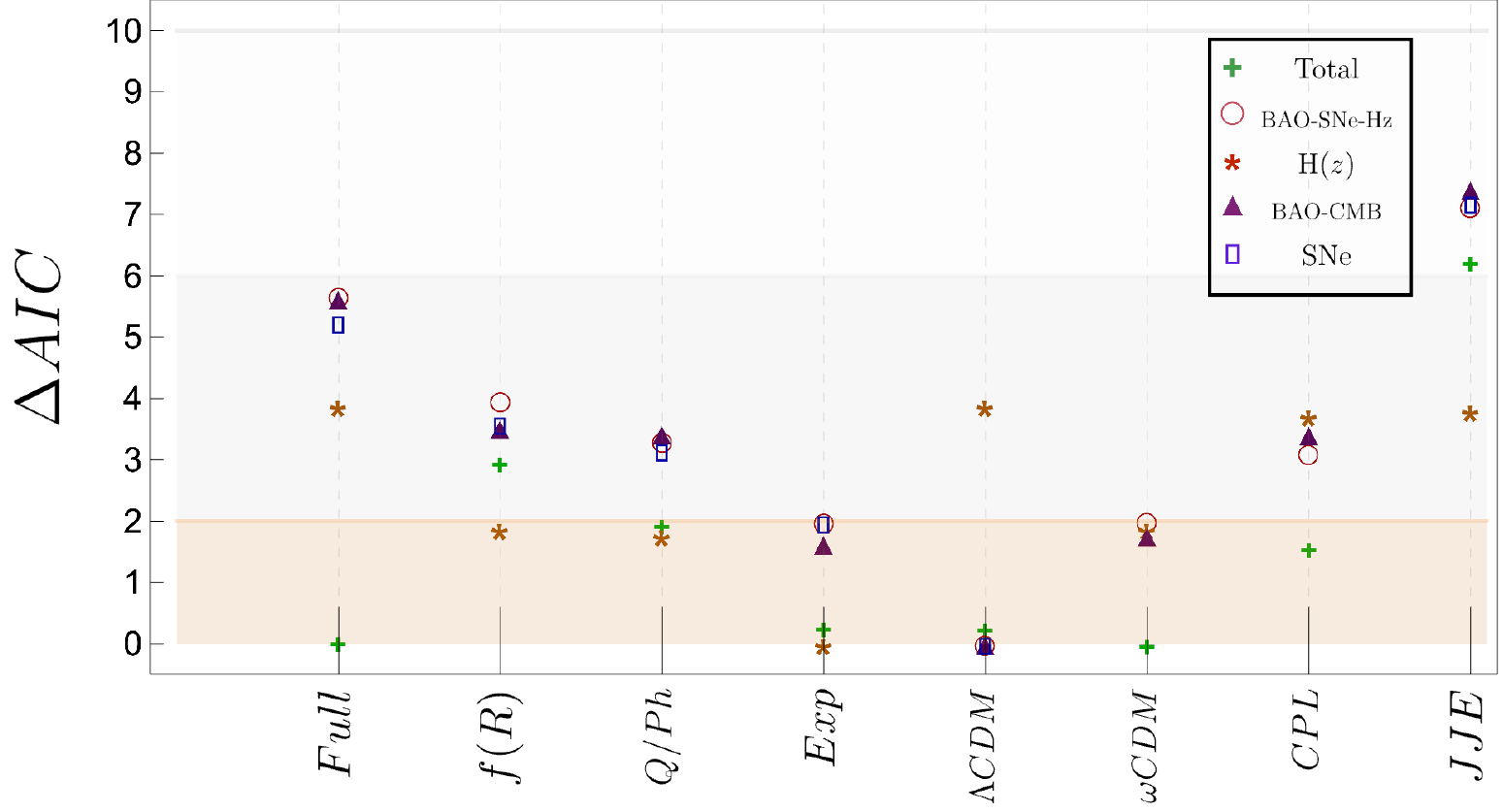}
        \caption{$\Delta$AIC for all the models disregarding their physical behavior}
        \label{fig:DeltaAICagnostic}
    \end{subfigure}
   
     \begin{subfigure}[t]{0.48\textwidth}
        \includegraphics[width=\linewidth]{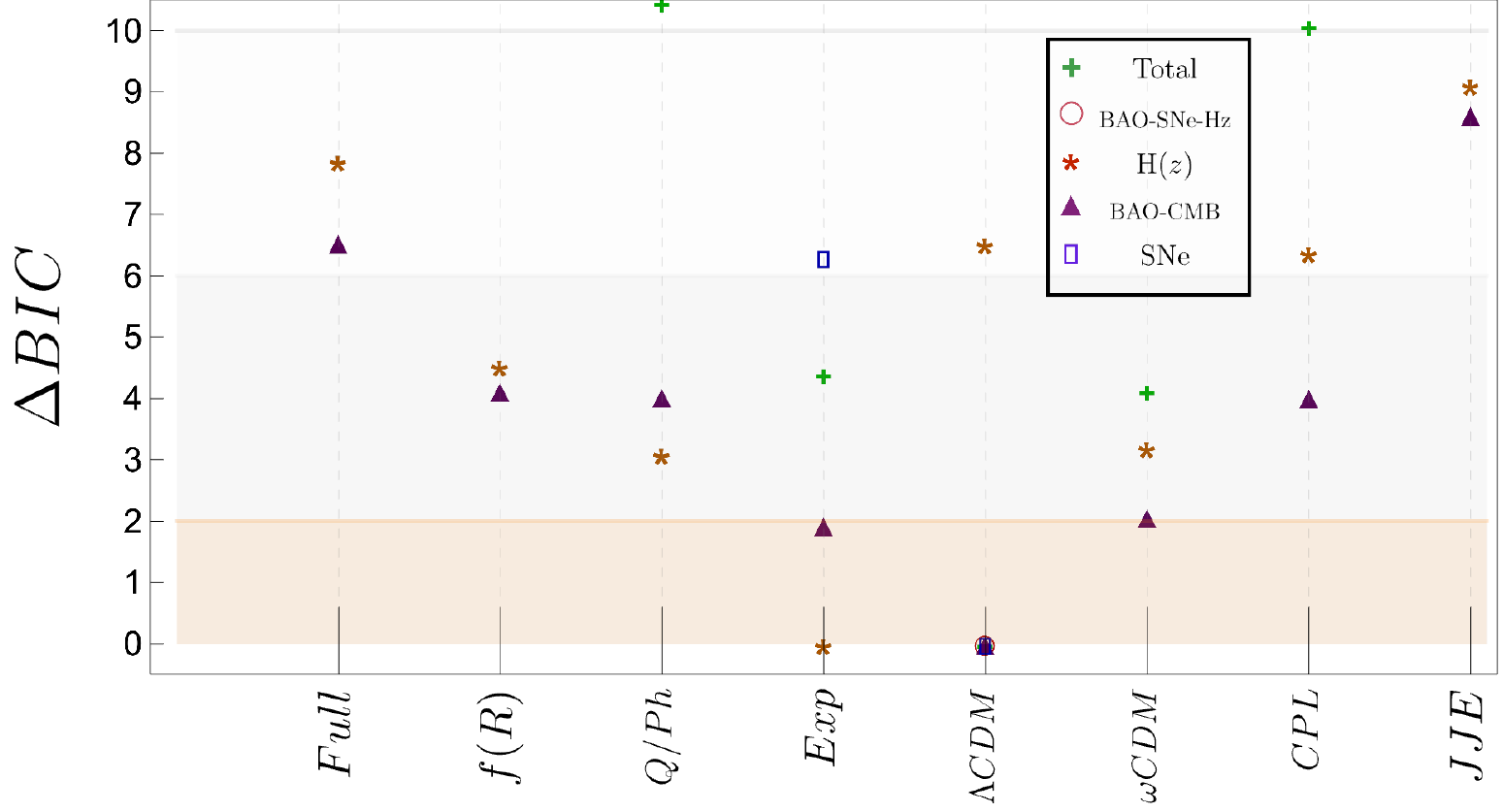}
        \caption{$\Delta$BIC for all the models disregarding their physical behavior.}
        \label{fig:DeltaBICagnostic}
    \end{subfigure}

    \caption{$\Delta$AIC and $\Delta$BIC values for eight parameterized models disregarding their physical behavior. The Akaike information criterion is shown in figure (a), and the Bayesian information criterion (b). }
    \label{fig:DeltaAIC-BIC_agnostic}
\end{figure}

Figure \ref{fig:DeltaAIC-BIC_agnostic} shows the results for all models considered when the different sets of observations are considered: $H(z)$, SNe, BAO-CMB, BAO-SNe-$H(z)$ and the Total likelihood.  
From figure \ref{fig:DeltaAICagnostic}, depicting the values of $\Delta AIC$, we see that all our models are equally competitive under different data sets, including also the cases of wCDM  and CPL.
Importantly, we note that  under the total likelihood, the Full Model was ranked even better than a cosmological constant solution. 
The only model that was penalized in light of this criterion was the JJE parameterization, which mimics with sub-percent precision the dynamics from viable $f(\mathcal{R})$ theories, but  scored poorly in this comparison.

In figure \ref{fig:DeltaBICagnostic} we find the results coming for the BIC values. 
It is well known that BIC penalizes more stringently the complexity of models. 
In this regard, our results are consistent. 
From the models compared, although the cosmological constant solution has the lowest BIC value, the exponential (equation \eqref{CaseI}) and wCDM parameterizations, both with a single extra parameter, lay within the same range as $\Lambda CDM$. 
Unlike $\Lambda CDM$ or wCDM, for the exponential  model, it is possible to have a physical connection to a background model as in a quintessence framework \cite{Roy:2018nce}.  Within this agnostic comparison, the complexity of JJE was severely penalized as it is not even shown in the plot within the Jeffrey' scale. 

On the other hand, if we are not completely agnostic about the underlying theory or model for accelerated cosmic expansion, but we have a physical theory guiding us, we might look at how different proposals compare. 
To perform this analysis, and given the different nature of the models included in our parameterization, we contrast the evidence amongst models with similar dynamics: according to the number of crossings around the phantom dividing line, $\omega=-1$,\textbf{(a)} those with two crossings, \textbf{(b)}those with a single crossing, and  \textbf{(c)} the constant models which do not cross the phantom line. 

In category \textbf{(a)}, for the oscillatory cases, we have the JJE parameterization, the case we coined $f(R)$ within our present proposal (eqn. \ref{CasefR}), and  the general case of our model (eqn. \ref{eq:full-model}). 
Category \textbf{(b)} includes the cases that can be associated with scalar fields: the exponential model (eqn. \ref{CaseI}), quintessence/phantom cases (eqn. \ref{Fig:CaseIIa}), and the CPL parametric equation of state. Lastly, category \textbf{(c)} includes the cases of constant equation of state such as $\Lambda CDM$ and $wCDM$ ($\omega=$constant).

\begin{figure}
    \centering
    \begin{subfigure}[t]{0.48\textwidth}
        \includegraphics[width=\linewidth]{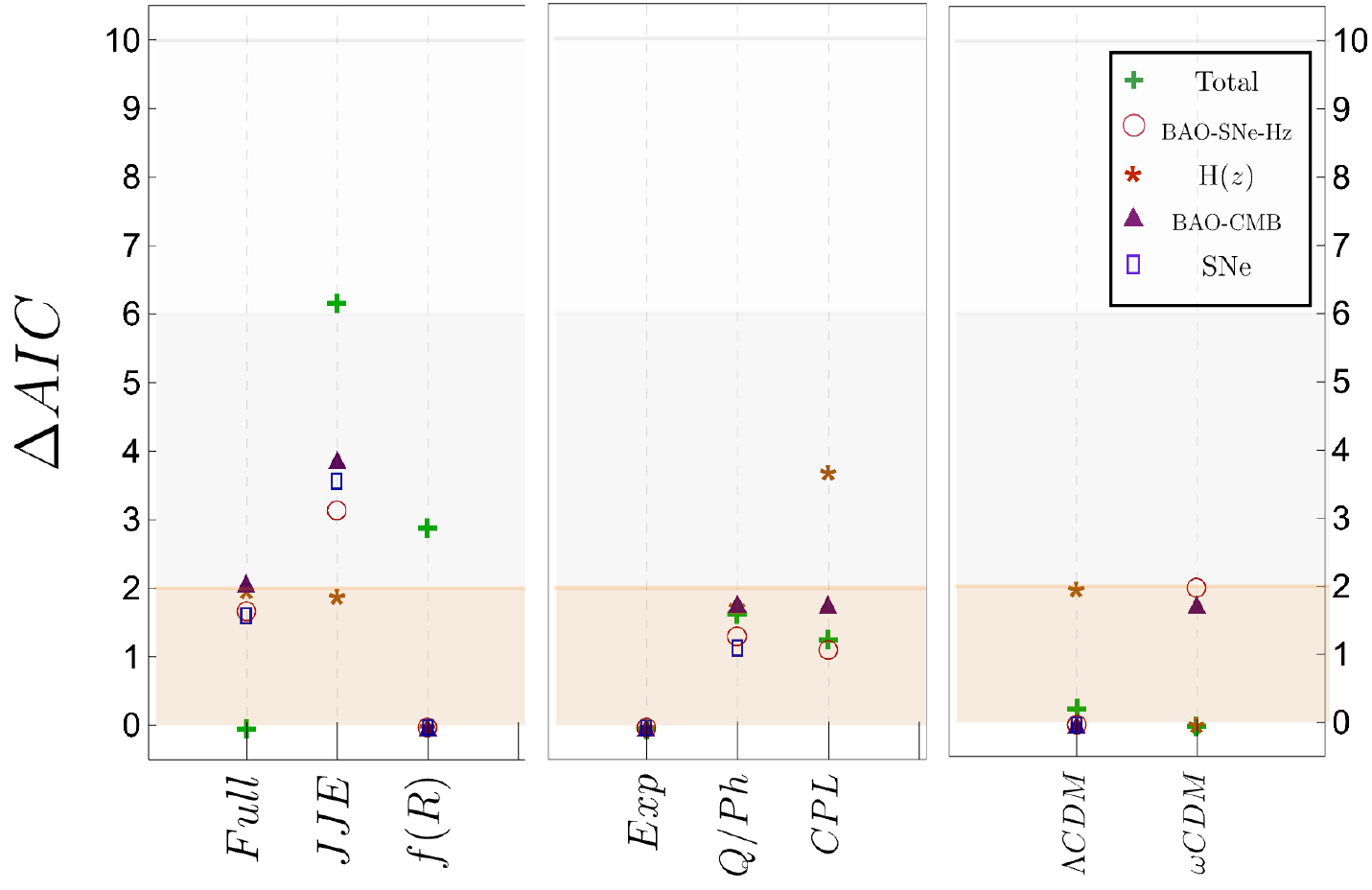}
        \caption{$\Delta$AIC for models grouped by the oscillatory or physical behavior. Values, for CPL, and $\omega$CDM, both using SNe data, are out of the plot, with values: 53.80, and 57.79, respectively.}
        \label{fig:DeltaAIC}
    \end{subfigure}
   
     \begin{subfigure}[t]{0.48\textwidth}
        \includegraphics[width=\linewidth]{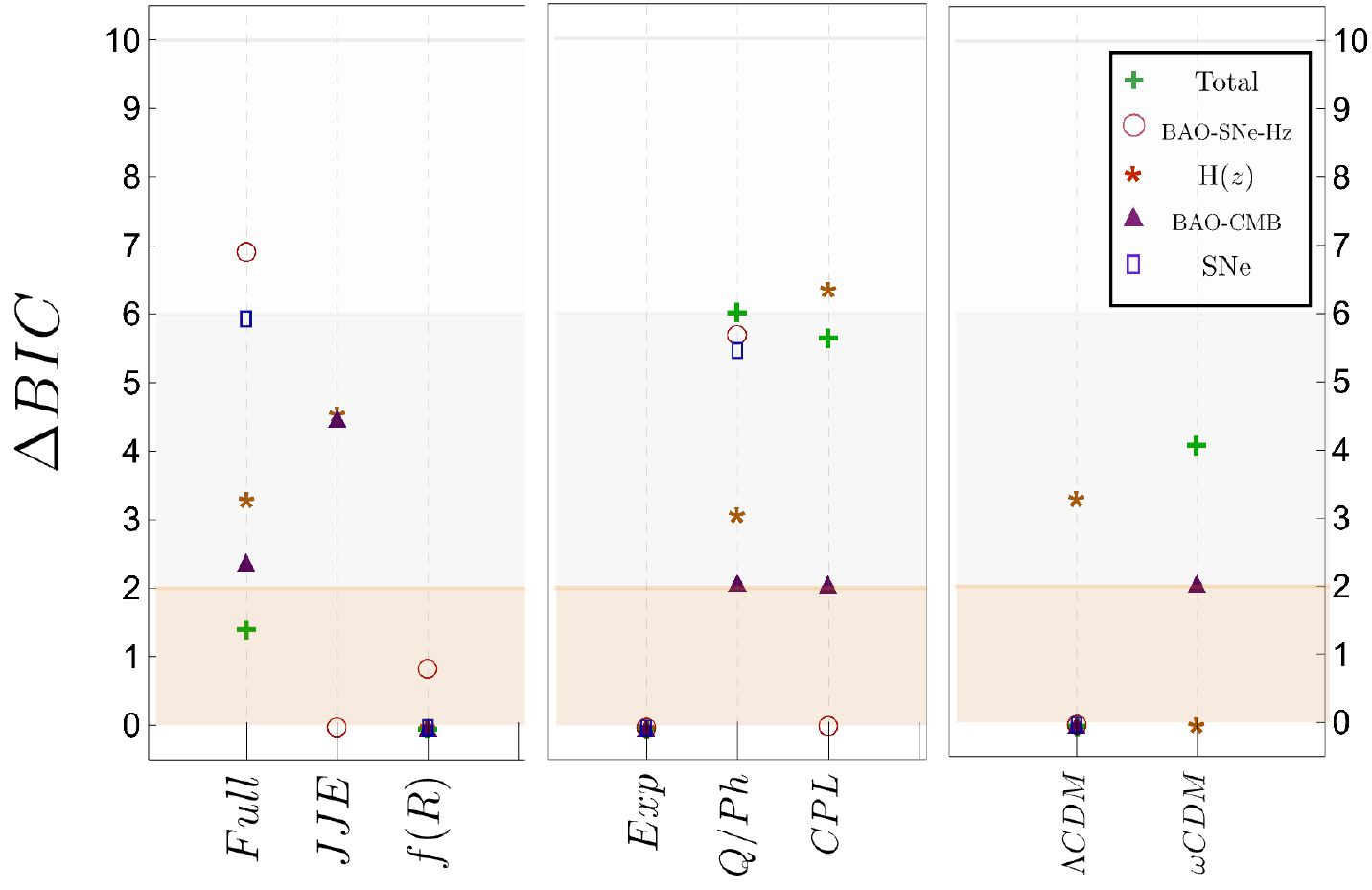}
        \caption{$\Delta$BIC for models grouped by the oscillatory or physical behavior. Values for JJE  using SNe ($\Delta$BIC=12.25), Total ($\Delta$BIC=12.05), CPL with SNe ($\Delta$BIC=58.13),  and $\omega$CDM with SNe ($\Delta$BIC=62.11) and BAO-SNe-H(z) ($\Delta$BIC=583.74) are out of the plots.}
        \label{fig:DeltaBIC}
    \end{subfigure}

    \caption{$\Delta$AIC and $\Delta$BIC values for eight parameterized models grouped accordingly to the oscillatory or physical behavior. The Akaike information criterion is shown in figure (a), and the Bayesian information criterion (b). The left panel compares the parameterization allowing the three free parameters to vary, labeled `Full'; the parameterization with $c=0$, labeled `$f(R)$'; and the JJE parameterization. Central panel compares the parameterization with $n=1$, labeled `Exp', with $n=0$ for Quintessence/Phantom, and the CPL parameterization. Right panel compares $\Lambda$CDM with $\omega$CDM.}
    \label{fig:DeltaAIC-BIC}
\end{figure}

Figure  \ref{fig:DeltaAIC} shows the relative values of AIC. The left panel shows the oscillatory models. The middle panel shows the cases that could be associated with a scalar field. The right panel shows the constant cases. Figure \ref{fig:DeltaBIC} relative values for the BIC. As before,  we show the values for the criteria when the different sets of observations are considered: $H(z)$, SNe, BAO-CMB, BAO-SNe-$H(z)$ and the Total likelihood.

Lets us discuss the information criteria in this sense, where we prioritize the physics of the EoS.

We can notice that the other two oscillatory models are highly disfavored by this criterion. When we take observations separately, the favored model is the $f(R)$ case for all the other sets of data. The JJE case is disfavored under this criterion for all the data sets. For the Bayesian Information Criterion, when the total data collection is considered, the favored oscillatory model is the $f(R)$ case, with an important difference with the JJE case where the value is out of the figure. The full model seems disfavored for all data sets. The only case where the model JJE is favored is for the BAO-SNe-$H(z)$ data set.

For the non-oscillatory and non-constant cases (middle panel of figures \ref{fig:DeltaAIC-BIC} (a) and (b)), the AIC favored case is the exponential model for all the data sets. It is worth noticing that the CPL case is highly disfavored under this comparison for the SNe data. The case Quin/Phantom of the present proposal, even when it is not the favored model, the values of the criterion are in the middle region.

The main conclusions from this informed model selection analysis are: (a) if we take an oscillatory behavior for $\omega(z)$, the f(R) class within our parameterization is the preferred model, (b) if the underlying theory comes from a scalar field theory, the exponential case is the preferred model, and (c) the cosmological constant is the preferred model for a constant equation of state prescription.


\section{Conclusions and discussion}
\label{sec:conclusions}
We present a new parameterization that can reproduce the generic behavior of the most widely used physical models for accelerated expansion with infrared corrections. 
Our mathematical form for $\omega(z)$ has at most three free parameters which can be mapped back to specific archetypal models for dark energy. 
We analyze in detail how different data combinations can constrain the specific cases embedded in our form for $\omega(z)$ and report: the confidence intervals, individual uncertainties, resulting dynamics, and statistical indicators of the goodness of the fit. 
We show that different observational data sets can constrain the parameters and that all cases were good fits to the data. 

With only one free parameter ($n=C=1$, $A$ free), we can parameterize the expansion rate of the variety of models described in \cite{Roy:2018nce}. 
We call this case, Exponential model. With this, we obtain not only a much richer dynamics for dark energy than the simplified  $\omega = \text{const.}\neq-1$ model, but also, a good fit to the data sets we employ, with $\chi^2_{red}=1.764$ in the worst case, and $\chi^2_{red}=0.988$, in the best.

With two free parameters ($n=1$, $A$ and $C$, free), we are able to reproduce the generic expansion rate of minimally coupled scalar fields, such as quintessence. Depending on the sign of $\omega(z)$, the generic behavior of the so-called Phantom models could be described with this subset of parameters. This comprises one of the most explored models for DE, which we can model with the same number of free parameters as in the widely used CPL \cite{Chevallier2001, Linder2003} parameterization.
Our fits to the data are competitive, and both parameters can be simultaneously constrained.  
 
Using a different subset of only two free parameters from our model ($C=0$, $A$, and $n$ free), we are able to mimic the generic expansion rate provided by cosmologically viable $f(R)$ theories of gravity (as shown in \cite{Jaime:2013zwa}, and previously attempted in \cite{Jaime:2018ftn}). 
For this case, in particular, referred to as $f(R)$-like and characterized by an oscillatory behavior for the EoS around $\omega=-1$, we find the best fit of the whole sample, BAO-SNe-CC-CMB, with a $\chi^2_{red}=0.991$.

Equally important is that we can constrain our free parameters in all the cases studied without degeneracy, divergent evolution at high redshift, rapidly oscillatory behavior, or other mathematical misbehaves.

When we let the three parameters vary freely, testing the general form of $\omega(z)$, we can answer which dynamical behavior is favored by observations. In this case, we find, as a result, an EoS which oscillates around the phantom-dividing line, and, with over 99$\%$ of confidence, the cosmological constant solution is disfavored. 

The strength of our proposal lies in its independence from a specific theoretical model.
Hence, even when we argued that the simplest, theoretically-sustained, explanation behind an oscillatory profile for $\omega(z)$, could arise in the context of $f(R)$ theories of gravity, as opposed to a convoluted mixture of scalar fields, a tantalizing alternative to this could come from an unknowingly biased radial selection of the extra-galactic targets in the samples we use. 

We analyze in detail how different combinations of data can constrain the specific cases embedded in our form for $\omega(z)$, and report the confidence intervals, individual uncertainties, resulting dynamics, and statistical indicator of the goodness of our fits, as well as a comparison against the required increase in precision for observations of the cosmic distances to be able to differentiate among particular cases. We find that all cases are good fits to the data. 

It is interesting to note that our best fit values for $H_0$ lie in between the values to be known in tension.

When we performed an agnostic model comparison, we found that all cases included within our proposal are competitive under the Akaike Information Criterion (AIC), and the case of exponential model was as good as the cosmological constant solution within the Bayesian Information Criterion (BIC). We discussed the relevance of an informed model selection. We propose using the dynamical features of $\omega(z)$ as a discerning tool to group and further compare models with a shared physical behavior.

Given that our parameterization is not tied to a specific model, the perturbations should be performed  case-by-case, depending on the physics of the chosen model. Such  analysis is beyond the scope of the present paper, and it will be presented in a posterior work.

To summarise, in this work, we have presented a single equation that can reproduce a variety of well-motivated physical scenarios for cosmic expansion at late times. We probed its adequacy to be implemented to data and aim to provide the community with a simple framework to incorporate physically-motivated models into surveys and clustering analyzes and better link observational phenomena and theoretical hypotheses for testing the nature of cosmic acceleration.

\section*{Acknowledgements}
The authors thank E. Almaraz and M. Rodr\'iguez-Meza for helpful discussions and B. Roukema for helpful suggestions to improve this document. 
GA acknowledges the postdoctoral fellowship from DGAPA-UNAM.
MJ acknowledges the support of the Polish Ministry of Science and Higher Education MNiSW grant DIR/WK/2018/12, as well as the research project “VErTIGO” funded by the National Science Center, Poland, under agreement number 2018/30/E/ST9/00698. 
Part of this work was supported by the ``A next-generation worldwide quantum sensor network with optical atomic clocks'' project, which is carried out within the TEAM IV program of the Foundation for Polish Science co-financed by the European Union under the European Regional Development Fund.
LGJ thanks the financial support of SNI, CONACyT-140630, and the hospitality of the ININ.
GA and LGJ acknowledge the support from PAPIIT IN120620.

\section*{Data Availability}

All the observational data used in this work is of public knowledge.

The cosmic chronometers sample we used can be found in \cite{Farooq:2013hq}, as a compiled table of $z$, $H(z)$, and the related errors, $\sigma_{H(z)}$. 

Our BAO data points can be found in the respective reference. For the six-degree-field galaxy survey (6dFGS) \cite{Beutler:2011hx}, the Sloan Digital Sky Survey Data Release 7 (SDSS DR7) \cite{Ross:2014qpa},  the reconstructed value SDSS(R) presented in \cite{Padmanabhan2pc}, and  the uncorrelated values of the complete BOSS sample SDSS DR12 are reported in \cite{Alam:2016hwk}.  The measurement of the auto and cross-correlation of the Lyman-$\alpha$ Forest (Ly$\alpha$-F) measurements from quasars of the 11th Data Release of the Baryon Oscillation Spectroscopic  (BOSS DR11) can be found in \cite{Delubac:2014aqe, Font-Ribera:2013wce}. 

The compressed CMB likelihood with Planck \emph{TT+lowP} values can be found in \cite{planck15DE}, and we have given the full form of the reduced matrix in section \ref{subsec:data}.

Our chosen Supernovae compilation was Union 2.1, presented in \cite{Suzuki_2012} and which can be downloaded from \texttt{http://supernova.lbl.gov/Union/}.

 Our numerical implementation will be made publicly available in the repository  \texttt{https://github.com/oarodriguez/cosmostat/}, but a version of the code can be shared earlier upon reasonable request to the authors.


\bibliography{bestiario-bib}

\appendix

\section{Propagation of uncertainties in the EoS}\label{app:deltaomega}

From equation (\ref{eq:eos}) we note that $\omega(z)$ is a function of parameters $A$, $n$, and $C$, and of redshift, $z$. From here, it follows that the uncertainty  $\delta\omega(z)$ depends, in the same way, on the parameters, and their uncertainties $\delta A$, $\delta n$, and $\delta C$.

When the uncertainties are independent of each other,  $\delta\omega(z)$ has a broader dispersion around the central point at every $z$.

From the individual errors, $\delta A$, $\delta n$, and $\delta C$, we estimate the propagated uncertainty in the resulting $\omega(z)$, computed as:

\begin{equation}\label{eq:deltaomega}
\delta \omega(z)=(\omega +1)\left[\frac{\delta A}{A}-\frac{\delta C}{(z^n-z-C)}+\frac{z^n\rm{ln}(z)\,\delta n}{(z^n-z-C)}\right].
\end{equation}

We use  equation (\ref{eq:deltaomega}), $\delta \omega(z)$, considering the uncertainties parameters $\delta A$, $\delta n$, and $\delta C$ as independent of $z$.

In the case that the uncertainties depend explicitly on $z$, equation (\ref{eq:deltaomega}) will be an overestimation of $\delta\omega(z)$, which guarantees that the dynamical range for each EoS lies inside our estimated errors. 

Using the uncertainties for $A$, $n$, and $C$, reported in column 5 from table \ref{Table:best-fit-values}, into equation (\ref{eq:deltaomega}) for  $\delta\omega(z)$, we calculated the  99.7$\%$ CL in figures of table \ref{Table:Plots-EOS-dataset}.

\section{Numerical code}\label{app:code}

\begin{figure}[h!]
        \includegraphics[width=\linewidth]{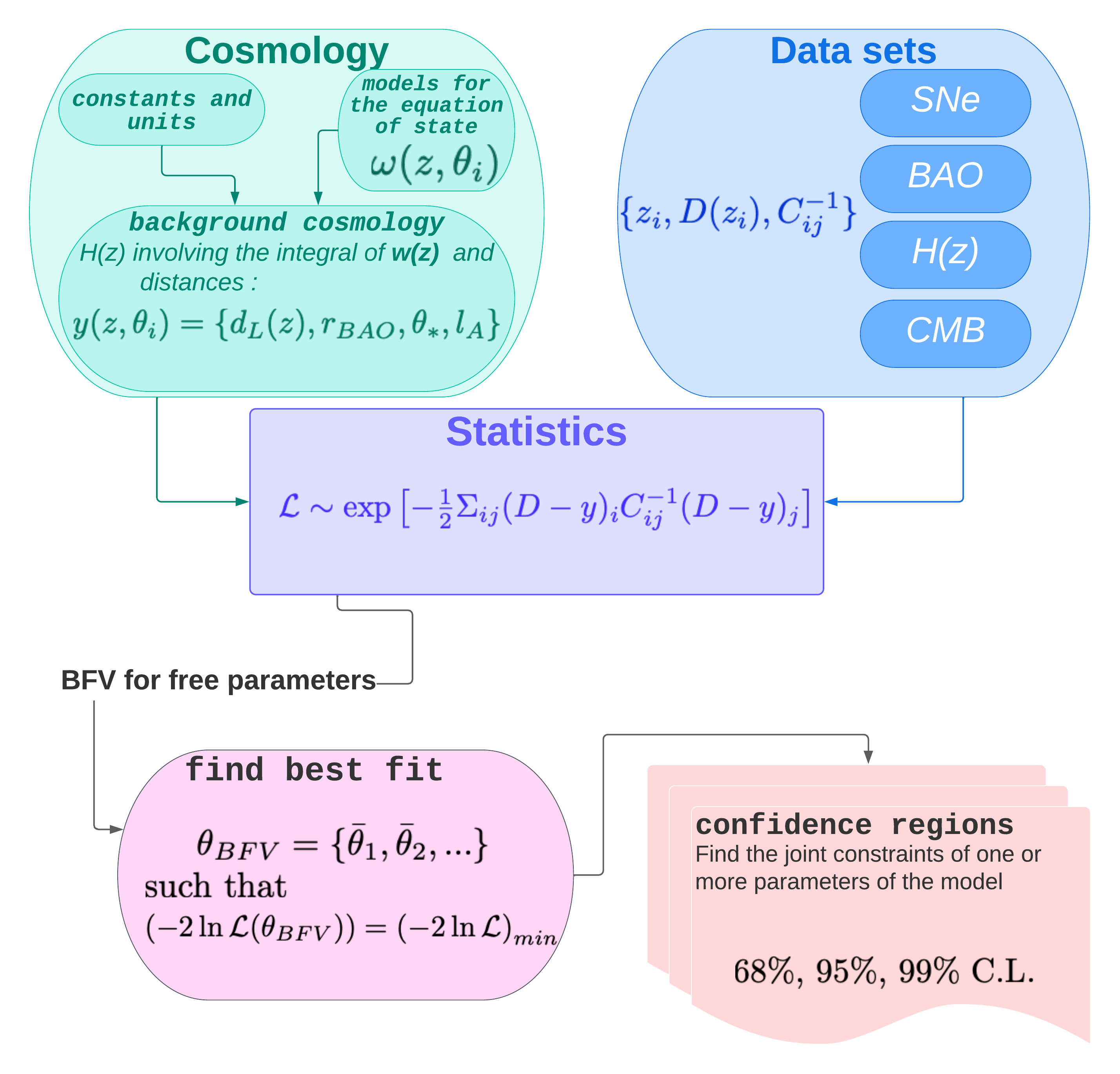}
        \caption{Flux diagram of the code used to produce our results. The code has been structured in a modular way, allowing the user to add models for the equation of state $\omega(z)$, or data sets to the current version. In the core, it performs a numerical optimization of the likelihood function to find the best fitting parameters of each model, at the same time that it returns the AIC and BIC values, according to one or more data sets. Subsequently, it finds the compact regions around the best fit value in parameters space. As customary, we choose to report $68.3\%$, $95.4\%$, and $99.7\%$ confidence levels.}
        \label{fig:code}
\end{figure}

Our statistical treatment is based on the multivariate Gaussian approximation of the likelihood function (see for instance \cite{Heavens:2009nx}), this is, we assume gaussian-distributed errors on the measurements. 
\begin{equation}
    \mathcal{L}(\theta) =\frac{\sqrt{(2\pi)^{-m}}}{|\text{det}\,C|}\exp\left[-\frac{1}{2}\Sigma_{ij}(D-y(\vec{\theta}))_iC^{-1}_{ij}(D-y(\vec{\theta}))_j \right],
\end{equation}
where $C_{ij}$ is the covariance matrix, and $\vec{\theta}$ (of length $m$), are the parameters of the given model. 
Then, our best fit value parameters are those that optimize the merit function, 
\begin{equation}
   -2\ln\mathcal{L}(\vec{\theta}_{bfv}) = \left[-2\ln\mathcal{L}\right]_{min}
\end{equation}

From where we see that $\vec{\theta}_{bfv}$ are those parameters which minimize the $\chi^2$ estimator. Hence, the reduced chi-squared, $\chi^2_{red}=\chi^2_{min}/(d.o.f.)$, is our statistical measure of the goodness of the fit.  

The scheme shown in figure \ref{fig:code} describes  our numerical implementation, based on \texttt{python}.
For optimising the parameters, since it is not possible to simply step through the parameter space (due to the high dimensionality of the problem), we perform instead a constrained optimization on the parameters space, $\vec{\theta}$, and iteratively explore it from a starting point. 
The final point then is taken as the most likely parameter vector, $\vec{\theta}_{BFV}$.
\\
We use a numerical implementation of the Differential Evolution algorithm  included and documented in the \texttt{python scipy} library. 
This algorithm is a stochastic population-based method and it is particularly useful for global optimization problems, such as the one at hand. More details about the algorithm can be found in \cite{diffevol}.
\\
This step is performed in the \texttt{find best fit} part of the code (see the diagram of figure \ref{fig:code}).
\\
The results from this step are then used to estimate the information criteria, AIC, and BIC as defined in \ref{sec:aicbic}.\\
Instead of reporting the full probability distribution of errors, we calculate the confidence regions in parameter space that contain a certain percentage of the total probability distribution around the best fit value. As customary, we choose to report $68.3\%$, $95.4\%$, and $99.7\%$ confidence levels.

\end{document}